\documentclass[11pt]{article}
\pdfoutput=1

\usepackage{cite}
\usepackage{booktabs}
\usepackage[english]{babel}
\usepackage{mathtools}
\usepackage{amsmath,amssymb,amsbsy,amstext, amsthm, simplewick, amsfonts}
\usepackage{hyperref}
\hypersetup{
    colorlinks = true, 
    urlcolor = rosewood, 
    linkcolor = rosewood, 
    citecolor = pyblue 
  }
  
\usepackage{graphicx}
\usepackage{amsfonts}
\usepackage{amssymb}
\usepackage[small]{caption}
\usepackage{upgreek}
\usepackage[svgnames,dvipsnames,x11names,table]{xcolor}
\usepackage{multirow}
\usepackage{geometry}
\usepackage[hang,flushmargin]{footmisc}
\usepackage{bm}
\usepackage{braket}
\usepackage{subcaption}
\usepackage{mathtools}
\usepackage{setspace}
\usepackage{cleveref}
\usepackage{comment}
\usepackage{scalerel}
\usepackage[normalem]{ulem}
\usepackage{slashed}
\usepackage{tensor}
\usepackage{enumitem}
\usepackage{ifthen}
\usepackage{appendix}
\usepackage{floatrow}
\usepackage{siunitx}
\usepackage{framed}
\usepackage{float}
\usepackage{geometry}
\usepackage{fontawesome5}
\usepackage{wrapfig}
\usepackage{stackengine, xcolor}
\usepackage{tikz}
\usetikzlibrary{decorations.markings}
\usetikzlibrary{decorations.pathmorphing}
\usetikzlibrary{shapes.misc}
\usetikzlibrary{arrows}
\usetikzlibrary{shapes.geometric}
\usetikzlibrary{plotmarks}

\usepackage{makecell}

\usepackage{wasysym}

\makeatletter
\g@addto@macro\bfseries{\boldmath}
\makeatother

\numberwithin{equation}{section} 

\renewcommand{\thefootnote}{\fnsymbol{footnote}}

\newcolumntype{M}[1]{>{\centering\arraybackslash}m{#1}}
\definecolor{rosewood}{rgb}{0.4, 0.0, 0.04}
\definecolor{pyblue}{RGB}{31, 119, 180}
\definecolor{pyorange}{RGB}{255, 127, 14}
\definecolor{pygreen}{RGB}{44, 160, 44}
\definecolor{pyred}{RGB}{214, 39, 40}
\definecolor{pypurple}{RGB}{148, 103, 189}
\definecolor{lightgray}{gray}{0.9}

\hypersetup{
    colorlinks=true,
    linkcolor={Purple!80!black},
    citecolor={pyblue},
    urlcolor={Purple!50!black}
}

\usepackage{colortbl}

\makeatletter
\newlength{\apb@width}
\newcommand{\autoparbox}[2][c]{\settowidth{\apb@width}{#2}\parbox[#1]{\apb@width}{#2}}

\makeatother

\usepackage[framemethod=default]{mdframed}
\newmdenv[skipabove=7pt,
skipbelow=7pt,
rightline=false,
leftline=false,
topline=false,
bottomline=false,
backgroundcolor=gray!10,
linecolor=gray,
innerleftmargin=5pt,
innerrightmargin=5pt,
innertopmargin=5pt,
innerbottommargin=5pt,
leftmargin=0cm,
rightmargin=0cm,
linewidth=4pt]{eBox}

\usepackage[most]{tcolorbox}
\tcbset{colback=white, colframe=black,
        highlight math style= {enhanced, 
            colframe=red,colback=red!10!white,boxsep=0pt}
        }

\crefname{table}{Table}{Tables}
\crefname{equation}{Eq.}{Eqs.}
\crefname{appendix}{App.}{Apps.}
\crefname{section}{Section}{Secs.}
\crefname{figure}{Fig.}{Figs.}

\def \A {\mathcal{A}}
\def \B {\mathcal{B}}
\def \C {\mathcal{C}}
\def \D {\mathcal{D}}

\def \F {\mathcal{F}}
\def \G {\mathcal{G}}
\def \H {\mathcal{H}}

\def \N {\mathcal{N}}
\def \O {\mathcal{O}}
\def \P {\mathcal{P}}

\def \S {\mathcal{S}}
\def \T {\mathcal{T}}

\def \d {\mathrm{d}}
\def \k {\bm{k}}
\def \x {\bm{x}}
\def \kp {k_{\rm p}}

\def \Res{\, \mathrm{Res} \,}
\def \Re{\, \mathrm{Re} \,}
\def \Im{\, \mathrm{Im} \,}
\def \arg {\text{arg}}

\def \arccosh{\, \mathrm{arccosh} \,}
\def \arctanh{\, \mathrm{arctanh} \,}
\def \arccoth{\, \mathrm{arccoth} \,}

\def \aa{\mathsf{a}}
\def \bb{\mathsf{b}}

\def \fnl {f_{\text{NL}}}

\def\pmm{\mathbin{\ensurestackMath{\abovebaseline[-3.4pt]{%
  \stackunder[-3.5pt]{\color{pyblue}+}{\color{pyred}-}}}}}

\def\ppm{\mathbin{\ensurestackMath{\abovebaseline[-3.4pt]{%
  \stackunder[-3.5pt]{\color{pyred}+}{\color{pyblue}-}}}}}

\newmuskip\pFqmuskip
\newcommand*\pFq[6][8]{%
  \begingroup 
  \pFqmuskip=#1mu\relax
  \mathcode`\,=\string"8000
  \begingroup\lccode`\~=`\,
  \lowercase{\endgroup\let~}\pFqcomma
  {}_{#2}F_{#3}{\left[\genfrac..{0pt}{}{#4}{#5};#6\right]}%
  \endgroup
}
\newcommand{\pFqcomma}{\mskip\pFqmuskip}

\newcommand*\ptFq[6][8]{%
  \begingroup 
  \pFqmuskip=#1mu\relax
  \mathcode`\,=\string"8000
  \begingroup\lccode`\~=`\,
  \lowercase{\endgroup\let~}\pFqcomma
  {}_{#2}\tilde{F}_{#3}{\left[\genfrac..{0pt}{}{#4}{#5};#6\right]}%
  \endgroup
}

\newcommand*\pcalFq[6][8]{%
  \begingroup 
  \pFqmuskip=#1mu\relax
  \mathcode`\,=\string"8000
  \begingroup\lccode`\~=`\,
  \lowercase{\endgroup\let~}\pFqcomma
  {}_{#2}\mathcal{F}_{#3}{\left[\genfrac..{0pt}{}{#4}{#5};#6\right]}%
  \endgroup
}

\newcommand*\pregFq[6][8]{%
	\begingroup 
	\pFqmuskip=#1mu\relax
	\mathcode`\,=\string"8000
	\begingroup\lccode`\~=`\,
	\lowercase{\endgroup\let~}\pFqcomma
	{}_{#2}\tilde{F}_{#3}{\left[\genfrac..{0pt}{}{#4}{#5};#6\right]}%
	\endgroup
}

\def\pmm{\mathbin{\ensurestackMath{\abovebaseline[-3.4pt]{%
  \stackunder[-3.5pt]{\color{pyblue}+}{\color{pyred}-}}}}}

\def\ppm{\mathbin{\ensurestackMath{\abovebaseline[-3.4pt]{%
  \stackunder[-3.5pt]{\color{pyred}+}{\color{pyblue}-}}}}}

\allowdisplaybreaks[1]
\begin{document}

\begin{titlepage}
\setcounter{page}{1} \baselineskip=15.5pt
\thispagestyle{empty}
$\quad$
\vskip 0 pt

\vspace*{0cm}

\begin{center}
{\fontsize{18}{18} \bf The Exact and Approximate Tales of}\\[10pt] 
{\fontsize{18}{18} \bf Boost-Breaking Cosmological Correlators}
\end{center}

\vskip 12pt
\begin{center}
\noindent
{\fontsize{12}{18}\selectfont 
Zhehan Qin,$^{1, 2}$ 
Sébastien Renaux-Petel,$^3$ 
Xi Tong,$^2$ 
\\[5pt]
Denis Werth,$^3$ 
and 
Yuhang Zhu$^4$ 
}
\end{center}

\begin{center}
$^1$ \textit{Department of Physics, Tsinghua University, Beijing 100084, China} \\[5pt]
$^2$ \textit{Department of Applied Mathematics and Theoretical Physics,\\ Cambridge University, Cambridge, CB3 0WA, UK} \\[5pt]
$^3$\textit{Institut d'Astrophysique de Paris, Sorbonne Université, CNRS, Paris, F-75014, France} \\[5pt]
$^4$ \textit{Cosmology, Gravity and Astroparticle Physics Group,\\
Center for Theoretical Physics of the Universe,\\
Institute for Basic Science, Daejeon 34126, Korea}
\end{center}

\vspace{0.2cm}
\begin{center}{\bf Abstract}
\end{center}

Cosmological correlators offer a remarkable window into the high-energy physics governing Universe's earliest moments, with the tantalising prospect of discovering new particles. However, extracting new physics from these observables requires both precise theoretical predictions of inflationary theories and accurate, analytical templates suitable for data analysis throughout parameter and kinematic spaces. In this paper, we extend the current analytical results by computing the most general boost-breaking seed correlator mediated by the tree-level exchange of a massive spinning particle. We derive the result using two complementary approaches, bootstrapping from boundary differential equations, and direct spectral integration. Both representations are packaged as a single partially resummed series that converges in all physical kinematics. Computing this correlator marks a milestone for carving out the space of all boost-breaking correlators, and therefore completes the tree-level catalogue. We then introduce a general procedure to obtain accurate approximations for cosmological collider signals based on the saddle-point method. This approach allows for a clear physical intuition of various signals hidden in correlators, as the bulk physics is made manifest through the location of these saddles in the complex time plane, which depend on the external kinematics. Evaluating the time integrals at these saddles yields results given as elementary functions that remain valid beyond soft limits and provide intuitive control over both the signal shape and amplitude. We demonstrate the power of this method in both de Sitter-invariant and boost-breaking scenarios, and uncover novel refined waveform and strength dependence for oscillatory signals from massive fields. We provide a complete cosmological collider shape template capturing all boost-breaking effects for upcoming cosmological surveys.
\end{titlepage}

\renewcommand*{\thefootnote}{\arabic{footnote}}
\setcounter{footnote}{0}

\setcounter{page}{2}

\restoregeometry

\begin{spacing}{1.2}
\newpage
\setcounter{tocdepth}{3}
\tableofcontents
\end{spacing}

\setstretch{1.1}
\newpage

\section{Introduction}

The early Universe serves as a unique laboratory for testing the fundamental laws of Nature at energy scales far beyond the reach of terrestrial experiments. In particular, by magnifying quantum fluctuations from microscopic to macroscopic scales, cosmic inflation allows to probe quantum physics through cosmological observations. Among the most promising avenues for uncovering new physics are higher-point cosmological correlators of primordial fluctuations, commonly referred to as primordial non-Gaussianities. These detection channels are a central target of current and next-generation experiments, including observations of the cosmic microwave background and large-scale structure surveys, see~\cite{Meerburg:2019qqi, Achucarro:2022qrl} for reviews and the references therein. 

\vskip 4pt
A close interplay has emerged between theoretical high-energy physics, which aims to compute and classify the space of cosmological correlators, and observational cosmology, which seeks to constrain them with increasing precision. This synergy is embodied in the ``cosmological collider program''~\cite{Chen:2009we,Chen:2009zp, Baumann:2011nk,Noumi:2012vr, Arkani-Hamed:2015bza}, 
see also~\cite{Gong:2013sma, Chen:2015lza,Chen:2016uwp,Lee:2016vti,An:2017hlx,Lu:2019tjj,Hook:2019zxa,Liu:2019fag, Kumar:2019ebj,Wang:2019gbi,Wang:2020ioa,Wang:2020uic,Pinol:2021aun,Cui:2021iie,Reece:2022soh,Chen:2022vzh,Qin:2022lva,Werth:2023pfl,Jazayeri:2023xcj,Jazayeri:2023kji,Pinol:2023oux,Ema:2023dxm,Tong:2023krn,Chakraborty:2023eoq,Werth:2024aui,Aoki:2024jha,Wu:2024wti,Craig:2024qgy,McCulloch:2024hiz,Cabass:2024wob,Sohn:2024xzd,Pajer:2024ckd,Wang:2025qww}. Much like in ground-based colliders, where particle interactions are inferred from scattering data, cosmological correlators encode the signatures of particles that were spontaneously produced during inflation and decayed into curvature perturbations. Since the dynamics of the inflationary bulk is ultimately encoded in the spatial patterns of density fluctuations observed at late times, analysing the kinematic dependence of cosmological correlators enables a form of particle spectroscopy---revealing the masses, spins, dispersion relations and couplings of particles that were active in the primordial Universe. 

\vskip 4pt
To advance this program, it is essential to develop predictions for cosmological correlators that are {\it precise}, {\it efficient}, and {\it complete}, in the sense of covering wide regions of parameter space and being valid across all kinematic regimes. In this paper, we address these needs by completing the catalogue of boost-breaking cosmological correlators and by developing a systematic framework for constructing approximate solutions. Our approach ultimately provides refined templates for cosmological collider signals, facilitating their use in data analysis. 

\subsection{Exact Boost-Breaking Catalogue}

Significant progress has been made in our theoretical understanding of cosmological correlators, with exact analytical results now available for a wide range of physical processes. This progress has been driven by the development of powerful computational techniques, including solving boundary differential equations in kinematic space~\cite{Arkani-Hamed:2018kmz, Pimentel:2022fsc, Jazayeri:2022kjy, Qin:2022fbv, Qin:2023ejc, Aoki:2024uyi, Liu:2024str,Chen:2024glu}, the use of Mellin-space representations of massive mode functions exploiting approximate dilatation invariance~\cite{Sleight:2019mgd, Sleight:2019hfp, Qin:2022lva, Qin:2022fbv, Qin:2023bjk,Qin:2023nhv, Xianyu:2023ytd, Qin:2024gtr}, as well as spectral or dispersive methods~\cite{Xianyu:2022jwk, Werth:2024mjg, Liu:2024xyi}. 
As a result, we now have access to fully analytical expressions for the shape of some cosmological correlators across all kinematic configurations. Nevertheless, carving out most of the theory space achieving a systematic classification of the full range of possible signatures remains a major challenge.

\vskip 4pt
Building novel computational techniques has been closely guided by symmetries. A key insight is that, to first order in the mixing with dynamical gravity, inflationary fluctuations can be well approximated by weakly coupled quantum fields in a fixed (quasi) de Sitter background. Exploiting the full de Sitter symmetry group---importantly including de Sitter boosts---are known to completely fix many correlators and enabled the derivation of the first exact, closed-form solutions for cosmological correlators involving massive particle exchange~\cite{Maldacena:2011nz, Creminelli:2011mw, Kehagias:2012pd, Mata:2012bx, Ghosh:2014kba, Kundu:2014gxa, Arkani-Hamed:2015bza, Arkani-Hamed:2018kmz}. These developments closely parallel how conformal symmetry constrains correlators in conformal field theories, and how Poincaré symmetry shapes scattering amplitudes in flat space. Introducing a mild breaking of boost symmetry allowed these results to be extended to slow-roll inflationary scenarios, leading to predictions for primordial non-Gaussianities with amplitudes comparable to the gravitational floor in equilateral configurations, $\fnl^{\rm eq} = \mathcal{O}(\epsilon, \eta)$, where $\epsilon, \eta \sim 10^{-2}$ denote the slow-roll parameters~\cite{Maldacena:2002vr}. Notably, this predicts a cosmological collider signal lying in soft limits $k_L/k_S \ll1$---where $k_L$ (resp.~$k_S$) denotes some long (resp.~short) wavelength momentum---whose amplitude is highly suppressed i.e. $\fnl^{\rm CC} = \O(e^{-\pi\mu})$, where $\mu$ denotes the mass of the exchanged heavy particle in Hubble units.\footnote{This is sometimes called a Boltzmann suppression since the Boltzmann factor corresponding to a de Sitter space of Gibbons-Hawking temperature $T=H/(2\pi)$ is $e^{-m/T}\approx e^{-2\pi \mu}$, where $\mu^2=m^2/H^2-9/4$.} However from observations, we expect primordial fluctuations to be statistically homogeneous, isotropic and approximately scale-invariant, but not necessarily de Sitter boost-invariant. In fact, we generally expect the rest frame of the inflaton background to single out a preferred direction of time and break de Sitter boosts. This distinction is crucial: phenomenologically, the size of primordial non-Gaussianities is directly related to the amount of symmetry breaking, and rightfully relaxing certain symmetries not only enlarges the space of inflationary models, but also opens up new regions of kinematic space. Since the rolling of the inflaton background automatically breaks boosts, the most natural way to introduce the boost-breaking effect is starting from a de Sitter-covariant theory, and taking (some of) the inflaton fields to the background value, inducing operators that break boosts. While these boost-breaking operators could be treated perturbatively if the theory is weakly coupled, more nontrivial effects arise when they become non-perturbative. In particular, if these operators are quadratic, resumming them will in general modify the dispersion relation of the field content.

\vskip 4pt
To second-order in (spatial) derivatives in the linear theory, all possible boost-breaking effects can be captured by two key parameters: (i) a non-unit speed of sound $c_s \neq 1$, which distinguishes the sound cone from the light cone, and (ii) a non-zero helical chemical potential $\kappa \neq 0$, which arises when coupling matter fields through the rolling inflaton background. These parameters modify the dispersion relation of the massless field and the massive particle, which schematically take the form $\omega_\varphi^2 \sim c_s^2 \kp^2$ and
$\omega_\sigma^2 \sim \kp^2 + \kappa \kp + m^2$, with $\kp\equiv k/a$ being the physical (redshifting) momentum, and $m$ the mass of the heavy field.\footnote{Without loss of generality for our purposes, we rescaled the sound speed of the heavy field to unity. However, note that a non-unit sound speed necessarily comes with cubic interactions with the massless Goldstone field, with important observational consequences in general, see~\cite{Pinol:2023oux}.} Taken individually, the effects of these boost-breaking parameters are well understood. On one hand, a reduced sound speed $c_s<1$ for the curvature perturbation enhances equilateral-type non-Gaussianities $\fnl^{\rm eq} = \O(c_s^{-2})$~\cite{Alishahiha:2004eh ,Chen:2006nt,Cheung:2007st}, and amplifies the cosmological collider signal, $\fnl^{\rm CC} = \O(e^{-\pi\mu/2})$ in ultra-soft configurations $k_L/k_S \ll c_s$~\cite{Lee:2016vti,Jazayeri:2023xcj}. Notably, strong boost breaking via subluminal sound speed allows for the novel mildly-soft kinematic region ($k_L/k_S \approx c_s m/H$) to emerge, where a new cosmological low-speed collider resonance appears~\cite{Jazayeri:2022kjy,Jazayeri:2023xcj}. Closed-form solutions for the corresponding cosmological correlators were derived in~\cite{Pimentel:2022fsc, Jazayeri:2022kjy}. On the other hand, a non-zero chemical potential is known to enhance particle production~\cite{ Sou:2021juh,Adshead:2015kza}, see also~\cite{Chua:2018dqh,Tong:2022cdz,Bodas:2020yho,Bodas:2024hih,Wang:2019gbi,Wang:2020ioa,Qin:2022fbv,Qin:2022lva,Chen:2018xck,Hook:2019zxa,Chen:2023txq,Tong:2023krn,An:2025mdb}. A direct consequence is the exponential amplification of the cosmological collider signal $\fnl^{\rm CC} = \O(e^{-\pi(\mu-\kappa)})$~\cite{Wang:2019gbi}. The chemical potential is sensitive to the helical structure of the exchanged massive spinning field and therefore also leads to parity violation in cosmological correlators~\cite{Liu:2019fag, Jazayeri:2023kji, Stefanyszyn:2023qov}. The exact analytical form of helical cosmological correlators have been derived in~\cite{Qin:2022fbv,Qin:2023ejc}.

\vskip 4pt
Here, we combine the leading boost-breaking effects and compute the four-point seed correlator of external legs having a reduced sound speed, mediated by the tree-level exchange of massive and helical spinning particles. We present two complementary representations of this correlator: one obtained by the bootstrap approach, and the other obtained by extending the recently developed spectral method. Both techniques yield partially resummed single-series solutions that converge across all kinematic configurations. This result completes the tree-level catalogue of boost-breaking cosmological correlators.\footnote{While we mainly focus on the trispectrum, it is convenient to derive the bispectrum, which is more observationally relevant, by taking suitable soft limits.}

\subsection{Saddling Up for Approximate Correlators}

Even at tree level, cosmological correlators involving massive particle exchange are notoriously complex. Their exact expressions typically involve hypergeometric functions and infinite series that must be analytically continued to cover the full kinematic domain. Despite their generality and precision, these analytical solutions are nevertheless complicated to manipulate and the physical picture often remains concealed by obscure mathematics. 
For instance, it remains conceptually unclear why the amplitudes of cosmological collider signals are always proportional to certain fractional powers of the characteristic Boltzmann factor, i.e. $f_{\rm NL}^{\rm CC}=\mathcal{O}(e^{-\pi\mu/n})$, with $n=1,2$.
More critically, these expressions pose serious challenges for practical applications. Accurate numerical evaluation is computationally expensive, limiting their use in data analysis pipelines. Two main obstacles hinder the detection and constraint of cosmological collider signals: (i) the {\it exact} waveforms for the oscillatory features from massive particle production involve complicated special functions that are slow to evaluate compared to elementary functions; (ii) the portion of the signal that is observationally accessible lies in the mildly-soft regime---as realistic observations cannot probe arbitrarily soft modes due to cosmic variance---where it is dwarfed by large equilateral contaminations from curvature perturbation self-interactions or more generally effective field theory contributions. These ``background signals'' must be subtracted with high precision, yet they too are typically expressed as slowly converging infinite series of special functions.

\begin{figure}[h!]
    \centering
    \includegraphics[width=0.8\linewidth]{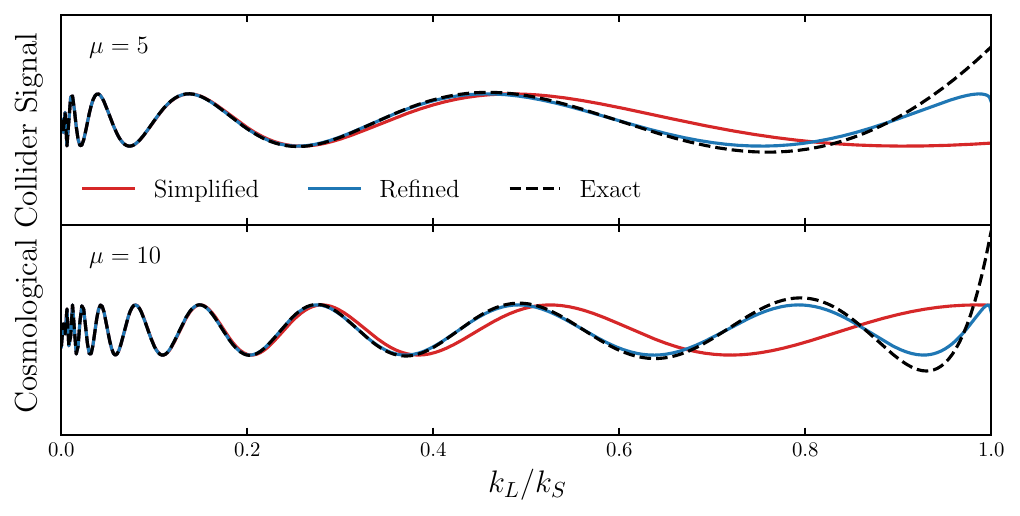}
   \caption{Illustration of the simplified, refined, and exact cosmological collider waveform templates (see Tab.~\ref{tab: cosmological collider signals}) as function of the momentum ratio $k_L/k_S$ in the soft limit, varying the mass parameter.}
  \label{fig: Template Intro}
\end{figure}

\vskip 4pt
Fortunately, the signature of massive particles in cosmological correlators are among the most transparent. Because particles of different de Broglie wavelengths born and decay at different times during inflation, their dynamics becomes imprinted in specific kinematic configurations of late-time correlators. In particular, the soft limit isolates a regime where the massive mode can be expanded at late times: it oscillates with both positive and negative frequency components due to particle production while being redshifted by the expanding background. This clear physical picture enables the construction of {\it simplified} waveform templates for cosmological collider signals, as illustrated in Fig.~\ref{fig: Template Intro}. These signals typically oscillate as a function of the logarithm of the momentum ratio, with a frequency set by the particle’s mass: $(k_L/k_S)^{+i\mu} + {\rm c.c.} \sim \cos[\mu\log(k_L/k_S)]$, providing a striking observational signature of heavy fields during inflation. Yet, such a simple template is only valid in the ultra-soft limit due to the limitation of late-time approximation of the mode functions. As one moves away from the ultra-soft limit, significant dephasing error starts to appear, as shown in Fig.~\ref{fig: Template Intro}.

\vskip 4pt
In search for a clearer bulk physics interpretation and better boundary templates, in this paper, we shall introduce the saddle point method as a powerful tool to classify and gain physical insight into various cosmological collider signals. We show that the production and decay moments correspond to resonant saddles in the time integrals of cosmological correlators. This means that most of the signal is captured by evaluating the time integrals on these saddle points, yielding an accurate expression for the waveform. This approach relies solely on analysing the location of saddle points in the complex time plane, providing a natural explanation for the behaviour of these signals by tracking the motion of these saddles driven by external kinematics. Surprisingly, we find the characteristic rational powers of the Boltzmann factor for cosmological collider signals are solely determined by the arc angle $\theta_n$ touring around the zero of time to reach the complex saddle point, i.e. $f_{\rm NL}^{\rm CC}=\mathcal{O}(e^{-\theta_n \mu})$, $\theta_n=\pi/n$, with $n=1,2$. By evaluating the time integrals using the saddle point method, we obtain approximate solutions for a range of correlators, expressed entirely in terms of elementary functions. These simplified expressions are not only easy to implement but also offer practical {\it refined} waveform templates for observational searches. Comparison with other templates is summarised in Tab.~\ref{tab: cosmological collider signals} and illustrated in Fig.~\ref{fig: Template Intro}. The novel templates are valid across almost all kinematic and parameter space, reaching a perfect balance between precision, efficiency and completeness.

\begin{table}[h!]
\centering
\renewcommand{\arraystretch}{2} 
\setlength{\tabcolsep}{10pt} 
\begin{tabular}{|>{\columncolor[gray]{0.9}}>{\centering}p{2cm} | c | c | c |} 
 \hline 
 \rowcolor[gray]{0.9}
   & {\bf Simplified} & {\bf Refined} & {\bf Exact} \\
 \hline
 {\bf Template} & 
 $\cos\left[\mu\, \log (k_L/k_S)\right]$ & $\cos\left[\mu \arccosh (k_S/k_L)\right]$ & $\pFq{2}{1}{\tfrac{1}{2}-i\mu, \tfrac{1}{2}+i\mu}{1}{\frac{1-k_S/k_L}{2}}$
  \\
  {\bf Method} & late-time expansion & saddle point & exact evaluation \\
  {\bf Validity} & $k_L/k_S \ll 1$ & $k_L/k_S \lesssim 1$ & $k_L/k_S \leq 1$ \\
 {\bf Efficiency} &  {\LARGE \textcolor{pygreen}{\smiley}} & {\LARGE \textcolor{pygreen}{\smiley}} & {\LARGE \textcolor{pyred}{\frownie}} \\
 {\bf Insight} & {\LARGE \textcolor{pygreen}{\smiley}} & {\LARGE \textcolor{pygreen}{\smiley}} & {\LARGE \textcolor{pyred}{\frownie}} \\
 \hline
\end{tabular}
\caption{Summary of the key properties---including the template schematic form, kinematic regime of validity, computational cost, and physical intuition---for the three levels of complexity of templates describing the de Sitter invariant cosmological collider signal in the soft limit of cosmological correlators. The kinematic dependence of each template schematic form is expressed in terms of $k_L/k_S$, where $k_L$ represents a long-wavelength mode and $k_S$ a short-wavelength mode. In this work, we propose complete boost-breaking versions of these templates.}
\label{tab: cosmological collider signals}
\end{table}

\subsection{Summary of Main Results}

For the reader's convenience, our main results are summarised below:

\begin{itemize}
    \item \textit{Exact tale}: We provide the exact closed-form solutions for the helical boost-breaking seed correlator mediated by the tree-level exchange of massive and spinning particles. This takes into account the chemical potential effect for the massive mode, and a reduced sound speed for external legs. The correlator is derived using two complementary approaches---the bootstrap and the spectral methods---and is expressed as a single, partially resummed series that converges rapidly across the entire kinematic domain. Both representations are complementary in terms of computational cost and convergence.
    \item \textit{Approximate tale}: We introduce a novel general method for constructing approximate solutions to cosmological collider signals, detailed in Sec.~\ref{subsec: Extracting and Approximating Non-Analyticities}. Based on the resonant production/decay picture of bulk evolution, this approach combines a Wentzel–Kramers–Brillouin (WKB) approximation for the massive internal mode with a saddle-point evaluation of the time integrals, providing an efficient and physically transparent way to capture and classify the non-analytic features of correlators associated with particle production.
    \item \textit{Phenomenology}: Using the saddle-point approach, we uncover several key phenomenological features arising from boost symmetry breaking, which we summarize below:
        \begin{itemize}
            \item We find a novel cosmological collider waveform template that extends up to equilateral-type configurations. At the level of the bispectrum dimensionless shape function, it takes the form
            \begin{equation}
                S^{\rm CC} \sim \cos\left[\mu \arccosh(k_S/k_L) + \delta\right]\,.
            \end{equation}
            \item In the case of a reduced sound speed, we find the following novel interpolated expression for the amplitude of the cosmological collider signal
            \begin{equation}
                \fnl^{\rm CC} \sim \exp\left[-\pi\mu/2 - \mu \arcsin c_s\right]\,,
            \end{equation}
            that interpolates between the usual de Sitter invariant case ($e^{-\pi\mu}$), and the ultra-reduced sound speed ($e^{-\pi\mu/2}$).
            \item In the presence of a non-zero chemical potential, we highlight
            a novel signature dubbed {\it transient cosmological collider signal}, which takes the form
            \begin{equation}
                S^{\rm CC} \sim \cos\left[c_s \kappa \, k_L/k_S + \delta\right]\,,
            \end{equation}
            oscillating linearly in the momentum ratio with a frequency fixed by the sound speed and the chemical potential.
            \item We provide a complete refined template for the bispectrum shape capturing all boost-breaking effects in~\eqref{eq: bispectrum shape boost-breaking template}.
        \end{itemize}
\end{itemize}

\begin{center}
    ***
\end{center}

\paragraph{Outline.} The outline of the paper is as follows: In Sec.~\ref{sec: Bootstrapping Boost-Breaking Cosmological Correlators}, we start by briefly reviewing the physics of broken boosts, and introduce the boost-breaking helical seed correlator (Sec.~\ref{subsec: Boost-Breaking Physics & Helical Seed Correlator}). We then compute this correlator using two different approaches: (i) the bootstrap method (Sec.~\ref{subsec: Via Boundary Kinematic Differential Equation}), and (ii) the spectral approach (Sec.~\ref{subsec: Via Spectral Representation}). Both results are packaged into a single partially resummed series that converges in all physical kinematic configurations. Eventually, we benchmark both representations in terms of convergence rate, computational cost, and analytic continuation. In Sec.~\ref{sec: Classifying non-Analyticities}, we introduce the saddle point method supplemented by the WKB expansion to obtain approximate yet remarkably accurate expressions for the non-analytic signals hidden in soft limits of cosmological correlators (Sec.~\ref{subsec: Extracting and Approximating Non-Analyticities}). We then apply this method to a number of concrete cases with increasing complexity (Sec.~\ref{subsec: Application to Tree-Level Exchange Correlators}), and find refined simple templates for cosmological collider signals. Finally, we state our conclusions and discuss future directions in Sec.~\ref{sec: Conclusions}. Appendix~\ref{app: Mathematical Interlude} provides mathematical details about multi-valued complex functions and Riemann surfaces, along with a pedagogical introduction to the saddle point method for complex integrals. It includes detailed computations, whose results are referenced in the main text, ensuring the paper remains self-contained.

\paragraph{Notation and Conventions.} We use the mostly-plus signature for the metric $(-, +, +, +)$. Throughout most of the paper, we work in the Poincaré patch of de Sitter space in cosmic time $t$ and conformal time $\tau$ with the metric
\begin{align}
    \d s^2 =-\d t^2+a^2(t)\d \bm{x}^2= a^2(\tau)(-\d\tau^2+\d \bm{x}^2)~,
\end{align}
where $a(t)=e^{Ht}=a(\tau) = -(H\tau)^{-1}$ is the scale factor and $H$ is the Hubble parameter. To avoid the cluttering of symbols, we will often set the Hubble scale to unity i.e. $H=1$. Spatial three-dimensional vectors are written in boldface $\bm{k}$ (with $\hat{\k}\equiv \k/|\k|$), and spatial indices are denoted with Latin letters $i, j, \ldots$. Removal of the boldface denotes taking the magnitude i.e. $k=|\bm k|$. Momenta in plain font with multiple labels denote their sum, e.g. $k_{12}\equiv k_1+k_2$ and $k_{34}\equiv k_3+k_4$. The cross-ratios of momenta are defined by
\begin{equation}
    u \equiv \frac{s}{c_sk_{12}}\,, \quad v \equiv \frac{s}{c_sk_{34}}\,,
\end{equation}
and
\begin{equation}
    \tilde{u} \equiv \frac{2s}{c_sk_{12}+s} \,, \quad \tilde{v} \equiv \frac{2s}{c_sk_{34}+s} \,,
\end{equation}
with $\bm s\equiv \bm{k}_1+\bm{k}_2$ the $s$-channel momentum and $s$ its magnitude. We use \textsf{sans serif} letters $(\sf{a, b, \ldots})$ for Schwinger-Keldysh indices. A prime on correlators denotes the stripping of the momentum conserving $\delta$-function,
\begin{align}
    \left\langle\mathcal{O}(\{\bm k\})\right\rangle\equiv \left\langle\mathcal{O}(\{\bm k\})\right\rangle' \, (2\pi)^3\delta^3\left(\sum \bm k\right)~.
\end{align}
Additional definitions will be introduced when needed in the main text.

\section{Bootstrapping Boost-Breaking Cosmological Correlators}
\label{sec: Bootstrapping Boost-Breaking Cosmological Correlators}

The central object of our interest is the boost-breaking four-point seed correlator that corresponds to massive-particle exchange at tree level. In this section, we bootstrap this seed using two different approaches. First, we solve the boundary differential equation in external kinematics satisfied by the seed correlator and fix unconstrained coefficients using physics inputs---in our case, by matching the internal soft limit. Second, we leverage the recently developed spectral representation method to compute this seed. Both methods yield a single-series representation that converges in all physical kinematic configurations. We eventually compare both representations in terms of evaluation speed and convergence.

\subsection{Boost-Breaking Physics \emph{\&} Helical Seed Correlator}
\label{subsec: Boost-Breaking Physics & Helical Seed Correlator}

We begin by introducing the physics of broken boosts. In this work, we allow external fields to have a reduced speed of sound $c_s < 1$, and consider the exchange of a massive (spinning) particle that contains linear terms in its dispersion relation $\omega^2 \supset \kp$, where $\kp \equiv k/a$ is the physical momentum. This is achieved by coupling the massive field to a helical chemical potential, whose physical origin we review here. We end this subsection by introducing the helical seed correlator, which is the most general boost-breaking correlator that contains all possible boost-breaking effects up to the second order in (spatial) derivatives. 

\paragraph{Spinning helical fields.} Massive spinning particles during inflation can be described and classified based on the unbroken (three-dimensional) rotational symmetry group~\cite{Bordin:2018pca}, see also~\cite{Stefanyszyn:2023qov} for further details. This approach is particularly well-suited for studying the effects of boost-breaking, providing a more general framework than classifications based on the full de Sitter isometry group, as considered in~\cite{Lee:2016vti}. Massive spinning fields with integer spin $S$ are defined as totally-symmetric and traceless 3-tensors, $\sigma_{i_1 \ldots i_S}$ with $i_1, \ldots, i_S$ running over all spatial dimensions. This object contains ${S+2 \choose 2} - {S \choose 2}=2S+1$ degrees of freedom. The most general quadratic action up to two derivatives is
\begin{equation}
    \begin{aligned}
        S = \frac{1}{2S!} \int \d t\d^3x a^3 &\left[\dot{\sigma}_{i_1 \ldots i_S}^2 - c^2 \frac{(\partial_j \sigma_{i_1 \ldots i_S})^2}{a^2} - \delta c^2 \frac{(\partial_j \sigma_{j i_2 \ldots i_S})^2}{a^2} \right.\\
        &\left.- m^2 \sigma_{i_1 \ldots i_S}^2 - 2S \xi \epsilon_{ijk} \sigma_{il_2 \ldots l_S}\frac{\partial_j}{a}\sigma_{k l_2 \ldots l_S}\right] \,,
    \end{aligned}
\end{equation}
where we have introduced the four free parameters $c, \delta c, m^2$ and $\xi$, which respectively correspond to two sound speeds, the (squared) mass, and the chemical potential. Note that the spatial indices of $\sigma_{i_1 \ldots i_S}$ can be raised with $\delta_{ij}$ as this object lives on the flat constant-time hypersurface. The field $\sigma_{i_1 \ldots i_S}(t, \x)$ can be expanded into its different helicity components in Fourier space
\begin{equation}
    \sigma_{i_1 \ldots i_S}(t, \x) = \int \frac{\d^3k}{(2\pi)^3} \sigma_{i_1 \ldots i_S}(t, \k) e^{i\k \cdot \x}\,, \quad \text{with} \quad \sigma_{i_1 \ldots i_S}(t, \k) = \sum_{\lambda=-S}^S \sigma^\lambda_k(t) \, \varepsilon_{i_1 \ldots i_S}^\lambda(\hat{\k})\,,
\end{equation}
where $\sigma^\lambda_k(t)$ are the mode functions and $\varepsilon^\lambda_{i_1 \ldots i_S}(\hat{\k})$ are the polarisation tensors satisfying the following reality of the field in real space condition and the normalisation choice
\begin{equation}
    [\varepsilon^\lambda_{i_1 \ldots i_S}(\hat{\k})]^* = \varepsilon^\lambda_{i_1 \ldots i_S}(-\hat{\k})\,, \quad \sum_{i_1 \ldots i_S} \varepsilon^\lambda_{i_1 \ldots i_S}(\hat{\k})[\varepsilon^{\lambda'}_{i_1 \ldots i_S}(\hat{\k})]^* = S! \, \delta_{\lambda \lambda'}\,.
\end{equation}
The longitudinal polarisation tensor $\bm{\varepsilon}^0(\hat{\k})$ is a function of $\hat{\k}$ only, whereas the (partially) transverse tensors $\bm{\varepsilon}^\lambda(\hat{\k},\hat{\bm e}^\pm)$ with $\lambda\neq 0$ in addition depend on the transverse polarisation vectors
\begin{align}
\hat{\bm e}^\pm=\frac{\hat{\bm{n}}-(\hat{\bm{n}}\cdot\hat{\bm{k}})\hat{\bm{k}}\pm i \,\hat{\bm{k}}\times\hat{\bm{n}}}{\sqrt{2[1-(\hat{\bm{n}}\cdot\hat{\bm{k}})^2]}}~,
\end{align}
with $\hat{\bm n}$ an arbitrary unit reference vector satisfying $\hat{\bm n}\times \hat{\bm k}\neq 0$. The transverse polarisation vectors are constructed such that $\hat{\bm k}\cdot \hat{\bm e}^\pm=0$. 
Note that the fully transverse tensors depend only on the transverse vectors i.e.  $\bm{\varepsilon}^{\pm S}(\hat{\k},\hat{\bm e}^\pm)=\bm{\varepsilon}^{\pm S}(\hat{\bm e}^\pm)$. We refer the readers to \cite{Stefanyszyn:2023qov} for the detailed form of these polarisation tensors.

In the helical basis, the mode functions decouple from each other and the quadratic action in Fourier space becomes
\begin{equation}
    S^{(\lambda)} = \frac{1}{2} \sum_{\lambda = -S}^{+S}\int\d t \,a^3 \int \frac{\d^3k}{(2\pi)^3} \left[(\dot{\sigma}_k^\lambda)^2 - c_{S, \lambda}^2 \, \frac{k^2}{a^2}\,(\sigma_k^\lambda)^2 - m^2 (\sigma_k^\lambda)^2 - 2 \lambda \xi \, \frac{k}{a} \, (\sigma_k^\lambda)^2\right]\,,
\end{equation}
where $c_{S, \lambda}^2 \equiv c^2 + \tfrac{S^2-\lambda^2}{S(2S-1)}\delta c^2$. The boost-breaking physics is now clear. Each helicity mode can propagate with a different sound speed, and the parity-violating chemical potential term affects all modes with $\lambda \neq 0$ in an asymmetric way. 
Varying this action yields the equation of motion
\begin{equation}
\label{eq: helical EOM}
    \left(\frac{\partial^2}{\partial\tau^2} - \frac{2}{\tau} \frac{\partial}{\partial\tau} + c_{S, \lambda}^2k^2 - 2\lambda \kappa \,\frac{k}{\tau} + \frac{m^2}{H^2 \tau^2}\right)\sigma^\lambda_k = 0\,,
\end{equation}
where we have switched to conformal time $\tau$, defined by $\d\tau\equiv \d t/a$, and $\kappa \equiv \xi/H$ is the dimensionless chemical potential in Hubble units. Up to an irrelevant overall phase, and imposing Bunch-Davies initial conditions, the solution is
\begin{equation}
\label{eq: Whittaker mode function}
    \sigma_k^\lambda(\tau) = -\frac{H\tau}{\sqrt{2c_{S, \lambda}k}}  e^{-\pi\lambda \kappa/2}\,  W_{i\lambda \kappa, i\mu}(2ic_{S, \lambda}k\tau)\,.
\end{equation}
For later convenience, we also define the tilded symbol $\tilde{\sigma}$ by absorbing one scale factor, i.e. $\tilde{\sigma}_k\equiv a^{-1}(\tau)\sigma_k$, so that
\begin{align}
    \tilde{\sigma}_k^\lambda(\tau) = \frac{e^{-\pi\lambda \kappa/2}}{\sqrt{2c_{S, \lambda}k}}  \,  W_{i\lambda \kappa, i\mu}(2ic_{S, \lambda}k\tau)\,,
    \label{eq:rescaled-sigma}
\end{align}
where $W_{i\lambda \kappa, i\mu}$ is the Whittaker-$W$ function, and $\mu \equiv \sqrt{m^2/H^2-9/4}$ is the dimensionless mass of the $\sigma$ field. For the purpose of studying generic boost-breaking effects, we will focus on the transverse modes ($\lambda\neq 0$) in the following. For a fixed spin $S$ and given that transverse helicities
$\pm\lambda$ that propagate with the same effective sound speed---as $c_{S, \lambda}^2 = c_{S, -\lambda}^2$---, we can rescale spatial coordinates to absorb $c_{S, \lambda}$ in the sound speed of the external field $c_s$, i.e.~$c_s \to c_s/c_{S, \lambda}:=c_s$. Notice that this choice is degenerate with setting $\delta c=0$ and $c=1$, as far as the helicity-$\pm\lambda$ modes are concerned. 
This way, $c_s<1$ means that the external field propagates slower than the field $\sigma_{i_1 \ldots i_S}$, and the opposite for $c_s>1$, which by no means implies superluminality. We shall see later that the chemical potential exponentially enhances the cosmological collider signals but could lead to dangerous particle overproduction if it is larger than the mass of the particle \cite{Sou:2021juh}. Therefore, in what follows, we restrict ourselves to $\kappa \lesssim \mu$ to avoid such an instability.

\paragraph{Helical seed integral.} The observables of interest are expectation values of two massless fields, the scalar field inflaton $\varphi$ (or equivalently, the curvature perturbation $\zeta$) and the tensor field graviton $\gamma_{ij}$. In many applications involving helical chemical potentials, it is sufficient to compute a single helical seed integral. The corresponding cosmological correlators can then be easily obtained, as we will demonstrate with an explicit example below. Using the Schwinger-Keldysh (SK) diagrammatic rules~\cite{Chen:2017ryl}, one can define the helical seed integral as \cite{Qin:2023ejc}
\begin{equation}
\label{eq: helical seed correlator}
    F_{\aa \bb}^{(\lambda)} \equiv - \aa\bb s\int_{-\infty^\aa}^0\frac{\d\tau'}{\tau'} e^{i\aa c_s k_{12}\tau'} \int_{-\infty^\bb}^0\frac{\d\tau''}{\tau''} e^{i\bb c_s k_{34}\tau''} \G_{\aa\bb}^\lambda(s; \tau', \tau'')\,,
\end{equation}
where $s = |\k_1+\k_2|$ is the magnitude of the exchanged momentum in the $s$-channel, and $\aa, \bb = \pm$ denote SK indices. We have added an overall factor $s$ to ensure the seed integral is dimensionless and scale-invariant. The bulk-bulk SK propagators for the helicity-$\lambda$ component of the massive spinning field, $\G_{\aa\bb}^\lambda$, are constructed out of the mode functions~\eqref{eq: Whittaker mode function} and the Heaviside $\Theta$-function,
\begin{equation}
    \begin{aligned}
        \G_{-+}^\lambda(k; \tau', \tau'')  &= \tilde{\sigma}_k^\lambda(\tau')\tilde{\sigma}_k^{\lambda*}(\tau'') \,, \quad \G_{+-}^\lambda(k; \tau', \tau'') = \G_{-+}^{\lambda*}(k; \tau', \tau'')\,, \\
        \G_{\pm\pm}^\lambda(k; \tau', \tau'') &= \G_{\mp\pm}^\lambda(k; \tau', \tau'')\Theta(\tau'-\tau'') + \G_{\pm\mp}^\lambda(k; \tau', \tau'')\Theta(\tau''-\tau')\,,
    \end{aligned}    
\end{equation}

\paragraph{Spin-1 example.} Starting from the boost-breaking helical seed integral~\eqref{eq: helical seed correlator}, we can derive a wide range of correlators involving the exchange of massive spinning fields with non-unit sound speeds and chemical potentials. Consider the spin-1 case \cite{Jazayeri:2023kji} as a particularly interesting example, the four-point correlator of inflaton fluctuations $\varphi$ is generated by the interaction 
\begin{equation}
\label{eq: eg helical interaction}
    S_{\text{int}}=\int \d\tau\d^3 x \left(\frac{a^{-1}}{\Lambda^3}\partial_j\varphi'\partial_i\partial_j\varphi \sigma_i\right) \,.
\end{equation}
The corresponding four-point exchange correlator can be expressed as
\begin{equation}
    \begin{aligned}
        \braket{\varphi_{\k_1} \varphi_{\k_2} \varphi_{\k_3} \varphi_{\k_4}}' &= -\frac{1}{16\Lambda^6} \sum_{\lambda=\pm 1} \sum_{\sf{a}, \sf{b} = \pm} (\hat{\k}_1\cdot \hat{\k}_2)(\hat{\k}_3\cdot \hat{\k}_4) \, \left[\hat{\k}_1 \cdot \bm{\varepsilon}^\lambda(\hat{\bm{s}})\right]^*\left[\hat{\k}_3 \cdot \bm{\varepsilon}^\lambda(\hat{\bm{s}})\right] \\
        &\times\frac{1}{k_1 k_3 s^7} (1-k_1 \partial_{k_1}) (1-k_3 \partial_{k_3}) [\D_{k_{12}}^{\sf{a}}]^3[\D_{k_{34}}^{\sf{b}}]^3F^{(\lambda)}_{\sf{ab}} + 3\,\text{perms.} \\
        &+ (t\text{-channel}) + (u\text{-channel}) \,,
    \end{aligned}
\label{eq:example-trispectrum-specific-interactions}    
\end{equation}
where $\D_k^{\sf{a}} \equiv i s/({\sf{a}}c_s) \partial_{k}$ and a prime on a correlator denotes that we have dropped the momentum conserving delta function 
This interaction comes from the operator $\nabla_\mu \delta g^{00}\delta K_\nu^\mu \sigma^\nu$ within the EFT of inflation framework~\cite{Cheung:2007st}. The polarisation factor can be further explicitly written as
\begin{equation}
    [\hat{\k}_1\cdot \bm{\varepsilon}^{\lambda}(\hat{\bm{s}})]^*[\hat{\k}_3\cdot \bm{\varepsilon}^{\lambda}(\hat{\bm{s}})] = \frac{1}{2}\left[\hat{\k}_1\cdot\hat{\k}_3-(\hat{\k}_1\cdot\hat{\bm{s}})(\hat{\k}_3\cdot\hat{\bm{s}})+i\lambda \hat{\bm{s}}\cdot(\hat{\k}_1\times\hat{\k}_3)\right] \,.\label{4ptkinematicFactor}
\end{equation}
It is clear from the expression of the kinematic factor \eqref{4ptkinematicFactor} that the parity-even sector of the four-point correlator is purely real whereas the parity-odd sector is purely imaginary. In the case without a chemical potential, the seed function $F_{\sf ab}^{(\lambda)}$ that encodes dynamical information does not depend on helicity and the parity-odd part in the kinematic factor cancels between positive and negative helicities, leaving the result parity-even. However, a non-zero chemical potential creates an imbalance between the positive and negative helicity modes in their dynamics i.e. $F_{\sf ab}^{(\lambda)}\neq F_{\sf ab}^{(-\lambda)}$, thereby offsetting the parity-odd part of the correlator from zero.

\subsection{Via Bootstrap Equation}
\label{subsec: Via Boundary Kinematic Differential Equation}

We now compute the helical seed correlator using the bootstrap approach~\cite{Arkani-Hamed:2018kmz, Pimentel:2022fsc, Jazayeri:2022kjy, Qin:2022fbv, Qin:2023ejc, Aoki:2024uyi}. As we will see, this method yields a double-series representation that only converges in a restricted region of the kinematic space, due to a reduced sound speed $c_s< 1$. However, by resumming one layer of the series, we are able to analytically continue the solution beyond its region of validity to the full physical kinematic region. 

\paragraph{Boundary equation \emph{\&} kinematics.} The helical seed correlator satisfies the following boundary differential equations~\cite{Qin:2023ejc}
\begin{equation}
\label{eq: helical bootstrap eq}
    \left[\Delta_{\pm, \tilde{u}}^{(\lambda)} + \left(\mu^2 + \frac{1}{4}\right)\right] F_{\pm \mp}^{(\lambda)} = 0 \,, \quad 
    \left[\Delta_{\pm, \tilde{u}}^{(\lambda)} + \left(\mu^2 + \frac{1}{4}\right)\right] F_{\pm \pm}^{(\lambda)} = \frac{1}{2}\frac{\tilde{u} \tilde{v}}{\tilde{u} + \tilde{v} - \tilde{u}\tilde{v}} \,,
\end{equation}
with the differential operator for different helicities defined by
\begin{equation}
    \Delta_{\pm, \tilde{u}}^{(\lambda)} \equiv \tilde{u}^2(1 - \tilde{u})\partial_{\tilde{u}}^2 - \left[(1 \pm i \lambda \kappa)\tilde{u}^2 \right]\partial_{\tilde{u}} \,,
\end{equation}
and similar equations for $\tilde{u} \leftrightarrow \tilde{v}$. We have defined the following dimensionless kinematic variables
\begin{equation}
    \tilde{u} \equiv \frac{2s}{c_sk_{12}+s} \,, \quad \tilde{v} \equiv \frac{2s}{c_sk_{34}+s} \,.
\end{equation}
Note that these variables differ from the ones introduced in~\cite{Arkani-Hamed:2018kmz,Pimentel:2022fsc, Jazayeri:2022kjy} 
\begin{equation}
\label{eq: original kinematic variables}
    u \equiv \frac{s}{c_sk_{12}}\,, \quad v \equiv \frac{s}{c_sk_{34}}\,,
\end{equation}
and are simply related to the tilted variables by the map $\tilde{u} = 2u/(1+u)$ and $\tilde{v} = 2v/(1+v)$. The inverse map is given by $u=\tilde{u}/(2-\tilde{u})$ and $v=\tilde{v}/(2-\tilde{v})$. In the case of a unit sound speed, these two variables are always constrained to be less than unity due to triangle inequality, $0< \tilde{u}, \tilde{v}< 1$. The solution to \eqref{eq: helical bootstrap eq} in this unit-sound-speed case has been found in~\cite{Qin:2022fbv}. However, for a reduced sound speed, $c_s< 1$, $\tilde{u}, \tilde{v}$ may exceed unity i.e. $0< \tilde{u}, \tilde{v}< 2/(c_s+1)$, which renders the original series solution of~\cite{Qin:2022fbv} superficially divergent. 

\paragraph{Singularity structure.}
One advantage of expressing the seed integral \eqref{eq: helical seed correlator} as a solution to the bootstrap equations \eqref{eq: helical bootstrap eq} is that we can readily identify all possible singularities of the correlator in the complex plane, even before deriving its explicit expressions. Since the seed function is symmetric under the exchange  $\tilde{u}\leftrightarrow\tilde{v}$, it suffices to fix $\tilde v$ in the physical region with $0<\tilde v<2/(c_s+1)$, and consider the $\tilde u$-complex plane. Kinematical singularities can only arise in the following two cases:
\begin{itemize}
    \item The coefficient of the leading differential operator in $\Delta_{\pm,\tilde u}^{(\lambda)}$ vanishes or diverges:
    \begin{equation}
    \label{eq_sing1}
        \tilde u^2-\tilde u^3=0~~\text{or}~~\infty \quad \Rightarrow \quad s=0 ~~\text{or}~~c_sk_{12}\pm s=0~.
    \end{equation}
    The internal soft singularity at $s=0$ corresponds to the branching point of the (non-local) cosmological collider signals. The other two singularities at $c_s k_{12}+s$ and $c_s k_{12}-s$ correspond to the partial-energy and folded poles, respectively.\footnote{Under the symmetrisation $k_{12}\leftrightarrow k_{34}$, we immediately deduce the presence of the other partial energy singularity at $c_sk_{34}+s=0$ and folded singularity at $c_sk_{34}-s=0$.}
    \item The source term diverges:
    \begin{equation}
    \label{eq_sing2}
        \tilde u+\tilde v-\tilde u\tilde v=0 \quad \Rightarrow\quad  k_{12}+k_{34}=0,
    \end{equation}
    which maps to the total-energy singularity.
\end{itemize}
Notice that \eqref{eq_sing1} and \eqref{eq_sing2} are only necessary conditions for a singular point of the seed integral. As we will see below, the folded poles $c_sk_{12}-s=0$ (and also $c_sk_{34}-s=0$) will be eliminated by boundary conditions. Surprisingly, although the sound speed modifies the location of the partial energy singularities and folded singularities in a trivial manner ($k_i\to c_s k_i$ for external fields), neither the chemical potential nor the sound speed alters the singularity structure of the correlator at the tree level. This is rather non-trivial, as both the chemical potential and the sound speed break the spacetime symmetry of the theory, and one might be led to expect the disruption of the analytic structure of correlators. It would be interesting to investigate the underlying reasons and to examine whether the same conclusion holds also at the loop level.

\paragraph{Homogeneous solution.} The homogeneous equation takes the same form as in the unit sound speed case. Here, we will adhere to the derivation in previous works and present only the key results relevant to our discussion. We refer the interested readers to \cite{Qin:2023ejc} for more technical details. There are two differential operators that are related to each other through the helicity reflection condition:
\begin{equation}
    \Delta_{-, \tilde{u}}^{(\lambda)} = \Delta_{+, \tilde{u}}^{(-\lambda)} \,.
\end{equation}
Since the equation is second order, each differential operator possesses two linearly independent solutions such that 
\begin{equation}
    \left[\Delta_{+, \tilde{u}}^{(\lambda)} + \left(\mu^2 + \frac{1}{4}\right)\right] \H_{\pm}(\tilde{u}) = 0 \,.
\end{equation}
The homogeneous solutions are explicitly given by \cite{Qin:2022fbv,Qin:2023ejc}
\begin{equation}
    \H_\pm(\tilde{u}) = \tilde{u}^{\tfrac{1}{2}\pm i\mu} {\Gamma(\tfrac{1}{2}\pm i\mu)\,\Gamma(\tfrac{1}{2}+i\lambda\kappa \pm i\mu)}\,\ptFq{2}{1}{\tfrac{1}{2}\pm i\mu, \tfrac{1}{2}+i\lambda\kappa \pm i\mu}{1\pm2i\mu}{\tilde{u}} \,,
\end{equation}
where $\pregFq{2}{1}{a, b}{c}{z} \equiv \tfrac{1}{\Gamma(c)} \pFq{2}{1}{a, b}{c}{z}$ is the regularised hypergeometric function. The hypergeometric functions exhibit a branch cut along $\tilde{u} \ge 1$. In the case where $c_s=1$, $\tilde{u}$ remains less than unity, ensuring that the branch point (i.e. the folded 
configuration) is never crossed in the physical region.
However, when $c_s<1$, the variable $\tilde{u}$ can exceed unity in certain kinematic regions, thereby crossing the branch point. Since  the evaluation of the hypergeometric function ${}_2F_1$ for $\tilde{u}>1$ is well-established, the homogeneous solution remains valid even in this regime. In other words, the homogeneous solution can be smoothly continued from the unit sound speed case to the small sound speed case, owing to the known analytic continuation properties of hypergeometric series. For the factorised contribution, the general solution is then expressed as 
\begin{equation}
    \begin{aligned}
        F_{+-}^{(\lambda)}(\tilde{u}, \tilde{v}) &= \A_{++}^{+-} \H_+(\tilde{u}) \H_-^*(\tilde{v}) + \A_{+-}^{+-} \H_+(\tilde{u}) \H_+^*(\tilde{v}) \\
        &+ \A_{-+}^{+-} \H_-(\tilde{u}) \H_-^*(\tilde{v}) + \A_{--}^{+-} \H_-(\tilde{u}) \H_+^*(\tilde{v}) \,,
    \end{aligned}
\end{equation}
and $\A$ represents some coefficients that will be determined through the boundary conditions later. We naturally have $F_{-+}^{(\lambda)} = F_{+-}^{(\lambda)*}$. Also for the nested contribution, we can express it as 
\begin{equation}
\label{eq: helical general nested solution}
    \begin{aligned}
        F_{++}^{(\lambda)}(\tilde{u}, \tilde{v}) &= \A_{++}^{++} \H_+(\tilde{u}) \H_+(\tilde{v}) + \A_{+-}^{++} \H_+(\tilde{u}) \H_-(\tilde{v})  \\
        &+ \A_{-+}^{++} \H_-(\tilde{u}) \H_+(\tilde{v}) + \A_{--}^{++} \H_-(\tilde{u}) \H_-(\tilde{v}) \\
        &+ \S_{+}^{(\lambda)}(\tilde{u}, \tilde{v}) \,,
    \end{aligned}
\end{equation}
where $\S_{+}^{(\lambda)}$ is a particular solution for the inhomogeneous differential equation which we solve now. As for the other nested seed function, it can be obtained using the relation $F_{--}^{(\lambda)} = F_{++}^{(\lambda)*}$.

\paragraph{Particular solution.} To find a particular solution, we assume $\tilde{u}<\tilde{v}$ without loss of generality. The result for the other region is obtained by swapping the relevant variables. Note that this procedure is highly non-trivial for higher-point correlators as one needs to analytically continue correlators to other kinematic regions, picking up homogeneous solutions along the way. In this discussion, we focus on $\S_+$ as an illustrative example, while the corresponding result for $\S_-$(the particular solution of $F^{(\lambda)}_{--}$) can be derived straightforwardly using the conjugate relation $\S_- = [\S_+]^*$. The source term in Eq.~\eqref{eq: helical bootstrap eq} can be systematically expanded as follows
\begin{equation}
    \frac{\tilde{u}\tilde{v}}{\tilde{u}+\tilde{v}-\tilde{u}\tilde{v}} =  \sum_{n=0}^\infty \tilde{u}^{n+1}\left(1 - \frac{1}{\tilde{v}}\right)^n = \tilde{u} \sum_{n=0}^\infty (-1)^n \left(\frac{\tilde{u}}{\tilde{v}}\right)^n(1-\tilde{v})^n \,.
\end{equation}
We propose the following ansatz for the particular solution, assuming a functional form that resembles the structure of the source:
\begin{equation}
    \S_+^{(\lambda)} = \sum_{m, n=0}^\infty \A_{m, n}^{(\lambda)} \, (-1)^n \, \tilde{u}^{m+1}\left(\frac{\tilde{u}}{\tilde{v}}\right)^n(1-\tilde{v})^n \,.\label{SplusBSeqSol}
\end{equation}
Substituting the proposed ansatz into the differential equation, we obtain a recurrence relation that governs the coefficients
\begin{equation}
    \A_{m+1, n}^{(\lambda)} = \frac{(m+n+1)(m+n+1+i\lambda \kappa)}{(m+n+1+\tfrac{1}{2})^2 + \mu^2} \, \A_{m, n}^{(\lambda)} \,, \quad
    \A_{0, n}^{(\lambda)} = \frac{1}{2} \frac{1}{(n+\tfrac{1}{2})^2+\mu^2} \,.
\end{equation}
Using this recurrence relation, the coefficients can be iteratively solved. Explicitly, the coefficients are given by \cite{Qin:2022fbv,Qin:2023ejc}
\begin{equation}
    \A_{m, n}^{(\lambda)} = \frac{1}{2}\frac{(n+1)_m (n+1+i\lambda \kappa)_m}{(n+\tfrac{1}{2}+i\mu)_{m+1} (n+\tfrac{1}{2}-i\mu)_{m+1}} \,,
\end{equation}
where $(a)_m\equiv\Gamma(a+n)/\Gamma(a)$ is the Pochhammer symbol. This series expansion is well-behaved only within the region $\tilde{u}<1$, indicating that the solution derived here appears to be valid exclusively for the case of unit sound speed. In the kinematic region where  $\tilde{u}>1$, the series fails to converge, suggesting that the ansatz employed here is unsuitable for this range. To address this problem, one could try to attack the equations \eqref{eq: helical bootstrap eq} with an alternative ansatz that manifestly converges in $\tilde{u}>1$ (for example, with negative powers of $\tilde{u}$). For the case without the chemical potential, such ansatz has been adopted and solved in \cite{Jazayeri:2022kjy}. However, in this work, we provide a more straightforward solution, via a direct resummation of \eqref{SplusBSeqSol} that smoothly facilitates its analytic continuation to the $\tilde{u}>1$ regions. To illustrate this process, we observe that the $m$-layer of the series can be resummed, and the particular solution becomes
\begin{equation}
\label{eq: helical particular solution resummed expression}
    \begin{aligned}
        \S_\pm^{(\lambda)} = \frac{\tilde{u}}{2} \sum_{n=0}^\infty \frac{(-1)^n}{(n+\tfrac{1}{2})^2+\mu^2} \left(\frac{\tilde{u}}{\tilde{v}}\right)^n (1-\tilde{v})^n \pFq{3}{2}{1 \,\, n+1, n+1\pm i\lambda \kappa}{\tfrac{3}{2}+n-i\mu \,\, \tfrac{3}{2}+n+i\mu}{\tilde{u}} \,.
    \end{aligned}
\end{equation}
Here, we have summed the $\tilde{u}$-series into a generalised hypergeometric function. This allows us to evaluate the series even for $\tilde{u}>1$. Given that we have assumed $\tilde{u}<\tilde{v}$ and by definition $0<\tilde{v}<2$, it can be explicitly verified that this resummed single-layer series converges throughout the entire kinematic region under consideration. One aspect we have not yet addressed is the branch cut of the hypergeometric series, where $\tilde{u}>1$ along the real axis. Taking $F_{++}^{(\lambda)}$ as an example, the appropriate branch involves selecting the lower-half plane prescription, specifically $\tilde{u}=\tilde{u}-i\epsilon$ with the infinitesimal $\epsilon>0$ (vice versa, for $F_{--}^{(\lambda)}$ the appropriate choice, consistent with the complex conjugate relation, is to take $\tilde{u}=\tilde{u}+i\epsilon$). This branch choice can be justified by analysing the corresponding bulk integral, which will be discussed in detail later. As an interesting aside, using the variables $(\tilde{u}, \tilde{v})$ instead of $(u, v)$ allows us to obtain a series representation that converges over a broader kinematic region than originally designed for. Indeed, writing $(\tilde{u}/\tilde{v})^n(1-\tilde{v})^n = \tilde{u}^n (\tfrac{1-\tilde{v}}{\tilde{v}})^n$, we obtain that the series is convergent for $|\tilde u(1-\tilde v)/\tilde v|<1$. This region is larger than $\tilde{u}<\tilde{v}$ and also extends beyond the unit circle for a reduced sound speed. A natural question is whether one can find another set of well-chosen kinematic variables so that the series converges in the entire physical kinematic domain, which would suggest that (both layers of) the series can be fully resummed into known hypergeometric series. We leave these thoughts for a future work. 

\begin{figure}[h!]
    \centering
    \begin{subfigure}{.5\textwidth}
        \centering
        \includegraphics[width=0.985\linewidth]{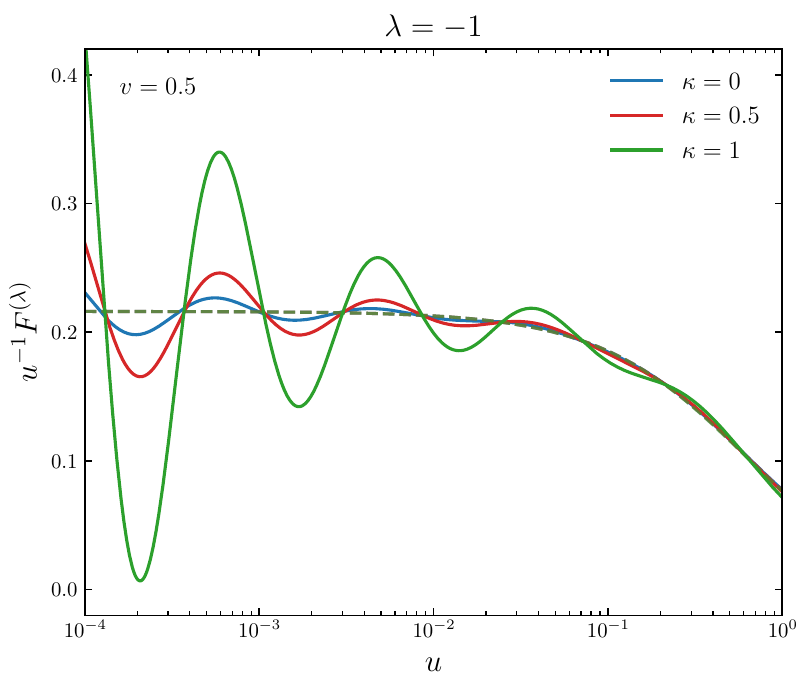}
    \end{subfigure}%
    \begin{subfigure}{.5\textwidth}
        \centering
        \includegraphics[width=1\linewidth]{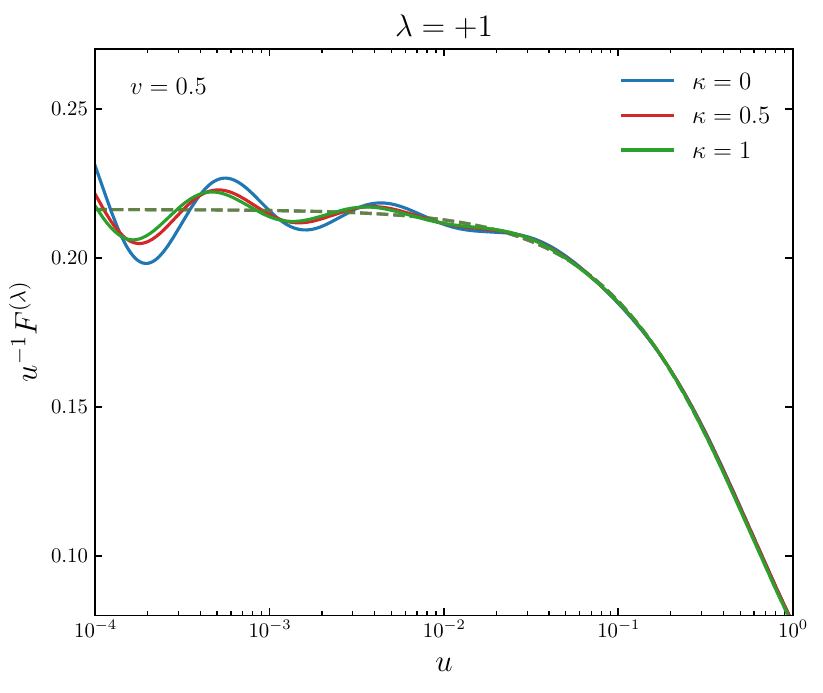}
    \end{subfigure}
   \caption{The (rescaled) boost-breaking helical seed correlator $F^{(\lambda)}(u, v)$, as function of the dimensionless kinematic variable $u$ with fixed $v=0.5$ and at fixed mass $\mu=3$ for $\kappa=0$, $\kappa=0.5$ and $\kappa=1$, for $\lambda=-1$ (\textit{left panel}) and $\lambda=+1$ (\textit{right panel}). We include terms in the series from $n = 1, \ldots, 10$. We show both the full correlator (\textit{solid line}), and the background series contribution (\textit{dashed line}) which is almost identical for all parameters. We considered kinematic variables inside the unit circle corresponding to a unit sound speed $c_s=1$.}
  \label{fig: Helical seed correlator bootstrap}
\end{figure}

\begin{figure}[h!]
    \centering
    \begin{subfigure}{.5\textwidth}
        \centering
        \includegraphics[width=1\linewidth]{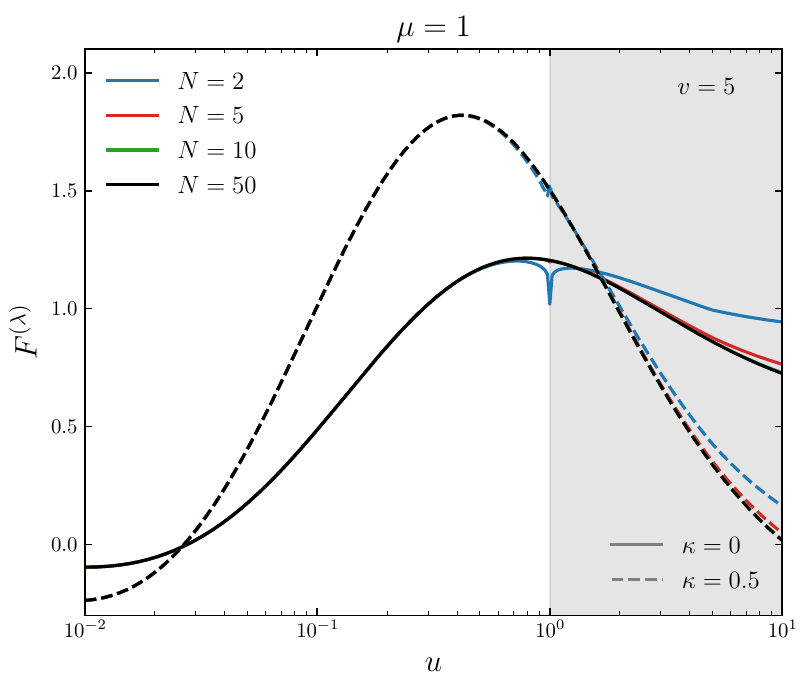}
    \end{subfigure}%
    \begin{subfigure}{.5\textwidth}
        \centering
        \includegraphics[width=1\linewidth]{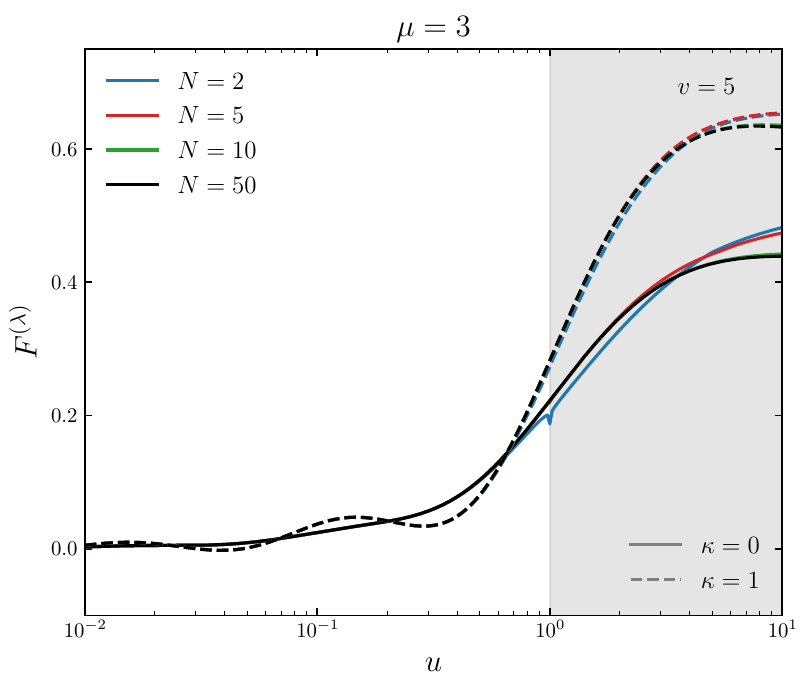}
    \end{subfigure}
   \caption{The full boost-breaking helical seed correlator $F^{(\lambda)}(u, v)$ solved using the bootstrap method, as function of the dimensionless kinematic variable $u$ with fixed $v=5$, for $\mu=1$ (\textit{left panel}) for $\kappa=0$ (\textit{solid line}) and $\kappa=0.5$ (\textit{dashed line}), and $\mu=3$ (\textit{right panel}) for $\kappa=0$ (\textit{solid line}) and $\kappa=1$ (\textit{dashed line}). We always consider the enhanced helicity $\lambda = -1$ when $\kappa\neq 0$. We include terms in the series from $n=1, \ldots, N$. The shaded gray region corresponds to kinematic configurations outside the unit circle which covers the case of a reduced sound speed (here $c_s=0.1$) for the external field.}
  \label{fig: Helical seed correlator bootstrap outside circle}
\end{figure}

\paragraph{Boundary conditions.} We have nearly completed the calculation of the analytical expressions, except for the unfixed coefficients $\A$. Several approaches can be used to impose the boundary conditions, such as requiring the absence of spurious poles or considering the factorised limit as discussed previously. For simplicity, here the coefficients are directly determined by matching the expression in the (hierarchical) soft limit $\tilde{u}\ll \tilde{v}\ll 1 $, where the bulk integral can be evaluated easily through the IR expansion of the integrand or using the Mellin-Barnes transformation. The final expressions are found to be \cite{Qin:2022fbv,Qin:2023ejc}:
\begin{equation}
    \begin{aligned}
        \A_{++}^{+-} &= -\A_{+-}^{+-} = -\A_{-+}^{+-} = \A_{--}^{+-} = \frac{-e^{-\pi\lambda \kappa}}{4\sinh^2(2\pi\mu)}[\cosh(2\pi\kappa) + \cosh(2\pi\mu)] \,, \\
        \A_{++}^{++} &= -\A_{+-}^{++} = \frac{-i\pi e^{-\pi(\lambda\kappa-\mu)}\cosh[\pi(\mu + \lambda\kappa)]}{2\sinh^2(2\pi\mu) \Gamma(\tfrac{1}{2}+ i\lambda \kappa+i\mu) \Gamma(\tfrac{1}{2}+ i\lambda \kappa-i\mu)} \,, \\
        \A_{--}^{++} &= -\A_{-+}^{++} = \frac{-i\pi e^{-\pi(\lambda\kappa+\mu)}\cosh[\pi(-\mu + \lambda\kappa)]}{2\sinh^2(2\pi\mu) \Gamma(\tfrac{1}{2}+ i\lambda \kappa+i\mu) \Gamma(\tfrac{1}{2}+ i\lambda \kappa-i\mu)} \,.
    \end{aligned}
\end{equation}
This concludes the derivation of the helical seed correlator involving two boost-breaking effects: (i) the sound speed, and (ii) the chemical potential. The full exact solution is stored in the \textsf{Mathematica} notebook available on the following Github repository [\href{https://github.com/deniswerth/Boost-Breaking-Cosmological-Correlators}{\faGithub}]. This result is valid for all physical kinematic regions and exhibits rapid convergence. In Fig.~\ref{fig: Helical seed correlator bootstrap}, we show the boost-breaking helical seed correlator at a fixed mass $\mu$ but varying the chemical potential $\kappa$ and for both helicities $\lambda =\pm1$. We note that the chemical potential essentially modifies the signal contribution, boosting the amplitude and slightly changing the phase of the cosmological collider signal, but barely affects the background signal. Also, note that the signal coming from one transverse helicity ($\lambda=-1$) is enhanced compared to the other one. We show in Fig.~\ref{fig: Helical seed correlator bootstrap outside circle} the seed correlator for kinematic configurations extending outside the unit circle $1 \leq u, v$, which confirms that the found solution is well convergent when the external field has a reduced sound speed. However we have noticed that the convergence rate is slower outside the unit circle than inside. We will come back to this point later when comparing with the spectral representation solution. The solution presents a slight kink around $u=1$ (from the discontinuity of the first derivative) that disappears swiftly as we dial up the number of terms included $N$, and that is less visible as we also increase the mass $\mu$. The characteristic peak for $\mu=1$ is the low-speed collider signal~\cite{Jazayeri:2022kjy, Jazayeri:2023xcj, Jazayeri:2023kji}. Also notice that the enhanced non-analytic contribution for the negative helicity ($\lambda=-1$) is well visible for $\mu=3$ and $\kappa=1$ and with an amplitude that increases as $\kappa$ approaches $\mu$. This is because the amplitude of this cosmological collider signal is given by the Boltzmann factor $\sim e^{-\pi(\mu+\lambda\kappa)}$. We will give more details and thoroughly study this signal in Sec.~\ref{sec: Classifying non-Analyticities}.

\paragraph{Discussion.} Let us make a few remarks concerning the full solution for the boost-breaking helical massive exchange correlator.

\begin{itemize}
    \item{\it Bulk cutting rules revisited}: Bulk cutting rules have been introduced in~\cite{Tong:2021wai} for tree-level diagrams and subsequently generalised to account for loop-level processes in~\cite{Qin:2023bjk,Qin:2023nhv,Ema:2024hkj} for extracting non-analytic processes. To briefly revisit these cutting rules, as an illustrative example, consider the integral $F^{(\lambda)}_{++}$. First, assuming that $k_{12} > k_{34}$, by manipulating one of the Heaviside theta functions, $\Theta(\tau' - \tau'') = 1 - \Theta(\tau'' - \tau')$, the bulk propagator can be decomposed into two distinct terms
    \begin{equation}
    \label{eq: helical bulk cutting rule}
        \G_{++}^\lambda(s; \tau', \tau'') = \G_{-+}^\lambda(s; \tau', \tau'') + \left[\G_{+-}^\lambda(s; \tau'', \tau') - \G_{-+}^\lambda(s; \tau', \tau'')\right]\Theta(\tau''-\tau')\,,
    \end{equation}
    where the first term is factorised in time, while the second term is still nested. Consequently, the integral $F^{(\lambda)}_{++}$ can be expressed as the sum of two components, $F^{(\lambda)}_{++} = F^{(\lambda)}_{\text{F}} + F^{(\lambda)}_{\text{N}}$, that can be schematically represented as
\begin{equation}
	\label{eq: illustration helical bulk cutting rule}
	\raisebox{0pt}{
		\begin{tikzpicture}[baseline={([yshift=-.5ex]current bounding box.center)}, line width=1. pt, scale=2]
			\draw[pyblue] (0, 0) -- (0.7, 0);
			\draw[fill=black] (0, 0) circle (.04cm);
			\draw[black] (0, 0) -- (-0.25, 0.35);
			\draw[black] (0, 0) -- (-0.25, -0.35);
			\draw[fill=white] ([xshift=-1pt,yshift=-1pt]-0.25, -0.35) rectangle ++(2pt,2pt);
			\draw[fill=white] ([xshift=-1pt,yshift=-1pt]-0.25, 0.35) rectangle ++(2pt,2pt);
			\draw[black] (0.7, 0) -- (0.95, 0.35);
			\draw[black] (0.7, 0) -- (0.95, -0.35);
			\draw[fill=black] (0.7, 0) circle (.04cm);
			\draw[fill=white] ([xshift=-1pt,yshift=-1pt]0.95, -0.35) rectangle ++(2pt,2pt);
			\draw[fill=white] ([xshift=-1pt,yshift=-1pt]0.95, 0.35) rectangle ++(2pt,2pt);
		\end{tikzpicture} 
	}\quad=\quad\raisebox{0pt}{
		\begin{tikzpicture}[baseline={([yshift=-.5ex]current bounding box.center)}, line width=1. pt, scale=2]
			\draw[pyblue] (0, 0) -- (0.3, 0);
			\draw[pyblue] (0.5, 0) -- (0.8, 0);
			\draw[fill=black] (0, 0) circle (.04cm);
			\draw[black] (0, 0) -- (-0.25, 0.35);
			\draw[black] (0, 0) -- (-0.25, -0.35);
			\draw[fill=white] ([xshift=-1pt,yshift=-1pt]-0.25, -0.35) rectangle ++(2pt,2pt);
			\draw[fill=white] ([xshift=-1.25pt,yshift=-1.25pt]-0.25, 0.35) rectangle ++(2pt,2pt);
			\draw[black] (0.8, 0) -- (1.05, 0.35);
			\draw[black] (0.8, 0) -- (1.05, -0.35);
			\draw[fill=black] (0.8, 0) circle (.04cm);
			\draw[fill=white] ([xshift=-1pt,yshift=-1pt]1.05, -0.35) rectangle ++(2pt,2pt);
			\draw[fill=white] ([xshift=-1pt,yshift=-1pt]1.05, 0.35) rectangle ++(2pt,2pt);
			\node at (0.4, 0) {\textcolor{pyblue}{$\times$}};
			\node at (0.4, 0.2) {\textcolor{pyblue}{F}};
		\end{tikzpicture} 
	}\quad+\quad\raisebox{0pt}{
		\begin{tikzpicture}[baseline={([yshift=-.5ex]current bounding box.center)}, line width=1. pt, scale=2]
			\draw[pyblue, double] (0, 0) -- (0.7, 0);
			\draw[fill=black] (0, 0) circle (.04cm);
			\draw[black] (0, 0) -- (-0.25, 0.35);
			\draw[black] (0, 0) -- (-0.25, -0.35);
			\draw[fill=white] ([xshift=-1pt,yshift=-1pt]-0.25, -0.35) rectangle ++(2pt,2pt);
			\draw[fill=white] ([xshift=-1pt,yshift=-1pt]-0.25, 0.35) rectangle ++(2pt,2pt);
			\draw[black] (0.7, 0) -- (0.95, 0.35);
			\draw[black] (0.7, 0) -- (0.95, -0.35);
			\draw[fill=black] (0.7, 0) circle (.05cm);
			\draw[fill=white] ([xshift=-1pt,yshift=-1pt]0.95, -0.35) rectangle ++(2pt,2pt);
			\draw[fill=white] ([xshift=-1pt,yshift=-1pt]0.95, 0.35) rectangle ++(2pt,2pt);
			\node at (0.35, 0.2) {\textcolor{pyblue}{N}};
		\end{tikzpicture} 
	} \,.
\end{equation}    
    Here, the first term on the right-hand side of~\eqref{eq: helical bulk cutting rule} is the {\it factorised} part, which is straightforward to evaluate due to the time-integral factorisation property. The second term is {\it nested}. The remarkable aspect of the bulk cutting rule is that the purest signals are often the simplest. As a direct consequence of microcausality \cite{Tong:2021wai,Ema:2024hkj}, these oscillatory cosmological collider signals, which include both local and non-local types, entirely stem from the factorised piece, while the nested term only contributes to the background, which is analytic in the momentum ratio.

    \vskip 4pt
    This cutting rule works perfectly in the case of a unit sound speed, even when the chemical potential effects are taken into account. However, it does not apply to certain kinematical regions when we have a small sound speed. To illustrate this more concretely, let us focus on the integral associated with the factorised part, which contains terms of the form
    \begin{equation}
        F_{\text{F}}^{(\lambda)} \supset \int_{-\infty^+}^0 \d \tau \, e^{i c_s k_{12} \tau} \sigma^\lambda_s(\tau) \sim \int_{-\infty^+}^0 \d\tau \, e^{i (c_s k_{12} - s) \tau} \,,
    \end{equation}
    where we have evaluated the UV behaviour of the massive mode function $\tau \to -\infty$. When $c_s = 1$, although the integral contains a spurious folded pole, which should be cancelled by the integral $F^{(\lambda)}_{\text{N}}$, the overall integral remains convergent in the physical kinematic regions due to the triangle inequality $k_{12} > s$. However, a small sound speed changes the situation significantly, as in certain kinematic regions, $c_s k_{12}$ can become smaller than $s$, leading to a divergence when $\tau \to -\infty^+$. What we are doing here is actually the analytic continuation of the bulk cutting rule. Since the term $F_{\text{F}}^{(\lambda)}$ is factorised, it must satisfy the homogeneous boundary equations and contains all the information about the cosmological collider signals. In contrast, $F_{\text{N}}^{(\lambda)}$ satisfies the inhomogeneous part, as the Heaviside step function generates the source term in the bootstrap differential equation. For regions where $c_s k_{12} > s$,~\eqref{eq: helical general nested solution} and~\eqref{eq: illustration helical bulk cutting rule} are in one-to-one correspondence, as all terms are convergent.  When $c_s k_{12} < s$, the resummation of the series allows us to analytically continue each term in~\eqref{eq: helical general nested solution} for a small sound speed, ensuring convergence. This is why we should choose the branch $\tilde{u}=\tilde{u}-i\epsilon$ (or equivalently, $s=s-i\epsilon$, $k_{12, 34}=k_{12, 34}-i\epsilon$) in the analytical expression when $\tilde{u}>1$. From the  bulk time integral, this corresponds to deforming the contour such that in the UV it behaves as $\sim e^{(s-c_s k_{12})\tau}$ when $\tau\to-\infty$, then it is convergent for cases with a small sound speed. In other words, this process generalises the bulk cutting rule by analytically continuing each term in the bulk time integrals, which can be represented schematically as:	
    \begin{equation}
            \A \cdot \H \cdot \H = \text{P.V.} \left(\raisebox{0pt}{
		\begin{tikzpicture}[baseline={([yshift=-.5ex]current bounding box.center)}, line width=1. pt, scale=2]
			\draw[pyblue] (0, 0) -- (0.3, 0);
			\draw[pyblue] (0.5, 0) -- (0.8, 0);
			\draw[fill=black] (0, 0) circle (.04cm);
			\draw[black] (0, 0) -- (-0.25, 0.35);
			\draw[black] (0, 0) -- (-0.25, -0.35);
			\draw[fill=white] ([xshift=-1pt,yshift=-1pt]-0.25, -0.35) rectangle ++(2pt,2pt);
			\draw[fill=white] ([xshift=-1.25pt,yshift=-1.25pt]-0.25, 0.35) rectangle ++(2pt,2pt);
			\draw[black] (0.8, 0) -- (1.05, 0.35);
			\draw[black] (0.8, 0) -- (1.05, -0.35);
			\draw[fill=black] (0.8, 0) circle (.04cm);
			\draw[fill=white] ([xshift=-1pt,yshift=-1pt]1.05, -0.35) rectangle ++(2pt,2pt);
			\draw[fill=white] ([xshift=-1pt,yshift=-1pt]1.05, 0.35) rectangle ++(2pt,2pt);
			\node at (0.4, 0) {\textcolor{pyblue}{$\times$}};
			\node at (0.4, 0.2) {\textcolor{pyblue}{F}};
		\end{tikzpicture} 
	} \right) \,, \quad
            \S = \text{P.V.} \left(\raisebox{0pt}{
\begin{tikzpicture}[baseline={([yshift=-.5ex]current bounding box.center)}, line width=1. pt, scale=2]
			\draw[pyblue, double] (0, 0) -- (0.7, 0);
			\draw[fill=black] (0, 0) circle (.04cm);
			\draw[black] (0, 0) -- (-0.25, 0.35);
			\draw[black] (0, 0) -- (-0.25, -0.35);
			\draw[fill=white] ([xshift=-1pt,yshift=-1pt]-0.25, -0.35) rectangle ++(2pt,2pt);
			\draw[fill=white] ([xshift=-1pt,yshift=-1pt]-0.25, 0.35) rectangle ++(2pt,2pt);
			\draw[black] (0.7, 0) -- (0.95, 0.35);
			\draw[black] (0.7, 0) -- (0.95, -0.35);
			\draw[fill=black] (0.7, 0) circle (.05cm);
			\draw[fill=white] ([xshift=-1pt,yshift=-1pt]0.95, -0.35) rectangle ++(2pt,2pt);
			\draw[fill=white] ([xshift=-1pt,yshift=-1pt]0.95, 0.35) rectangle ++(2pt,2pt);
			\node at (0.35, 0.2) {\textcolor{pyblue}{N}};
		\end{tikzpicture}  
}\right) \,,
    \end{equation}
    where the operation $\text{P.V.}$ represents extracting the finite part (Cauchy principal value) of the corresponding bulk integral.

    \item {\it Parity-odd correlator factorisation}: Exchanging massive spinning fields with a chemical potential can give rise to a large parity-odd four-point correlator~\cite{Jazayeri:2023kji}. Remarkably, given the constraints from unitarity and scale invariance, parity-odd parts of correlators must factorise, and the total-energy singularity ($k_T \equiv \sum_i k_i\to 0$) should be absent at tree-level~\cite{Stefanyszyn:2023qov,Stefanyszyn:2024msm,Stefanyszyn:2025yhq}. This result was derived from the properties of bulk-bulk propagators under a set of mild assumptions, which are applicable to spinning fields with a chemical potential and sound speed. With the complete analytical expressions now available, we are in a position to perform an explicit consistency check. Consider the example previously discussed, with the interaction~\eqref{eq: eg helical interaction}, the parity-odd component is proportional to
    \begin{equation}
        \braket{\varphi_{\k_1} \varphi_{\k_2} \varphi_{\k_3} \varphi_{\k_4}}'_{\text{PO}} \propto i \hat{\bm{s}} \cdot (\hat{\k}_1 \times \hat{\k}_3) (1-k_1\partial_{k_1}) (1-k_3\partial_{k_3}) 
         \sum_{\substack{{\sf{a}},{\sf{b}}=\pm \\ \lambda=\pm 1}} \lambda\, [\D_{k_{12}}^{\sf{a}}]^3[\D_{k_{34}}^{\sf{b}}]^3 F^{(\lambda)}_{\sf{ab}} \,.
    \end{equation}
    Given that the parity-odd part is purely imaginary, we have omitted certain real prefactors for the sake of simplicity. The homogeneous solutions are already factorised and do not exhibit any $k_T$-singularity. The theorem thus asserts that the contribution from the particular solution to the parity-odd correlator must vanish, as it originates from the fully connected part of the bulk integral, and also contains the $k_T$-singularity. To verify this, we note that the particular solution~\eqref{eq: helical particular solution resummed expression} satisfies the following property 
    \begin{equation}
        \S_+^{(\lambda)} = [\S_-^{(\lambda)}]^* = \S_-^{(-\lambda)}\,.
    \end{equation}
    Then, the summation over helicity gives 
    \begin{equation}
        \sum_{\substack{{\sf{a}}=\pm \\ \lambda=\pm 1}} \lambda \, \S_{\sf{a}}^{(\lambda)} = \S_+^{(+1)} + \S_-^{(+1)} - \S_+^{(-1)} - \S_-^{(-1)} = 0 \,,
    \end{equation}
    which is indeed consistent with the parity-odd  factorisation theorem.

    \item {\it Comparison with the cosmological low-speed collider result}: As mentioned above, there is an alternative approach to obtaining the convergent solution besides the resummation method employed here. For example, in previous studies involving reduced sound speed but not chemical potential, the ansatz for the particular solution was formulated as $\sum_{m,n=0}^{\infty}\left(a_{mn}+b_{mn}\log(u)\right)u^{-m}\left({u}/{v}\right)^n$ when $u<v$~\cite{Jazayeri:2022kjy}. The negative power of $u$ here manifestly ensures the convergence of series when $v>u>1$. We can adopt a modified ansatz similar to those used in previous studies. However, deriving and solving the recurrence relations for the coefficients proves to be non-trivial. This raises the question: how are these two different methods connected? To facilitate a comparison between both methods, we set the chemical potential to zero, $\kappa=0$. In this case, the solutions for the particular case with unit sound speed reads
    \begin{equation}
        \S_+=\sum_{m,n=0}^{\infty}\frac{(-1)^n(n+1)_{2m}}{2^{2m+1}(\frac{n}{2}+\frac{1}{4}+\frac{i\mu}{2})_{m+1}(\frac{n}{2}+\frac{1}{4}-\frac{i\mu}{2})_{m+1}}u^{2m+1}\left(\frac{u}{v}\right)^n \,.
    \end{equation}
    This series is convergent for $u<1$, but we can still perform the resummation along the $m$-direction. After resummation, the series becomes
    \begin{equation}
        \S_+ = 2u\sum_{n=0}^{\infty} \frac{(-1)^n}{(n+\tfrac{1}{2})^2+\mu^2}\left(\frac{u}{v}\right)^n \pFq{3}{2}{1 \,\, \tfrac{n+1}{2}, \tfrac{n}{2}+1}{\tfrac{5}{4}+\tfrac{n}{2}-\tfrac{i\mu}{2} \,\, \tfrac{5}{4}+\tfrac{n}{2}+\tfrac{i\mu}{2}}{u^2} \,.
    \end{equation}
    The two layers of infinite series are now reorganised into a single series involving a hypergeometric function. It can be explicitly checked that the result remains valid even for $c_s<1$. To connect to the results derived in~\cite{Jazayeri:2022kjy}, one can expand the hypergeometric function around $u\to \infty$. This reveals a structure consistent with $[a+b \log(u)]u^{-m}$, where the coefficients involve harmonic numbers. In summary, we started here with a series ansatz expansion around $u\to 0$ but after summation, the result converges at the significantly larger radius. The previously obtained result corresponds to the $u\to\infty$ expansion of this expression. The correspondence is quite natural, as the ansatz $[a+b \log(u)]u^{-m}$ was originally formulated based on an analysis of the bootstrap equations in the limit $u,v\to \infty$.
\end{itemize}

\subsection{Via Spectral Representation}
\label{subsec: Via Spectral Representation}

We now turn to the computation of the helical seed correlator using the spectral approach recently developed in~\cite{Werth:2024mjg}. Essentially, we extend the spectral representation of the time-ordered bulk-bulk SK propagator to include a helical chemical potential. This allows us to factorise the time integrals at the expense of introducing an additional spectral integration over the mass parameter of the exchanged field. Performing the remaining spectral integral naturally yields a single series that converges in all physical kinematic configurations, in particular outside the unit circle for $c_s< 1$. Unlike the bootstrap approach, which relies on a fortunate mathematical trick to resum a series layer and analytically continue the solution beyond its convergence disk, the spectral method achieves this continuation intrinsically.

\paragraph{Spectral representation of helical massive propagators.} The essence of the spectral approach lies in the ability to trade time ordering for a contour integral in the complex plane. Instead of implementing the spectral approach for the time-ordered SK propagator $\G_{++}^\lambda$, we introduce the following modified bulk-to-bulk propagator\footnote{This propagator is the one used in computing wavefunction coefficients, see e.g.~\cite{Anninos:2014lwa, Goon:2018fyu,Stefanyszyn:2023qov} for more details. 
For fields in the principal series, the term $\sigma_k^\lambda(\tau_0)/\sigma_k^{\lambda*}(\tau_0)$ oscillates as a function of $\tau_0$. We bypass this complication by first considering fields in the complementary series, for which  $\sigma_k^\lambda(\tau_0)/\sigma_k^{\lambda*}(\tau_0)$ becomes constant, and then analytically continue the results.
}
\begin{equation}
\label{eq: bulk-bulk pro def}
    \G^\lambda(k; \tau', \tau'') = \G_{++}^\lambda(k; \tau', \tau'') - \Delta\G^\lambda(k; \tau', \tau'')\,,
\end{equation}
where
\begin{equation}
    \Delta\G^\lambda(k; \tau', \tau'') \equiv \frac{\sigma_k^\lambda(\tau_0)}{\sigma_k^{\lambda*}(\tau_0)}\, \sigma_k^{\lambda*}(\tau')\sigma_k^{\lambda*}(\tau'')\,,
\end{equation}
with $\tau_0\to0$ being a small late-time cutoff. The additional piece $\Delta\G^\lambda$, symmetric under $\tau' \leftrightarrow \tau''$, enforces the correct boundary conditions as it makes $\G^\lambda$ vanish when either of its time arguments is taken to zero. Also note that it is already factorised in time, so that the resulting time integrals, once the propagator is inserted inside a correlator, are performed easily. After Wick rotating time by defining $\tau\equiv ie^{i\epsilon}\chi$ with $\epsilon>0$, the integral representation of the bulk-bulk propagator $\G^\lambda$ for a given helicity reads
\begin{equation}
\label{eq: spectral rep G propagator}
    \G^\lambda(k; \chi', \chi'') = \frac{1}{2k} \int_{-\infty}^{+\infty} \d\rho\, \N_{\rho, \kappa}^\lambda \frac{W_{-i\lambda\kappa, i\rho}(2k\chi') W_{-i\lambda\kappa, i\rho}(2k\chi'')}{\rho^2+\nu^2}\,,
\end{equation}
where we have defined
\begin{equation}
    \N_{\rho, \kappa}^\lambda \equiv \frac{\rho}{2\pi^2} \sinh(2\pi\rho) \Gamma(\tfrac{1}{2}+i\lambda\kappa-i\rho) \Gamma(\tfrac{1}{2}+i\lambda\kappa+i\rho)\,.
\end{equation}
Here, to simplify technicalities, we shall temporarily consider the spectral representation for light fields lying in the complementary series representation of the de Sitter group, where the dimensionless mass parameter is given by $\nu=\sqrt{9/4-m^2/H^2}\in \mathbb{R}$. Since the correlators are analytic functions of the mass parameter at $\nu=0$, we are allowed to analytically continue to heavy masses via $\nu\to i\mu$ at the end of the spectral calculation. In the limit $\kappa\to0$, we recover the de Sitter density of states $\N_{\rho, \kappa\to0}^\lambda = \tfrac{\rho}{\pi}\sinh(\pi\rho)$. The appearance of an imaginary component when $\kappa\neq 0$ reflects parity violation. The detailed derivation of Eq.~\eqref{eq: spectral rep G propagator} is given in the insert below. In the original spectral representation of the Feynman propagator for massive fields (with $\kappa=0$)~\cite{Werth:2024mjg}, the projection of incoming (Hankel function) states onto outgoing (Bessel) states---which encodes particle production---is effectively performed with a suitable $i\epsilon$ prescription. Interestingly, here, this projection is captured by combining the Feynman propagator $\G_{++}^\lambda$ with the boundary term $\Delta\G^\lambda$ and considering off-shell heavy states that are set on shell on the imaginary axis, yielding a propagator of light fields. 

\begin{framed}
{\small \noindent {\it Derivation.}---In this insert, we prove the spectral representation of the bulk-bulk propagator~\eqref{eq: spectral rep G propagator}, following~\cite{WhittakerRep}. We first consider the case $\chi'<\chi''$. Using the connection formula
\begin{equation}
    W_{-i\lambda \kappa, i\rho}(z) = \frac{\Gamma(-2i\rho)}{\Gamma(\tfrac{1}{2}-i\rho+i\lambda \kappa)}M_{-i\lambda \kappa, i\rho}(z) + \frac{\Gamma(2i\rho)}{\Gamma(\tfrac{1}{2}+i\rho+i\lambda \kappa)}M_{-i\lambda \kappa, -i\rho}(z)\,,
\end{equation}
to project the term $W_{-i\lambda \kappa, i\rho}(2k\chi')$ onto a linear combination of Whittaker-$M$ functions, along with the symmetry relation $W_{-i\lambda \kappa, i\rho}(z) = W_{-i\lambda \kappa, -i\rho}(z)$, the integral~\eqref{eq: spectral rep G propagator} can be written
\begin{equation}
    \frac{-i}{2k}\int_{-\infty}^{+\infty} \frac{\d\rho}{\pi} \frac{\rho}{\rho^2+\nu^2} \frac{\Gamma(\tfrac{1}{2}+i\lambda\kappa-i\rho)}{\Gamma(1-2i\rho)}M_{-i\lambda \kappa, -i\rho}(2k\chi') W_{-i\lambda \kappa, -i\rho}(2k\chi'')\,.
\end{equation}
We have also used the Euler's reflection formula $\Gamma(1-2i\rho)\Gamma(2i\rho) = -i\pi/\sinh(2\pi\rho)$ to arrive at this expression. We choose to close the integration contour in the upper-half complex plane, since the arc at infinity vanishes for $\chi'<\chi''$. The integrand exhibits two distinct types of poles: (i) on-shell poles located at $\rho = \pm i\nu$ coming from the term $1/(\rho^2+\nu^2)$, and (ii) a tower of helical poles located at $\rho = -i(n+\tfrac{1}{2})+\lambda \kappa$ coming from the term $\Gamma(\tfrac{1}{2}+i\lambda\kappa-i\rho)$, that are shifted away from the imaginary axis by the non-zero chemical potential. The analytic structure of the integrand in the complex $\rho$ plane is shown in the following figure

\begin{center}
    \centering
    \begin{tikzpicture}[scale = 2]
        \draw[black, ->] (-1.6,0) -- (1.6,0) coordinate (xaxis);
        \draw[black, ->] (0,-1.6) -- (0,1.6) coordinate (yaxis);
        \node at (1.9, 0) {$\Re\rho$};
        \node at (0, 1.7) {$\Im\rho$};
		
        \draw[pyred, fill = pyred] (0, 0.5) circle (.03cm);
        \node at (0.25, 0.5) {\textcolor{pyred}{$+i\nu$}};
        \draw[pyblue, fill = pyblue] (0, -0.5) circle (.03cm);
        \node at (-0.25, -0.5) {\textcolor{pyblue}{$-i\nu$}};
		
        \draw[pyblue, fill = pyblue] (-0.1, 0.2) circle (.03cm);
        \draw[pyblue, fill = pyblue] (-0.1, 0.4) circle (.03cm);
        \draw[pyblue, fill = pyblue] (-0.1, 0.6) circle (.03cm);
        \draw[pyblue, fill = pyblue] (-0.1, 0.8) circle (.03cm);
        \draw[pyblue, fill = pyblue] (-0.1, 1) circle (.03cm);
        \draw[pyblue, fill = pyblue] (-0.1, 1.2) circle (.03cm);
        \draw[pyblue, fill = pyblue] (-0.1, 1.4) circle (.03cm);
        \node at (-0.7, 0.7) {\textcolor{pyblue}{\tiny $+i\,(n+\tfrac{1}{2})- \lambda\kappa$}};
		
        \draw[pyred, fill = pyred] (0.1, -0.2) circle (.03cm);
        \draw[pyred, fill = pyred] (0.1, -0.4) circle (.03cm);
        \draw[pyred, fill = pyred] (0.1, -0.6) circle (.03cm);
        \draw[pyred, fill = pyred] (0.1, -0.8) circle (.03cm);
        \draw[pyred, fill = pyred] (0.1, -1) circle (.03cm);
        \draw[pyred, fill = pyred] (0.1, -1.2) circle (.03cm);
        \draw[pyred, fill = pyred] (0.1, -1.4) circle (.03cm);
        \node at (0.7, -0.7) {\textcolor{pyred}{\tiny $-i\,(n+\tfrac{1}{2})+ \lambda\kappa$}};
		
	\path[pyred, draw, line width = 0.8pt, postaction = decorate, decoration={markings,
			mark=at position 0.1 with {\arrow[line width=1pt]{>}},
			mark=at position 0.45 with {\arrow[line width=1pt]{>}}}] (-1.4, 0.05) -- (1.4, 0.05) arc (0:180:1.4) -- (1.4, 0.05);
		
        \node at (1.5, 1.2) {\textcolor{pyred}{$\chi'<\chi''$}};
        \path[pyblue, draw, line width = 0.8pt, postaction = decorate, 
		decoration={markings,
			mark=at position 0.2 with {\arrow[line width=1pt]{>}},
			mark=at position 0.45 with {\arrow[line width=1pt]{>}}}] (-1.4, -0.05) -- (1.4, -0.05) arc (0:-180:1.4) -- (1.4, -0.05);
	
        \node at (1.5, -1.2) {\textcolor{pyblue}{$\chi'>\chi''$}};
	\end{tikzpicture}
\end{center}

\noindent By Cauchy's residue theorem, since we close the contour upwards for $\chi'<\chi''$, we pick up only the pole $\rho=+i\nu$, which yields
\begin{equation}
\label{eq: spectral G after residue}
    \G^\lambda(k; \chi', \chi'') = \frac{1}{2k} \frac{\Gamma(\tfrac{1}{2}+\nu+i\lambda \tilde\kappa)}{\Gamma(1+2\nu)}\, M_{-i\lambda \kappa, \nu}(2k\chi') W_{-i\lambda \kappa, \nu}(2k\chi'')\,.
\end{equation}
This is not yet the desired expression. To connect with the bulk-bulk propagator $\G^\lambda$, we note that
\begin{equation}
\label{eq: DeltaG}
    \Delta\G^\lambda(k; \tau', \tau'') = -e^{i\pi(\nu+\frac{1}{2})}\frac{\Gamma(\tfrac{1}{2}+\nu+i \lambda \kappa)}{\Gamma(\tfrac{1}{2}+\nu-i \lambda \kappa)}\,\sigma_k^{\lambda*}(\tau')\sigma_k^{\lambda*}(\tau'')\,.
\end{equation}
This formula is strictly valid only for light fields as we have dropped the most decaying mode when approaching $\tau_0\to0$. For heavy fields $\mu=-i\nu>0$, both modes $(k\tau)^{\frac{1}{2}\pm i\mu}$ are decaying with the same rate while oscillating. We will see however that the derived final expression for the helical seed correlator can straightforwardly be analytically continued to the heavy case. In deriving~\eqref{eq: DeltaG}, we have used $W_{i\lambda\kappa, \nu}^*(z) = W_{-i\lambda\kappa, \nu}(z^*)$, valid for all $z\in \mathbb{C}\backslash\{\mathbb{R}_-\}$ away from the branch cut. By using the (inverse) connection formula
\begin{equation}
    \frac{1}{\Gamma(1+2\nu)} M_{i\lambda \kappa, \nu}(z) = \frac{e^{\pm i \pi(i\lambda \kappa-\nu-\frac{1}{2})}}{\Gamma(\tfrac{1}{2}+\nu+i \lambda \kappa)} W_{i\lambda \kappa, \nu}(z) + \frac{e^{\mp \pi\lambda \kappa}}{\Gamma(\tfrac{1}{2}+\nu-i \lambda \kappa)} W_{-i\lambda \kappa, \nu}(e^{\pm i\pi}z)\,,
\end{equation}
the bulk-bulk propagator for a given helicity reduces to
\begin{equation}
    \G^\lambda(k; \tau', \tau'') = \frac{1}{2k} \frac{\Gamma(\tfrac{1}{2}+\nu+i\lambda \tilde\kappa)}{\Gamma(1+2\nu)}\, M_{-i\lambda \kappa, \nu}(-2ik\tau') W_{-i\lambda \kappa, \nu}(-2ik\tau'')\Theta(\tau'-\tau'') + (\tau'\leftrightarrow\tau'')\,,
\end{equation}
where we have used the formula $M_{-i\lambda \kappa, \nu}(e^{\pm i\pi}z) = e^{\pm i\pi(\nu+\frac{1}{2})} M_{i\lambda\kappa, \nu}(z)$ to rotate the argument without crossing the branch cut which lies on the negative real axis. By defining $\tau\equiv i e^{i\epsilon}\chi$ with $\epsilon>0$, the propagator $\G^\lambda$ reads
\begin{equation}
\label{eq: Wick time G}
    \G^\lambda(k; \chi', \chi'') = \frac{1}{2k} \frac{\Gamma(\tfrac{1}{2}+\nu+i\lambda \tilde\kappa)}{\Gamma(1+2\nu)}\, M_{-i\lambda \kappa, \nu}(2k\chi') W_{-i\lambda \kappa, \nu}(2k\chi'')\Theta(\chi''-\chi') + (\chi'\leftrightarrow\chi'')\,,
\end{equation}
in the Wick rotated time domain, so that the arguments of the Whittaker functions lie on the real axis. This exactly boils down to the form~\eqref{eq: spectral G after residue} for $\chi'<\chi''$. For the case $\chi''<\chi'$, one needs to expand the second Whittaker term $W_{-i\lambda\kappa, i\rho}(2k\chi'')$ in~\eqref{eq: spectral rep G propagator} and close the contour in the lower-half complex plane. This completes the proof.
}
\end{framed}

\paragraph{Time integrals.} We have now the necessary ingredient to compute the helical seed correlator~\eqref{eq: helical seed correlator}. The non-time-ordered contribution $F_{+-}^{(\lambda)}$, for which the time integrals factorise, is given by\footnote{Here we have analytically continued back to heavy masses via $\nu\to i\mu$. We are guaranteed to land on a correct result since the correlators are analytic functions of the dimensionless mass parameter near the origin $\nu=0$.}
\begin{equation}
\label{eq: F+- spectral rep}
    F_{+-}^{(\lambda)} = \frac{\pi^2e^{-\pi\lambda  \kappa}}{2\cosh^2(\pi\mu)} \, \pregFq{2}{1}{\tfrac{1}{2}-i\mu, \tfrac{1}{2}+i\mu}{1+i \lambda  \kappa}{\frac{1}{2}-\frac{1}{2u}} \pregFq{2}{1}{\tfrac{1}{2}-i\mu, \tfrac{1}{2}+i\mu}{1-i\lambda \kappa}{\frac{1}{2}-\frac{1}{2v}}\,,
\end{equation}
where we have used the kinematic variables defined in~\eqref{eq: original kinematic variables}.  Naturally, we have $F_{-+}^{(\lambda)} = F_{+-}^{(\lambda)*}$. After Wick rotating time, the non-time ordered contribution reads
\begin{equation}
    \begin{aligned}
        F_{++}^{(\lambda)} &= -s \int_0^{+\infty}\frac{\d\chi'}{\chi'}\frac{\d\chi''}{\chi''}e^{-c_s k_{12}\chi'}  \G_{++}^\lambda(s; \chi', \chi'')e^{-c_s k_{34}\chi''} \\
        &\equiv F_\G^{(\lambda)} + F_{\Delta\G}^{(\lambda)}\,,
    \end{aligned}
\end{equation}
where we have split the integral according to~\eqref{eq: bulk-bulk pro def}. The piece $F_{\Delta\G}^{(\lambda)}$ coming from the difference between the Feynman propagator and the wavefunction bulk-bulk propagator is factorised in time and can be easily computed, yielding
\begin{equation}
\label{eq: boundary term spectral rep}
    F_{\Delta\G}^{(\lambda)} = \frac{i\pi^2e^{-\pi(\mu+\lambda  \kappa)}}{2\cosh^2(\pi\mu)} \frac{\Gamma(\tfrac{1}{2}+i \lambda \kappa+i\mu)}{\Gamma(\tfrac{1}{2}-i \lambda \kappa+i\mu)}\, \pregFq{2}{1}{\tfrac{1}{2}-i\mu, \tfrac{1}{2}+i\mu}{1+i \lambda  \kappa}{\frac{1}{2}-\frac{1}{2u}} \pregFq{2}{1}{\tfrac{1}{2}-i\mu, \tfrac{1}{2}+i\mu}{1+i\lambda \kappa}{\frac{1}{2}-\frac{1}{2v}}\,.
\end{equation}
Using the spectral representation~\eqref{eq: spectral rep G propagator} and after swapping the spectral and the time integrals, the remaining contribution reads
\begin{equation}
\label{eq: remaining spectral integral}
    F_\G^{(\lambda)} =\frac{-\pi^2}{2} \int_{-\infty}^{+\infty} \frac{\d\rho \,\N_{\rho, \kappa}^\lambda}{\cosh^2(\pi\rho)} \frac{\pregFq{2}{1}{\tfrac{1}{2}-i\rho, \tfrac{1}{2}+i\rho}{1+i \lambda  \kappa}{\frac{1}{2}-\frac{1}{2u}} \pregFq{2}{1}{\tfrac{1}{2}-i\rho, \tfrac{1}{2}+i\rho}{1+i \lambda  \kappa}{\frac{1}{2}-\frac{1}{2v}}}{\rho^2+\nu^2}\,.
\end{equation}
This spectral representation makes it manifest that the integral $F_\G^{(\lambda)}$ is obtained by gluing two off-shell three-point functions, and we naturally expect the solution to be of higher transcendentality than the factorised integral $F_{+-}^{(\lambda)}$ due to the additional spectral integration. 

\paragraph{Collecting poles.} Computing the helical seed correlator boils down to evaluating the spectral integral~\eqref{eq: remaining spectral integral}. The off-shell three-point function can be written in a compact form using the Legendre function\footnote{Note that this function is implemented in \textsf{Mathematica} by \textsf{LegendreP[$i\mu-1/2, -i\lambda\kappa, 3, z$]} where the ``3'' encodes the correct branch cut discontinuity.}
\begin{equation}
    P_{i\rho-1/2}^{-i\lambda\kappa}(z) \equiv \left(\frac{z+1}{z-1}\right)^{-i\lambda\kappa/2} \pregFq{2}{1}{\tfrac{1}{2}-i\rho, \tfrac{1}{2}+i\rho}{1+i \lambda  \kappa}{\frac{1}{2}-\frac{z}{2}}\,,
\end{equation}
so that $F_\G^{(\lambda)}$ can be written as
\begin{equation}
\label{eq: defining property spectral rep}
    \begin{aligned}
        F_\G^{(\lambda)} &=\frac{-\pi^2}{2}\left[\left(\frac{1-u}{1+u}\right) \left(\frac{1-v}{1+v}\right)\right]^{-\lambda i \kappa/2} \int_{-\infty}^{+\infty} \frac{\d\rho \,\N_{\rho, \kappa}^\lambda}{\cosh^2(\pi\rho)}\frac{P_{i\rho-1/2}^{-i\lambda\kappa}(u^{-1}) P_{i\rho-1/2}^{-i\lambda\kappa}(v^{-1})}{\rho^2+\nu^2} \\
        &\equiv \left[\left(\frac{1-u}{1+u}\right) \left(\frac{1-v}{1+v}\right)\right]^{-\lambda i \kappa/2} \tilde{F}_\G^{(\lambda)}\,,
    \end{aligned}
\end{equation}
where we have defined the ``stripped'' integral $\tilde{F}_\G^{(\lambda)}$. We finish this last integral using Cauchy residue's theorem. To this end, we need to understand the analytic structure of the meromorphic integrand, i.e.~identify poles, and disentangle the large-$\rho$ behaviour to close the contour accordingly. The Legendre-$P$ basis behaves well (meaning that it has a single mode) around the origin $\rho\approx 0$, whereas the associated Legendre-$Q$ basis behaves well at infinity $\rho\to\infty$. We therefore use the connection formula 
\begin{equation}
    \Gamma(\tfrac{1}{2}+i\rho + \lambda i \kappa)\Gamma(\tfrac{1}{2}-i\rho + \lambda i \kappa) P_{i\rho-1/2}^{-\lambda i \kappa}(x) = \frac{e^{-i \pi/2 + \pi\lambda \kappa}}{\sinh(\pi\rho)}\left[Q_{-i\rho-1/2}^{\lambda i \kappa}(x) - Q_{+i\rho-1/2}^{\lambda i \kappa}(x)\right]\,,
\end{equation}
to decompose the integral into two pieces $\tilde{F}_\G^{(\lambda)} = \tilde{F}_\G^{(\lambda), 0} + \tilde{F}_\G^{(\lambda), <, >}$, with (we have accounted for the shadow symmetry of the integrand $\mu\leftrightarrow-\mu$)
\begin{equation}
    \begin{aligned}
        \tilde{F}_\G^{(\lambda), 0} &\equiv \frac{2e^{2\pi\lambda\kappa}}{\pi} \int_{-\infty}^{+\infty}\frac{\d\rho}{2\pi} \frac{\textcolor{pygreen}{|\Gamma(1+i\rho)|^2} \textcolor{pyred}{|\Gamma(\tfrac{1}{2}+i\rho)|^2}}{\Gamma(\tfrac{1}{2}+i\lambda\kappa+i\rho) \Gamma(\tfrac{1}{2}+i\lambda\kappa-i\rho)} \frac{Q_{-i\rho-1/2}^{i\lambda\kappa}(u^{-1}) Q_{-i\rho-1/2}^{i\lambda\kappa}(v^{-1})}{\textcolor{pyblue}{\rho^2+\nu^2}}\,, \\
        \tilde{F}_\G^{(\lambda), <, >} &\equiv -\frac{2e^{2\pi\lambda\kappa}}{\pi} \int_{-\infty}^{+\infty}\frac{\d\rho}{2\pi} \frac{\textcolor{pygreen}{|\Gamma(1+i\rho)|^2} \textcolor{pyred}{|\Gamma(\tfrac{1}{2}+i\rho)|^2}}{\Gamma(\tfrac{1}{2}+i\lambda\kappa+i\rho) \Gamma(\tfrac{1}{2}+i\lambda\kappa-i\rho)} \frac{Q_{-i\rho-1/2}^{i\lambda\kappa}(u^{-1}) Q_{+i\rho-1/2}^{i\lambda\kappa}(v^{-1})}{\textcolor{pyblue}{\rho^2+\nu^2}} \,,
    \end{aligned}
\end{equation}
where we have used the Legendre duplication formula $\Gamma(z)\Gamma(\tfrac{1}{2}+z)=2^{1-2z}\sqrt{\pi}\,\Gamma(2z)$.\footnote{Note that whenever we write $|\Gamma(x+iy)|^2$ with both real $x$ and $y$, we actually mean $\Gamma(x+iy)\Gamma(x-iy)$ as there is an ambiguity when locating the poles.} Assuming $|u|<|v|$, we close the contour in the upper-half complex plane for both integrals as again the arcs at infinity vanish. The two integrands have several poles with different physical origins:
\begin{itemize}
    \item The \textcolor{pyblue}{particle production poles} coming from $1/(\rho^2+\nu^2)$ are located at \textcolor{pyblue}{$\rho = \pm i\nu$}. Picking up these poles extracts the non-analytic signal contributions, which read
    \begin{equation}
    \label{eq: collider signal contributions}
        \begin{aligned}
            \tilde{F}_\G^{(\lambda), 0} &\supset -\frac{ie^{2\pi\lambda \kappa}}{\pi\mu} \frac{|\Gamma(\tfrac{1}{2}+i\mu)|^2|\Gamma(1+i\mu)|^2}{\Gamma(\tfrac{1}{2}+i\lambda\kappa+i\mu) \Gamma(\tfrac{1}{2}+i\lambda\kappa-i\mu)}Q_{i\mu-1/2}^{i\lambda \kappa}(u^{-1})Q_{i\mu-1/2}^{i\lambda \kappa}(v^{-1}) \,, \\
             \tilde{F}_\G^{(\lambda), <, >} &\supset \frac{ie^{2\pi\lambda \kappa}}{\pi\mu} \frac{|\Gamma(\tfrac{1}{2}+i\mu)|^2|\Gamma(1+i\mu)|^2}{\Gamma(\tfrac{1}{2}+i\lambda\kappa+i\mu) \Gamma(\tfrac{1}{2}+i\lambda\kappa-i\mu)}Q_{i\mu-1/2}^{i\lambda \kappa}(u^{-1})Q_{-i\mu-1/2}^{i\lambda \kappa}(v^{-1})\,,
        \end{aligned}
    \end{equation}
    where we have set $\nu=i\mu$. These expressions are factorised in $u$ and $v$ as they should.
    \item The \textcolor{pygreen}{tower of density of states poles} coming from the factor $|\Gamma(1+i\rho)|^2$ are located on the entire imaginary axis \textcolor{pygreen}{$\rho = \pm i(n+1)$}. These two towers exactly cancel again each other, as also noted in~\cite{Werth:2024mjg}.
    \item The \textcolor{pyred}{EFT poles} coming from the factor $|\Gamma(\tfrac{1}{2}+i\rho)|^2$ are located at $\textcolor{pyred}{\rho=\pm i(n+\tfrac{1}{2})}$. We therefore pick up the poles in the upper-half complex plane and sum over $n\geq0$. Writing the Legendre-$Q$ function in its hypergeometric form\footnote{Note that the factor $\Gamma(\tfrac{1}{2}-i\rho+\lambda i\kappa)$ cancels against itself inside the integral. As such, this term does not bring additional tower of poles.}
    \begin{equation}
        Q_{-i\rho-1/2}^{i\lambda\kappa}(z) \equiv e^{-\pi\lambda \kappa} \sqrt{\frac{\pi}{2}} \Gamma(\tfrac{1}{2}-i\rho+\lambda i\kappa)(z^2-1)^{\lambda i \kappa/2} 2^{i\rho} z^{-\tfrac{1}{2}+i\rho -\lambda i \kappa}  \pregFq{2}{1}{\tfrac{3/2-i\rho +\lambda i \kappa}{2}, \tfrac{1/2-i\rho+\lambda i \kappa}{2}}{1-i\rho}{\frac{1}{z^2}}\,,
    \end{equation}
    the EFT contribution coming from the symmetric integral $F_\G^{(\lambda), 0}$ reads
    \begin{equation}
    \label{eq: EFT 1}
        \begin{aligned}
            &F_{\G, {\rm EFT}}^{(\lambda), 0} = -\pi \left[(u+1)(v+1)\right]^{\lambda i \kappa}\times \sum_{n=0}^{+\infty} \frac{(1+2n)}{(n+\tfrac{1}{2})^2+\mu^2}\left(\frac{v}{2}\right)^{n+1}\left(\frac{u}{2}\right)^{n+1}\\
            &\times\frac{\Gamma(1+n+i\lambda\kappa)}{\Gamma(-n+i\lambda\kappa)}\,\pregFq{2}{1}{\tfrac{1+n}{2}+\tfrac{\lambda i \kappa}{2}, 1+\tfrac{n}{2}+\tfrac{\lambda i \kappa}{2}}{\tfrac{3}{2}+n}{u^2} \, \pregFq{2}{1}{\tfrac{1+n}{2}+\tfrac{\lambda i \kappa}{2}, 1+\tfrac{n}{2}+\tfrac{\lambda i \kappa}{2}}{\tfrac{3}{2}+n}{v^2}\,.
        \end{aligned}
    \end{equation}
    This EFT series vanishes for $\kappa=0$ due to the factor $1/\Gamma(-n+i\lambda\kappa)$, in accordance with with~\cite{Werth:2024mjg}, meaning that this EFT series is a purely boost-breaking contribution to the full correlator. The second EFT series coming from $F_\G^{(\lambda), <, >}$ reads
    \begin{equation}
    \label{eq: EFT 2}
        \begin{aligned}
            F_{\G, {\rm EFT}}^{(\lambda),<,>} =& u\left[(u+1)(v+1)\right]^{\lambda i \kappa}\times \sum_{n=0}^{+\infty} \frac{(-1)^n}{(n+\tfrac{1}{2})^2+\mu^2}\left(\frac{u}{v}\right)^{n}\\
            &\times \pFq{2}{1}{\tfrac{1+n}{2}+\tfrac{\lambda i \kappa}{2}, 1+\tfrac{n}{2}+\tfrac{\lambda i \kappa}{2}}{\tfrac{3}{2}+n}{u^2}\,\pFq{2}{1}{\tfrac{1-n}{2}+\tfrac{\lambda i \kappa}{2}, -\tfrac{n}{2}+\tfrac{\lambda i \kappa}{2}}{\tfrac{1}{2}-n}{v^2} \,.
        \end{aligned}
    \end{equation}
    This expression reduces to the one found in~\cite{Werth:2024mjg} for $\kappa=0$. The total EFT series contribution is given by the sum of these two pieces: $F_{\G, {\rm EFT}}^{(\lambda)} = F_{\G, {\rm EFT}}^{(\lambda),0} + F_{\G, {\rm EFT}}^{(\lambda),<,>}$. The result for $|u|>|v|$ is simply found after swapping $u$ and $v$.
\end{itemize}

\begin{figure}[h!]
    \centering
    \begin{subfigure}{.5\textwidth}
        \centering
        \includegraphics[width=1\linewidth]{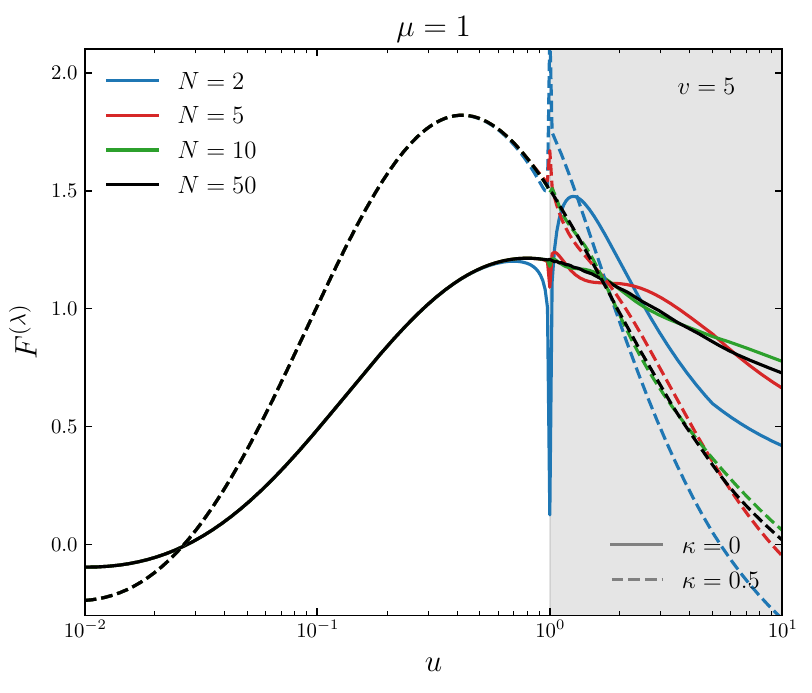}
    \end{subfigure}%
    \begin{subfigure}{.5\textwidth}
        \centering
        \includegraphics[width=1\linewidth]{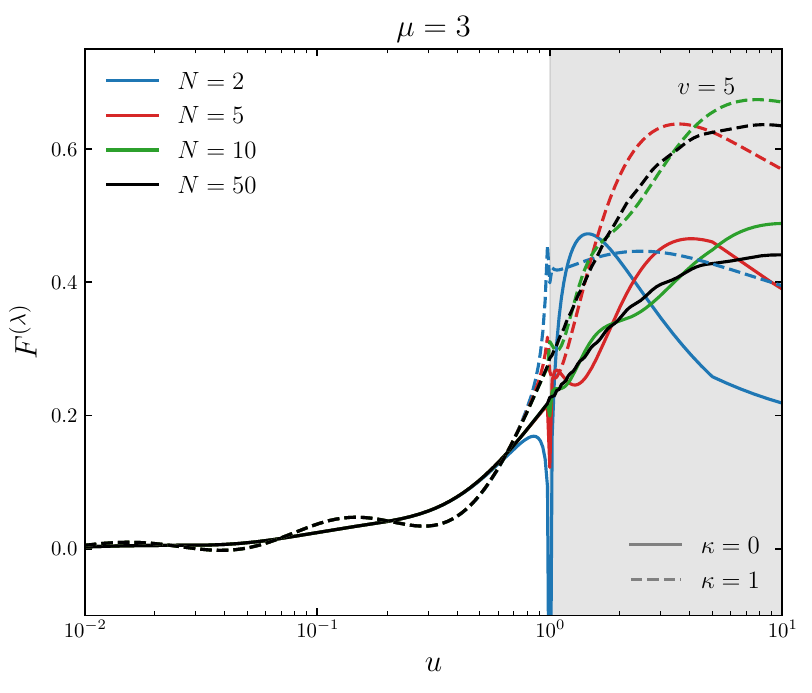}
    \end{subfigure}
   \caption{The full boost-breaking helical seed correlator $F^{(\lambda)}(u, v)$ solved using the spectral representation method, as function of the dimensionless kinematic variable $u$ with fixed $v=5$, for $\mu=1$ (\textit{left panel}) for $\kappa=0$ (\textit{solid line}) and $\kappa=0.5$ (\textit{dashed line}), and $\mu=3$ (\textit{right panel}) for $\kappa=0$ (\textit{solid line}) and $\kappa=1$ (\textit{dashed line}). We by default consider the enhanced helicity $\lambda = -1$ when $\kappa\neq 0$. We include terms in the series from $n=1, \ldots, N$. The shaded gray region corresponds to kinematic configurations outside the unit circle which covers the case of a reduced sound speed (here $c_s=0.1$) for the external field.}
  \label{fig: Helical seed correlator SpectralRep outside circle}
\end{figure}

\paragraph{Full result.} The full result is found by summing all contributions, namely the factorised piece~\eqref{eq: F+- spectral rep}, the boundary term~\eqref{eq: boundary term spectral rep}, the collider contributions from the time-ordered integrals~\eqref{eq: collider signal contributions} together with~\eqref{eq: defining property spectral rep}, as well as the two EFT series~\eqref{eq: EFT 1} and~\eqref{eq: EFT 2}. We have implemented the result in the \textsf{Mathematica} notebook available on the following Github repository [\href{https://github.com/deniswerth/Boost-Breaking-Cosmological-Correlators}{\faGithub}]. We have explicitly checked that this spectral representation exactly matches the expression found using the bootstrap approach, in all kinematic configurations. A noticeable feature of the spectral representation is that the solution is already partially resummed, yielding a single series of (a product of) hypergeometric functions. This automatically encodes the analytic continuation of the helical seed correlator outside the unit kinematic circle. In Fig.~\ref{fig: Helical seed correlator SpectralRep outside circle}, we show the correlator $F^{(\lambda)}$ as function of the kinematic variables extended outside the unit circle up to $u=10$ (which describes the case $c_s=0.1$). We notice that the solution converges well and stabilise when the series contains $N \sim \O(50)$ terms. However, compared to the bootstrap representation shown in Fig.~\ref{fig: Helical seed correlator bootstrap outside circle}, convergence is much slower, and the solution when summing just a few terms in the series features oscillating behaviours before converging to the exact result. We also notice a more pronounced discontinuity as the kinematic variable $u$ crosses the unit circle $u=1$, which fades away as we increase the number of terms in the series. 

\begin{figure}[h!]
    \centering
    \begin{subfigure}{.5\textwidth}
        \centering
        \includegraphics[width=1\linewidth]{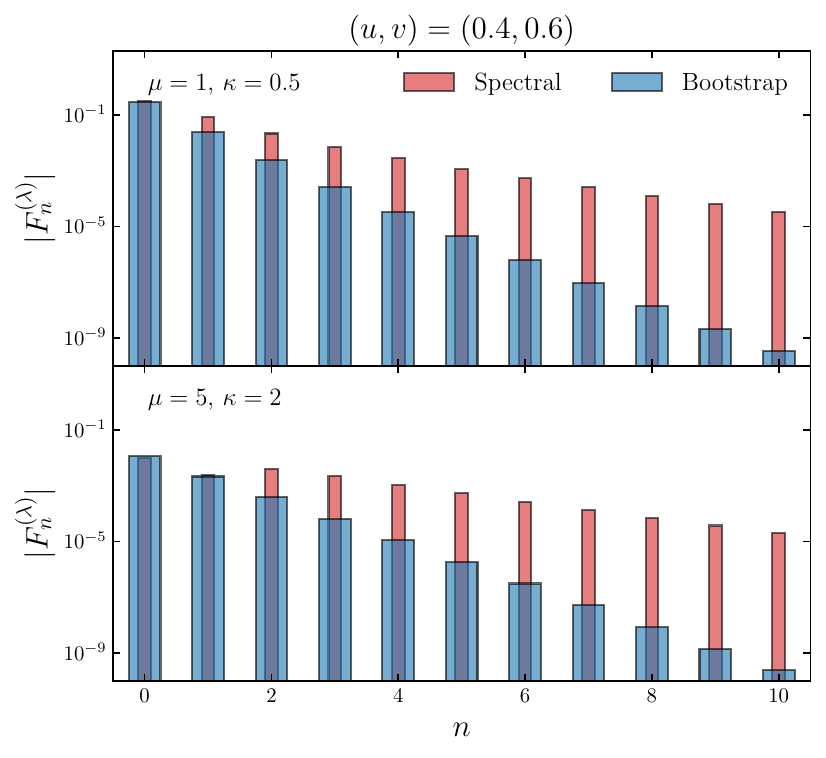}
    \end{subfigure}%
    \begin{subfigure}{.5\textwidth}
        \centering
        \includegraphics[width=1\linewidth]{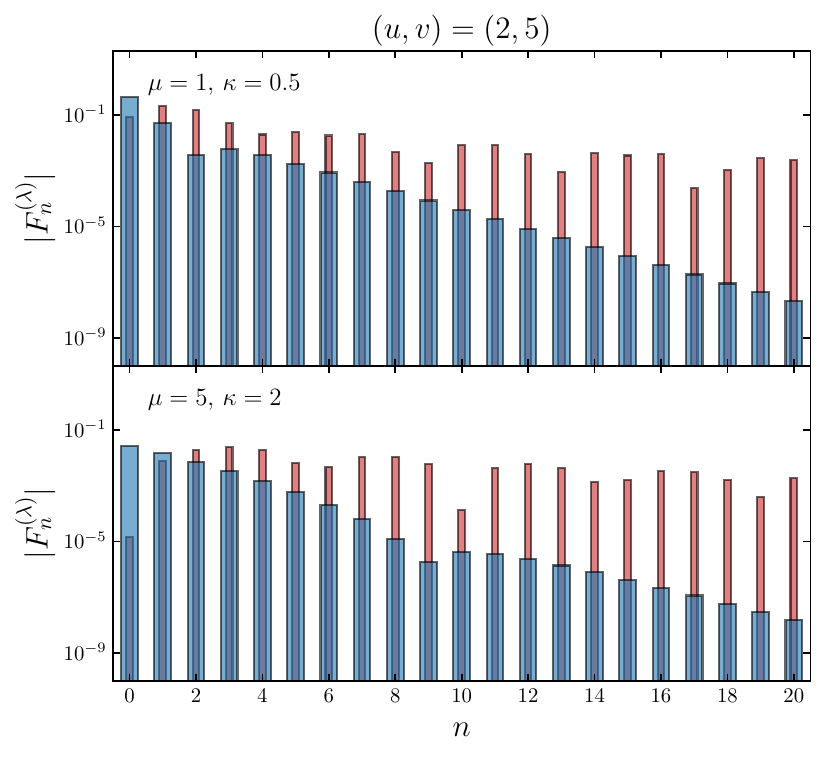}
    \end{subfigure}
    \caption{Amplitude of the series coefficients $|F^{(\lambda)}_n|$ to the boost-breaking helical seed correlator $\sum_n F^{(\lambda)}_n$ as function of $n$ for the bootstrap representation (in \textcolor{pyblue}{blue}) and the spectral representation (in \textcolor{pyred}{red}), both inside the unit kinematic circle (\textit{left panel}) and outside (\textit{right panel}). We have chosen $\mu=1, \kappa=0.5$ (\textit{upper panel}) and $\mu=5, \kappa=2$ (\textit{lower panel}).  We consider by default the negative helicity $\lambda = -1$ when $\kappa\neq 0$.}
    \label{fig: Convergence inside/outside}
\end{figure}

\paragraph{Discussion.} Further comments are in order.

\begin{itemize}
    \item \textit{Boost-breaking EFT series}: The spectral method naturally yields a partially resummed representation. This results in the two EFT series~\eqref{eq: EFT 1} and~\eqref{eq: EFT 2}. Physically, these series correspond to summing over the quasi-normal modes of the mediated field. To see this, let us expand the helical mode function~\eqref{eq: Whittaker mode function} in powers of $(-k\tau)$ at late times
    \begin{equation}
        \sigma_k^\lambda(\tau) = \frac{e^{-\pi\lambda \kappa/2}}{\sqrt{2k}} \sum_{n=0}^{+\infty}\sum_\pm c_n^\pm (-k\tau)^{\frac{1}{2}+n\pm i\mu}\,,
    \end{equation}
    with 
    \begin{equation}
    \label{eq: quasi-normal mode coeff}
        c_n^\pm \equiv \pm \frac{e^{i\frac{\pi}{2}(n+\frac{1}{2})}}{n!} \frac{\pi2^{\frac{1}{2}\pm i\mu}}{\Gamma(\tfrac{1}{2}\mp i\mu - i \lambda \kappa)} \pregFq{2}{1}{-n, \tfrac{1}{2}\pm i\mu - i \lambda \kappa}{1\pm 2i\mu}{2}\, \frac{e^{\pm \pi\mu/2}}{\sinh(2\pi\mu)}\,.
    \end{equation}
    This series is the gradient expansion of the mode function as $(-k\tau)^n = \kp^n$, with $\kp\equiv k/a$ being the physical redshifting momentum (we have set $H=1$). A noticeable feature is that this expansion contains odd powers of $\kp$, which reflects the breaking of boosts by the chemical potential. We can therefore interpret the EFT series $F_{\G, {\rm EFT}}^{(\lambda), 0}$ as summing over odd quasi-normal modes, and the EFT series $F_{\G, {\rm EFT}}^{(\lambda), <, >}$ as summing over the even quasi-normal modes. Therefore, when restoring boosts by setting $\kappa=0$, we naturally expect the entire tower of odd quasi-normal modes to vanish, and consequently $F_{\G, {\rm EFT}}^{(\lambda), 0}=0$ at the level of the helical seed correlator. This is consistent with the fact that when the chemical potential vanishes, there is no linear gradient term in the dispersion relation of the massive field, as can be seen from the equation of motion~\eqref{eq: helical EOM}. 
    Explicitly, setting $\kappa=0$ in the coefficients~\eqref{eq: quasi-normal mode coeff}, we have \cite{NIST:DLMF}
   \begin{equation}
        \pFq{2}{1}{-n, \tfrac{1}{2}\pm i\mu}{1\pm 2i\mu}{2} =\frac{\Gamma\left(\frac{n}{2}+\frac{1}{2}\right)\Gamma\left(1\pm i\mu\right)}{2\sqrt{\pi}\,\Gamma\left(1+\frac{n}{2}\pm i\mu\right)}\times \Big[(-1)^{n}+1\Big]\,,
    \end{equation}
    so that all odd powers of $(-k\tau)$ vanish.

\begin{figure}[h!]
    \centering
    \begin{subfigure}{.5\textwidth}
        \centering
        \includegraphics[width=0.99\linewidth]{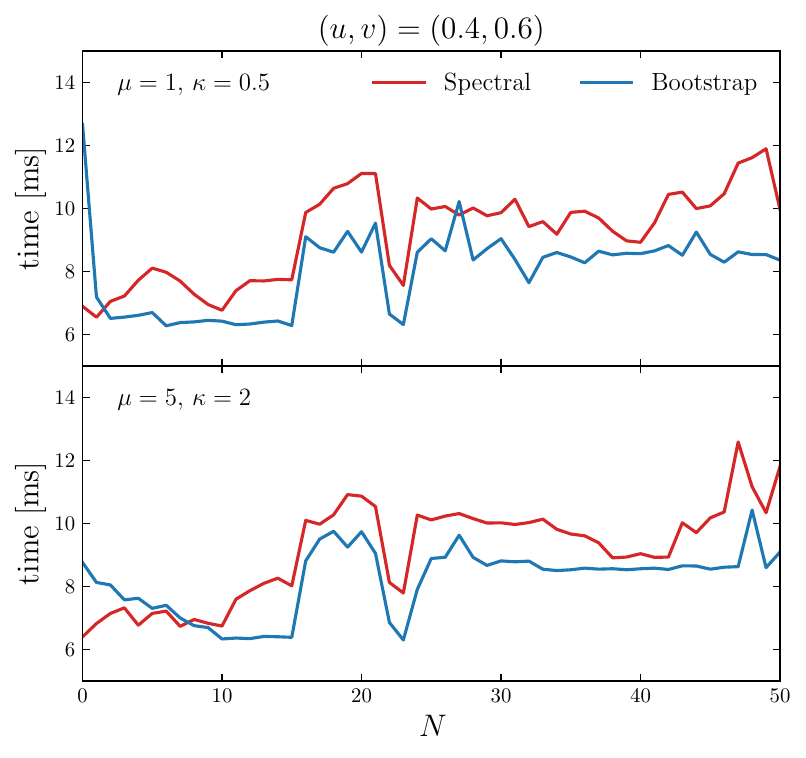}
    \end{subfigure}%
    \begin{subfigure}{.5\textwidth}
        \centering
        \includegraphics[width=1\linewidth]{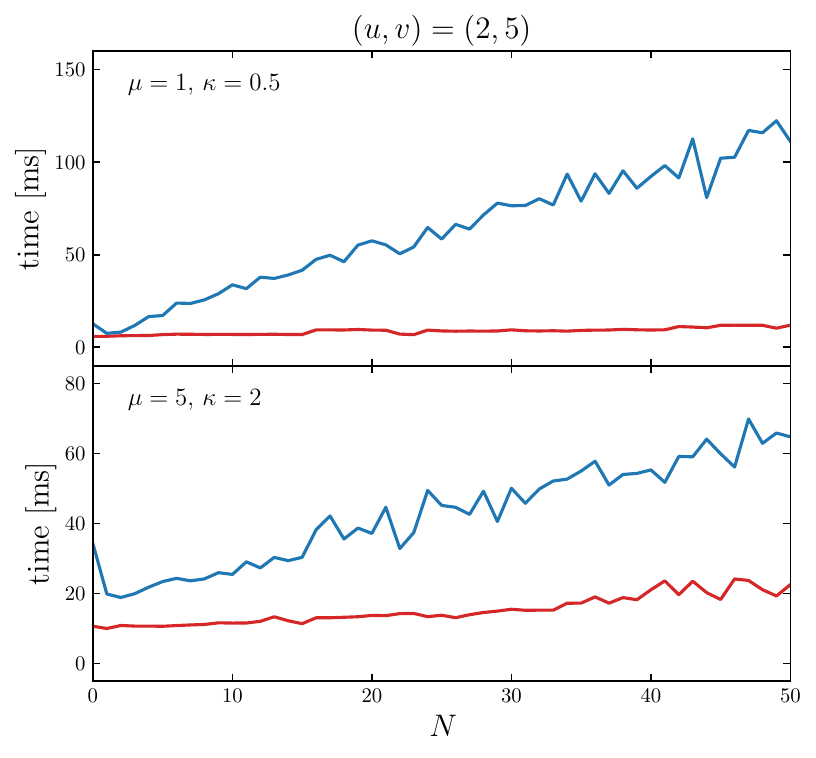}
    \end{subfigure}
   \caption{CPU time in milliseconds to evaluate the boost-breaking helical seed correlator $F^{(\lambda)}(u, v)$ including terms in the series from $n=1, \ldots, N$ as function of $N$ for the bootstrap representation (in \textcolor{pyblue}{blue}) and the spectral representation (in \textcolor{pyred}{red}), both inside the unit kinematic circle (\textit{left panel}) and outside (\textit{right panel}). We have chosen $\mu=1, \kappa=0.5$ (\textit{upper panel}) and $\mu=5, \kappa=2$ (\textit{lower panel}).  We consider by default the negative helicity $\lambda = -1$ when $\kappa\neq 0$. These computations were performed with \textsf{Mathematica} on a  Macbook Pro with M1 CPU and running MacOS 11.6 Big Sur.}
  \label{fig: Timing inside/outside}
\end{figure}

    \item \textit{Comparison with bootstrap result}: Both the spectral and the bootstrap methods yield a convergent series representation albeit with different convergence rates. In Fig.~\ref{fig: Convergence inside/outside}, we show the amplitude of the series coefficients $|F_n^{(\lambda)}|$ of the boost-breaking helical seed correlator $\sum_nF_n^{(\lambda)}$ as function of $n$. Both inside and outside the kinematic unit circle, the bootstrap series has a faster convergence rate. Keeping track of the mass parameter $\mu$ dependence only, this is because each term scales as $F_n^{(\lambda)} \sim 1/\mu^{2n}$, as can be seen from~\eqref{eq: helical particular solution resummed expression}, whereas the spectral series scales as $F_n^{(\lambda)} \sim 1/(n^2+\mu^2)$, see~\eqref{eq: EFT 2}. A noticeable feature of the spectral representation series is that the coefficient amplitude decrease $|F_n^{(\lambda)}|$ is not monotonic as $n$ increases outside the unit kinematic circle, but exhibits an (slowly decreasing) oscillating behaviour. This causes the solution to oscillate in the $u$-domain as, for a fixed number of terms in the series $n$, some coefficients at larger $n$ would be dominant. Interestingly, inside the unit circle, both series' evaluation times are of the same order of magnitude $\sim \O(10{\rm ms})$, as can be seen in Fig.~\ref{fig: Timing inside/outside}. However, beyond the unit kinematic circle, the computational cost of the bootstrap representation increases linearly, whereas the spectral representation remains stable. This difference arises because the analytic continuation of the $_2F_1$ hypergeometric function is significantly more efficient than that of $_3F_2$. However, since the bootstrap series converges more rapidly, an accurate result can be achieved with only a few terms, whereas the spectral representation requires summing substantially more terms to reach a comparable level of accuracy.
\end{itemize}

\section{Classifying Non-Analyticities}
\label{sec: Classifying non-Analyticities}

We now introduce the saddle point method supplemented with the WKB approximation, as a powerful tool to classify and gain physical insight into non-analytic signals of cosmological correlators. This approach relies solely on analysing the location of saddle points in the complex time plane---dominating the time integrals---, providing a natural explanation for the behaviour of these signals by tracking the motion of these saddles driven by external kinematics. By evaluating the time integrals using the saddle point method, we obtain approximate solutions for a range of correlators, expressed entirely in terms of elementary functions.

\vskip 4pt
In this section, we start by outlining the general strategy to classify non-analyticities of correlators and approximating cosmological collider signals. Then, we apply the method to concrete cases with increasing complexity, and provide new cosmological collider templates for future cosmological surveys.

\subsection{Extracting \emph{\&} Approximating Non-Analyticities}
\label{subsec: Extracting and Approximating Non-Analyticities}

Without loss of generality, we work with the exchange four-point seed correlators $F$ defined in Sec.~\ref{subsec: Boost-Breaking Physics & Helical Seed Correlator}, whose kinematics is described by external energies, $k_{12}$ and $k_{34}$, and the internal momentum $s$ in the $s$-channel. We include a reduced speed of sound for external legs and a helical chemical potential for the exchanged massive particle. These boost-breaking effects will be turned on in turn lately. Our approximation strategy consists of four steps.

\begin{itemize}
    \item {\it Apply bulk cutting rules}: Non-analytic scaling of cosmological correlators---and therefore cosmological collider signals---come from factorised contributions. We limit ourselves to the kinematic region $k_{12}>k_{34}$ and focus on the factorised contribution \cite{Tong:2021wai,Qin:2023bjk,Ema:2024hkj}
    \begin{equation}
        F_{\text{F}} = \int_{-\infty^+}^0 \d\tau'\, e^{ic_s k_{12} \tau'}\int_{-\infty^+}^0 \d\tau''\, e^{ic_s k_{34} \tau''} \, \G_{-+}(s; \tau', \tau'') \,.
    \end{equation}
    Here, we have removed powers of conformal time. We stress that this choice simplifies our technical discussion later on, but not at the cost of generality nor physicality. Note that we have also dropped overall kinematic factors that make the integral dimensionless. These prefactors can easily be restored at the end. Factorising the Wightman function $\G_{-+} \equiv \G_>$ into mode functions, we can separate the integral into two parts $F_{\text{F}} = F^{(3)}_L \times {F}^{(3)}_R$ with 
    \begin{equation}
        F^{(3)}_L = \int_{-\infty^+}^0 \d\tau'\,e^{ic_s k_{12} \tau'} \tilde{\sigma}_s(\tau') \,, \quad 
            {F}^{(3)}_R = \int_{-\infty^+}^0 \d\tau''\,e^{ic_s k_{34} \tau''} \tilde{\sigma}_s^*(\tau'') \,,
    \end{equation}
    where $\tilde{\sigma}_s$ is the rescaled mode function of the exchanged field, see~Eq.~\eqref{eq:rescaled-sigma}. In most of the cases, these integrals can be easily computed and result in hypergeometric functions. 
    
    \item {\it Apply the WKB approximation}: The rescaled mode function satisfies an equation of motion of the form 
    \begin{equation}
        \tilde{\sigma}_s'' + \omega^2_s(\tau)\tilde{\sigma}_s = 0 \,,
    \end{equation}
    where $\omega$ is an effective frequency for the $s$-mode that is theory-dependent. In particular, it depends on the momentum magnitude $s$ and other characteristics of the exchange field such as its mass and the chemical potential. Instead of directly solving the exact mode function, we write the solution as a time-dependent superposition of positive- and negative-frequency parts, both solved via the superadiabatic WKB approximation \cite{Berry1989,Li:2019ves,Sou:2021juh}
    \begin{equation}
    \label{eq: WKB approx mode function}
        \tilde{\sigma}_s(\tau) \approx \alpha_s(\tau) u_s(\tau) + \beta_s(\tau) u^*_s(\tau)\,, \quad \text{with} \quad u_s(\tau) \equiv \frac{e^{-i\int_{\tau_i}^\tau \d \tau' \omega_s(\tau')}}{\sqrt{2\omega_s(\tau)}} \,,
    \end{equation}
    where $\tau_i$ is an arbitrary reference (conformal) time. The integration contour follows the negative real axis with a slight shift towards the positive imaginary direction due to the $i\epsilon$-prescription. The time-dependent Bogoliubov  coefficients are approximately given by
    \begin{equation}
        \alpha_s(\tau) \approx 1 \,, \quad \beta_s(\tau) \approx -iS(\tau)e^{-F_s(\tau_i)} \,,
    \end{equation}
    where 
    \begin{align}
	    S(\tau)=\frac{1}{2}\left[1+\text{erf}\left(\frac{-\Im F_s(\tau)}{\sqrt{2|\Re F_s(\tau)|}}\right)\right]
	\end{align}
    is the known as the Stokes multiplier and $F_s$ is the Dingle's singulant defined by \cite{Berry1989}
	\begin{align}
	    F_s(\tau)=-2i\int_{\tau_c}^{\tau}\omega_s(\tau')d\tau'~,\label{DingleSingulantDef}
	\end{align}
   with $\tau_c$ being the turning points such that $\omega_s^2(\tau_c)=0$ with $\Im\tau_c<0$. In the early-time limit $\tau\to-\infty$, and the Stokes multiplier asymptotes to zero i.e. $S(-\infty)=0$, thereby recovering the Bunch-Davies initial condition. However, it asymptotes to unity in the late-time limit, $S(0^-)=1$, indicating the emergence of the negative-frequency mode and spontaneous particle production. The exact moment of particle production can be read from the Stokes line crossing time $\tau=\tau_*$ where $\Im F_s(\tau_*)=0$ and $S(\tau_*)=1/2$ \cite{Li:2019ves,Sou:2021juh}.
    
    \item {\it Use the saddle point method to find resonant times}: Plugging the WKB approximation~\eqref{eq: WKB approx mode function} in the bulk time integrals $F_L^{(3)}$ and ${F}_R^{(3)}$, all integrals are of the form
    \begin{equation}
        \int_\C \d\tau\, g(\tau) e^{\mu f(\tau)} \,,
    \end{equation}
    where $g$ and $f$ are theory-dependent functions of the complex conformal time $\tau$, and $\mu$ denotes the dimensionless mass parameter of the exchanged particle. Here, $\C$ denotes the specific contour on the complex plane that encodes the $i\epsilon$ prescription. In the large-mass regime $\mu\gg 1$, integrals of this type are known to be solvable as an asymptotic series in $\mu^{-1}$ via the saddle-point method. Such saddle-point approximations are capable of partially capturing effects exponentially suppressed by the mass $\mu$, and is therefore suitable for understanding non-perturbative physics beyond the local large-mass EFTs. We give details about this method in App.~\ref{app: Mathematical Interlude}. Solving for the saddle point $\tau_\bullet$ such that $f'(\tau_\bullet)=0$, the integration contour $\C$ is continuously deformed to a new contour $\C'$ that passes through the saddle point in the steepest direction. Note that in practice, whether or not $\C'$ has to pass exactly along the steepest descent direction is usually unimportant, since the integrand is often analytic around $\tau_\bullet$ and small deformations are allowed. Expanding the integrand around the saddle, the Gaussian integral can be finished, leading to
    \begin{equation}
        \int_\C \d\tau\, g(\tau) e^{\mu f(\tau)} \approx g(\tau_\bullet) e^{\mu f(\tau_\bullet)} \sqrt{\frac{-2\pi}{\mu f''(\tau_\bullet)}}\left[1 + \mathcal{O}\left(\mu^{-1}\right)\right] \,.
    \end{equation}
    Note that higher-order terms are small due to the small variance of the Gaussian. Therefore, at leading order, the resulting integral is fully controlled by the location of the saddle, which is a function of the external kinematics, i.e.~$\tau_\bullet = \tau_\bullet(k_{12}, k_{34}, s)$.
    
    \item {\it Classify cosmological collider signals}: To relate the mathematical notion of saddle points to the physical bulk evolution, we note that massive fields are constantly produced and decay during inflation. Due to the lack of time-translational symmetry and hence the non-conservation of energy, particles can be produced either in pairs from the propagator, or in singlet from the interaction vertices, before decaying. These production and decay events happen in the form of resonant transitions that manifest themselves as saddles on the complex time plane. The time spent by the particle between these saddles is thus recorded in the resultant cosmological collider signals. One can also estimate the size of the cosmological signal by inspecting the location of the saddle. The rule of thumb is
    \begin{equation}
    \label{eq: saddle rule of thumb}
        \begin{aligned}
	   &\text{Saddle points lying in the physical region i.e. $\tau_\bullet \in \mathbb{R}_-$} \to \text{$\mu^{-\#}$ suppression} \,, \\[10pt] 
	   &\text{Saddle points lying in the unphysical region i.e. $\tau_\bullet \notin \mathbb{R}_-$} \to \text{$e^{-\# \mu}$ suppression} \,,
	\end{aligned}
    \end{equation}
    where $\#$ denotes an order-one positive number. One can intuitively understand this as an ``on-shell'' versus ``off-shell'' effect. To summarise, by studying the kinematic dependence of the saddle point $\tau_\bullet = \tau_\bullet(k_{12}, k_{34}, s)$, we can classify cosmological collider signals into different categories with distinct shapes and sizes.
\end{itemize}

Through this procedure, we learn a beautiful correspondence between the \textcolor{pyblue}{observational}, the \textcolor{pygreen}{mathematical}, and the \textcolor{pyred}{physical} aspects of cosmological collider signals: \textcolor{pyblue}{(i)} the momentum ratio $k_L/k_S$, where $k_L$ is the momentum of a generic long-wavelength mode and $k_S$ is that of a generic short-wavelength one, reflects the kinematic control (more precisely the internal soft limit) of the boundary cosmological correlator on the observational side, \textcolor{pygreen}{(ii)} the location of the complex saddle point $\tau_\bullet$ dictates the dominant sources of the time integrals on the mathematical side, and \textcolor{pyred}{(iii)} the lifetime of the massive particle $\Delta\tau$ traces the origin of the cosmological collider signal on the physical side. Modifying any one of the three variables is equivalent to changing the other two, as represented in the picture below.

\begin{center}
\begin{tikzpicture}[scale = 3]
    \node at (0, 1) {\textcolor{pyblue}{$k_L/k_S$}};
    \node at (-0.5, 0) {\textcolor{pygreen}{$\tau_\bullet$}};
    \node at (0.5, 0) {\textcolor{pyred}{$\Delta \tau$}};

    \draw[<->, thick] (-0.05, 0.85) -- (-0.45, 0.15); 
    \draw[<->, thick] (0.05, 0.85) -- (0.45, 0.15);  
    \draw[<->, thick] (-0.35, 0) -- (0.35, 0);    
\end{tikzpicture}
\end{center}

\subsection{Application to Tree-Level Exchange Correlators}
\label{subsec: Application to Tree-Level Exchange Correlators}

We now apply the strategy outlined above to concrete examples with increasing complexity.

\subsubsection{de Sitter-Invariant Scenario}
\label{subsubsec: Unit Sound Speed & No Chemical Potential}

\begin{figure}[h!]
    \centering
    \centering
    \begin{subfigure}{.5\textwidth}
        \centering
        \includegraphics[width=0.8\linewidth]{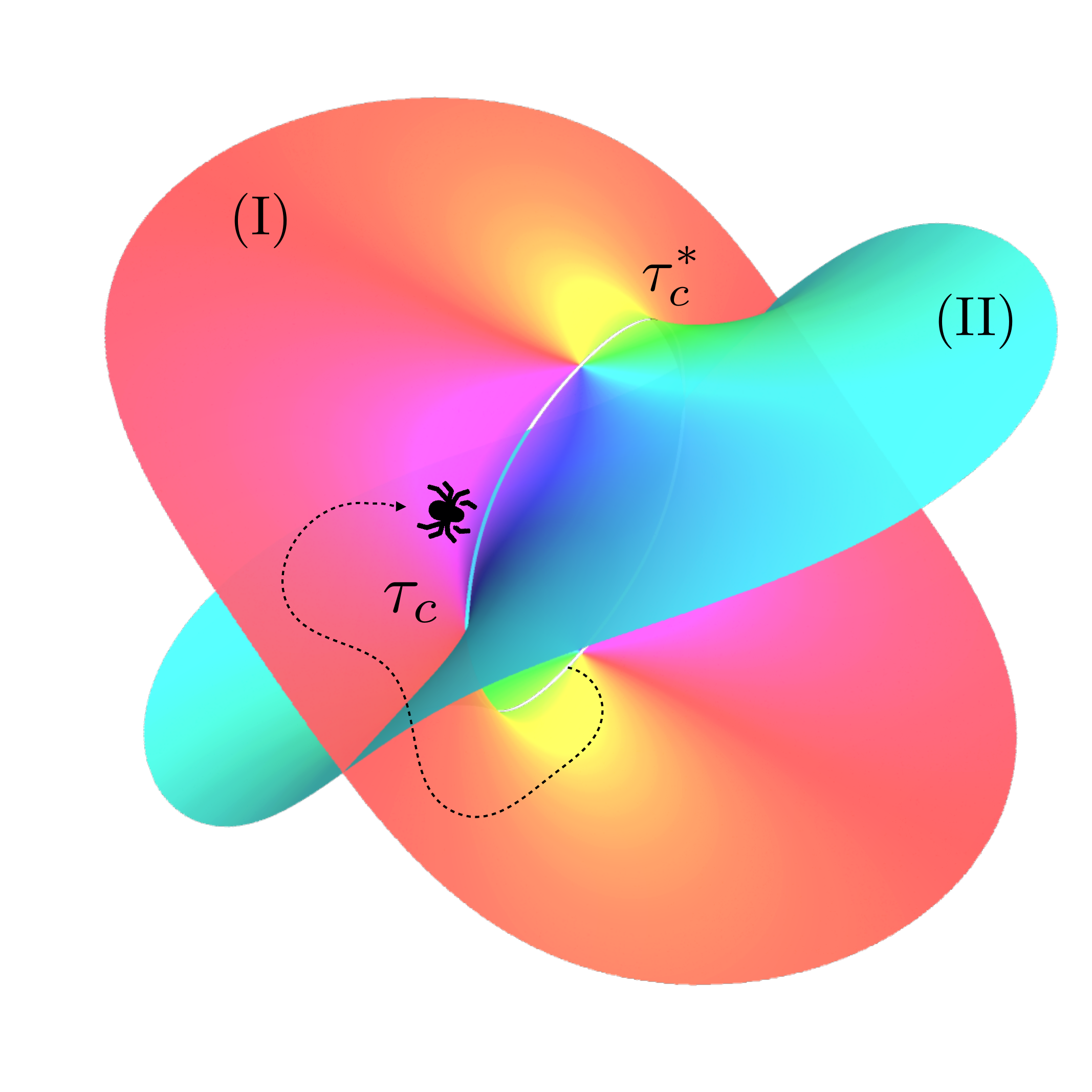}
    \end{subfigure}%
    \begin{subfigure}{.5\textwidth}
        \centering
        \includegraphics[width=0.75\linewidth]{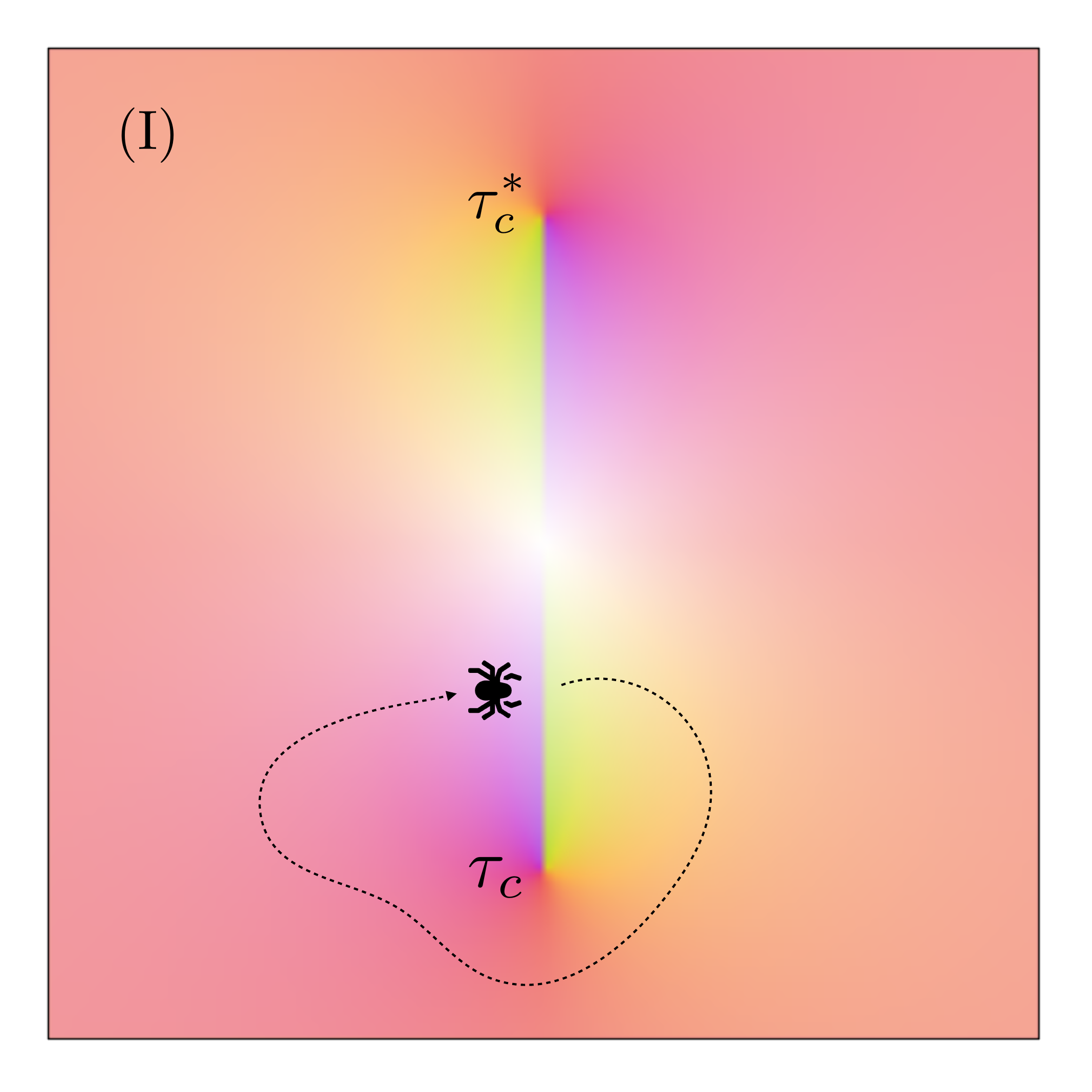}
    \end{subfigure}
    \caption{Illustration of the two-sheet Riemann surface for the effective frequency $\omega_s(\tau)$ in Eq.~\eqref{eq: canonical case effective frequency}. We have retained the topology of the Riemann surface while removing the pole at the origin $\tau=0$ for better visualisation. The branches that asymptote to \textcolor{pyred}{red} and \textcolor{cyan}{cyan} correspond to Sheet I and Sheet II, respectively, as defined in~\eqref{eq: sheet I and II def}. The curve connecting $\tau_c$ and $\tau^*_c$ is the branch cut, along which the two sheets are smoothly sewn together. However, discontinuity appears when crossing the branch cut and landing on the same sheet. We have cartooned the same ``journey of the spider \faIcon{spider}'' to better demonstrate the structure of the two-sheet Riemann surface.}
    \label{fig: 3D Riemann surface}
\end{figure}

Let us first consider the minimal canonical scenario of an exchanged massive scalar field with zero chemical potential $\kappa=0$ and with a trivial sound speed $c_s=1$. The effective frequency in this case is given by
\begin{equation}
\label{eq: canonical case effective frequency}
    \omega_s^2(\tau) = s^2 + \frac{\mu^2}{\tau^2} \,,
\end{equation}
which has a pole at $\tau=0$ and two simple zeros at $\tau_c, \tau_c^* = \mp i\mu/s$. Taking the square root gives a double-valued function defined on two Riemann sheets $\mathbb{C}_{\rm I, II}$ that we define by
\begin{equation}
\label{eq: sheet I and II def}
    \omega_s(\tau) = +\sqrt{s^2+\frac{\mu^2}{\tau^2}}\,, \,\,\, \tau \in \mathbb{C}_{\rm I}\,, \quad \text{and} \quad \omega_s(\tau) = -\sqrt{s^2+\frac{\mu^2}{\tau^2}}\,, \,\,\, \tau \in \mathbb{C}_{\rm II}\,,
\end{equation}
where $\sqrt{\,\cdot\,}$ denotes the standard principal-valued square root function. On either sheets, there is a branch cut that joins $\tau_c, 0$ and $\tau_c^*$, but there are no discontinuities between $\tau_c, 0$ and $\tau_c^*$ across different sheets. Instead, the discontinuities on the imaginary axis are from $\tau_c, \tau_c^*$ to $\mp i\infty$. The corresponding surface is represented in Fig.~\ref{fig: 3D Riemann surface}. We now follow the recipe outlined in the previous section. The first factorised time integral $F^{(3)}_L$ is given by
\begin{equation}
    \begin{aligned}
        F^{(3)}_L &\approx \int_{-\infty^+}^0 \frac{\d\tau'}{\sqrt{2\omega_s(\tau')}}\left[e^{i\left(k_{12}\tau'-\int_{\tau_i}^{\tau'} \d\tau\,\omega_s(\tau)\right)} + \beta_s(\tau')e^{i\left(k_{12}\tau'+\int_{\tau_i}^{\tau'} \d\tau\,\omega_s(\tau)\right)}\right] \\
        &\equiv F^{(3), \alpha}_L + F^{(3), \beta}_L \,, 
    \end{aligned}
\end{equation}
where the superscripts $\alpha$ and $\beta$ keep track of the positive- and negative-frequency parts of the massive mode functions. Here, we recall that $\tau_i$ is an arbitrary reference time which can be taken to be $\tau_i \to -\infty$ without loss of generality. To find the saddle points, we solve
\begin{equation}
\label{eq: saddle-point equation}
    \frac{\d}{\d\tau'} \left(k_{12} \tau' \mp \int_{\tau_i}^{\tau'}\d\tau \omega_s(\tau)\right) = 0 \,.
\end{equation}
For the integral $F^{(3), \alpha}_L$, the nearest saddle---which contributes the most---lies in the physical domain of $\mathbb{C}_{\rm I}$
\begin{equation}
    \tau_\bullet'^{(\alpha)} = \frac{-\mu}{\sqrt{k_{12}^2-s^2}} \in \mathbb{C}_{\rm I} \,.
\end{equation}
Although $\tau'_\bullet$ and $-\tau'_\bullet$ are both solutions of the saddle-point equation~\eqref{eq: saddle-point equation}, according to the rule of thumb, we always pick up the one closest to the physical region since its contribution is the largest. For the integral $F^{(3), \beta}_L$, there are no saddle points in the physical domain. Instead, the nearest saddle point lies in $\mathbb{C}_{\rm II}$
\begin{equation}
    \tau_\bullet'^{(\beta)} = \frac{+\mu}{\sqrt{k_{12}^2-s^2}} \in \mathbb{C}_{\rm II} \,.
\end{equation}
In addition, the Bogoliubov  coefficient is suppressed by $|\beta_s|=e^{-\pi\mu}$. Therefore, we can completely discard this term. In fact, one can show that this term evaluated at the saddle $\tau_\bullet'^{(\beta)}$ is of order $\O(e^{-2\pi\mu})$. Its cousin at $-\tau_\bullet'^{(\beta)}\in \mathbb{C}_{\rm II}$ is unreachable from the original contour on the first Riemann sheet and therefore does not contribute. For pedagogical reasons, we gather a pedagogical introduction to Riemann surfaces and the saddle point method (as well as concrete worked out examples) in App.~\ref{app: Mathematical Interlude}. Performing the saddle-point approximation for the positive-frequency integral, we obtain
\begin{equation}
\label{eq: saddle F3L}
    F^{(3)}_L \approx F^{(3), \alpha}_L \approx \frac{\sqrt{\pi\mu}}{(k_{12}^2-s^2)^{3/4}} e^{-i\pi/4} \exp\left[-i\mu \, \arccosh \frac{k_{12}}{s}\right] \,.
\end{equation}
We give details of the derivation in App.~\ref{app: Mathematical Interlude}, and represent the complex $\tau'$ plane in Fig.~\ref{fig: saddle FL no sound speed no chemical potential}. Note that keeping the dependence on $\tau_i$ gives an unphysical phase that will eventually be cancelled in the final result, after multiplying by ${F}^{(3)}_R$.
\begin{figure}[h!]
     \centering
     \begin{tikzpicture}
		[scale = 2]
        \node at (-1, 1) {\textcolor{pyblue}{$\mathbb{C}_{\rm I}$}};
        \node at (1, 1) {\textcolor{pyred}{$\mathbb{C}_{\rm II}$}};

        \draw[pyred, fill = pyred, opacity = 0.1] (0, 1.5) -- (0, -1.5) -- (1.5, -1.5) -- (1.5, 1.5) -- (0, 1.5);
        \draw[pyblue, fill = pyblue, opacity = 0.1] (0, 1.5) -- (0, -1.5) -- (-1.5, -1.5) -- (-1.5, 1.5) -- (0, 1.5);

        \draw[->] (0, -1.6) -- (0, 1.6) coordinate[label=below:$ $] (k1);
	\draw[->] (-1.6, 0) -- (1.6, 0) coordinate[label=below:$ $] (k1);
        \node at (0, 1.7) {$\Im\tau'$};
	\node at (1.9, 0) {$\Re\tau'$};

        \draw[-, double, decorate, decoration={segment length=6, amplitude=1.5}, black, line width=0.2mm] (0, 0.5) -- (0, 1.5);
        \draw[-, double, decorate, decoration={segment length=6, amplitude=1.5}, black, line width=0.2mm] (0, -0.5) -- (0, -1.5);
        \filldraw[black] (0, 0.5) circle (.03cm) node[left] {$\tau_c^*$};
        \filldraw[black] (0, -0.5) circle (.03cm) node[left] {$\tau_c$};

        \filldraw[black] (-1, 0) circle (.03cm) node[below] {$\tau_\bullet'^{(\alpha)}$};
        \filldraw[black] (1, 0) circle (.03cm) node[below] {$\tau_\bullet'^{(\beta)}$};;

        \path[gray, draw, line width = 1.2pt, postaction = decorate, decoration={markings,
			mark=at position 0.5 with {\arrow[line width=1pt]{>}}}] (-1.5, 0.1) -- (0, 0);
        \filldraw[black] (0, 0) circle (.03cm); 
    \end{tikzpicture}
    \caption{Analytic structure of the effective frequency $\omega_s$ in the complex $\tau'$ plane, together with both Riemann sheets $\textcolor{pyblue}{\mathbb{C}_{\rm I}}$ and $\textcolor{pyred}{\mathbb{C}_{\rm II}}$, for the case $c_s=1$ and $\kappa=0$ to compute $F^{(3)}_L$. Here and in subsequent figures, the double lines indicate the disconnection that cannot be crossed between the two sheets.} We also represent the turning points $\tau_c, \tau_c^*$, and the saddle points $\tau_\bullet'^{(\alpha)}, \tau_\bullet'^{(\beta)}$. The original integration contour is in \textcolor{gray}{gray}, which encodes the early-time $i\epsilon$ prescription.
    \label{fig: saddle FL no sound speed no chemical potential}
\end{figure}

\begin{figure}[h!]
     \centering
     \begin{tikzpicture}
		[scale = 2]
        \node at (-1, 1) {\textcolor{pyblue}{$\mathbb{C}_{\rm I}$}};
        \node at (1, 1) {\textcolor{pyred}{$\mathbb{C}_{\rm II}$}};

        \draw[pyred, fill = pyred, opacity = 0.1] (0, 1.5) -- (0, -1.5) -- (1.5, -1.5) -- (1.5, 1.5) -- (0, 1.5);
        \draw[pyblue, fill = pyblue, opacity = 0.1] (0, 1.5) -- (0, -1.5) -- (-1.5, -1.5) -- (-1.5, 1.5) -- (0, 1.5);

        \draw[->] (0, -1.6) -- (0, 1.6) coordinate[label=below:$ $] (k1);
	\draw[->] (-1.6, 0) -- (1.6, 0) coordinate[label=below:$ $] (k1);
        \node at (0, 1.7) {$\Im\tau''$};
	\node at (1.9, 0) {$\Re\tau''$};

        \draw[-, double, decorate, decoration={segment length=6, amplitude=1.5}, black, line width=0.2mm] (0, 0.5) -- (0, 1.5);
        \draw[-, double, decorate, decoration={segment length=6, amplitude=1.5}, black, line width=0.2mm] (0, -0.5) -- (0, -1.5);
        \filldraw[black] (0, 0.5) circle (.03cm) node[left] {$\tau_c^*$};
        \filldraw[black] (0, -0.5) circle (.03cm) node[left] {$\tau_c$};

        \path[pyblue, draw, line width = 1.2pt, postaction = decorate, decoration={markings,
			mark=at position 0.5 with {\arrow[line width=1pt]{>}}}] (-1.5, 0) -- (-0.2, 0);
        \node at (-0.75, 0.2) {\textcolor{pyblue}{$\C_1$}};

        \path[pygreen, draw, line width = 1.2pt, postaction = decorate, decoration={markings,
			mark=at position 0.5 with {\arrow[line width=1pt]{>}}}] (0.2, 0) -- (1, 0);
        \node at (0.75, 0.2) {\textcolor{pygreen}{$\C_3$}};

        \path[draw, line width = 1.2pt, pyred, postaction = decorate, decoration={markings,
			mark=at position 0.7 with {\arrow[line width=1pt]{>}}}] (-0.2, 0)  arc (180:0:0.2);
        \node at (0.3, 0.2) {\textcolor{pyred}{$\C_2$}};
            
        \filldraw[black] (-1, 0) circle (.03cm) node[below] {$\tau_\bullet''^{(\beta)}$};
        \filldraw[black] (1, 0) circle (.03cm) node[below] {$\tau_\bullet''^{(\alpha)}$};;
        \filldraw[black] (0, 0) circle (.03cm); 
    \end{tikzpicture}
    \caption{Analytic structure of the effective frequency $\omega_s$ in the complex $\tau''$ plane to compute ${F}^{(3)}_R$. For the positive-frequency integral ${F}^{(3), \alpha}_R$, the original integration contour is deformed to pass through the saddle point on the second Riemann sheet $\mathbb{C}_{\rm II}$, which results in three different portions.}
    \label{fig: saddle FR no sound speed no chemical potential}
\end{figure}
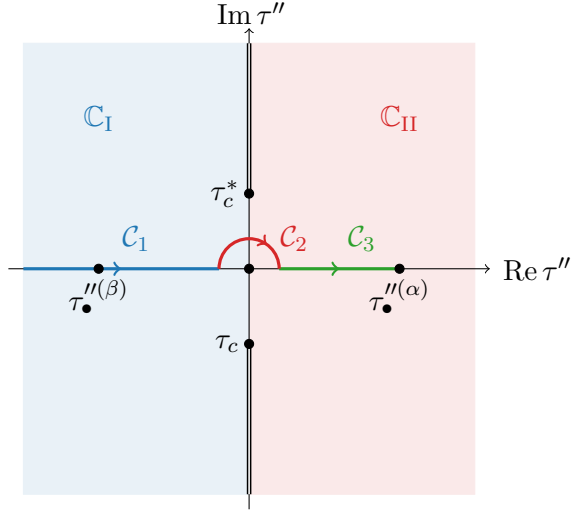

The factorised time integral ${F}^{(3)}_R$ turns out to be more complicated. Again, let us separate the integral into positive- and negative-frequency modes
\begin{equation}
    \begin{aligned}
        {F}^{(3)}_R &\approx \int_{-\infty^+}^0 \frac{\d\tau''}{\sqrt{2\omega_s(\tau'')}}\left[e^{i\left(k_{34}\tau''+\int_{\tau_i}^{\tau''} \d\tau\, \omega_s(\tau)\right)} + \beta_s^*(\tau'')e^{i\left(k_{34}\tau''-\int_{\tau_i}^{\tau''} \d\tau\, \omega_s(\tau)\right)}\right] \\
        &\equiv {F}^{(3), \alpha}_R + {F}^{(3), \beta}_R \,.
    \end{aligned}
\end{equation}
The saddle points now are solutions to the following equation
\begin{equation}
    \frac{\d}{\d\tau''} \left(k_{34} \tau'' \pm \int_{\tau_i}^{\tau''}\d\tau \,\omega_s(\tau)\right) = 0 \,.
\end{equation}
The integral ${F}^{(3), \beta}_R$, despite being suppressed by $|\beta_s|=e^{-\pi\mu}$, has a saddle in the physical domain 
\begin{equation}
    \tau_\bullet''^{(\beta)} = \frac{-\mu}{\sqrt{k_{34}^2-s^2}} \in \mathbb{C}_{\rm I} \,,
\end{equation}
while the saddle of the positive-frequency integral ${F}^{(3), \alpha}_R$ lies unphysically on the second Riemann sheet
\begin{equation}
    \tau_\bullet''^{(\alpha)} = \frac{+\mu}{\sqrt{k_{34}^2-s^2}} \in \mathbb{C}_{\rm II} \,.
\end{equation}
Consequently, we expect the contribution from both terms to be comparable, i.e.~both are of order $\O(e^{-\#\mu})$. Computing ${F}^{(3), \beta}_R$ is identical to that of $F^{(3), \alpha}_L$ and we directly write the result
\begin{equation}
\label{eq: saddle F3Rbeta}
    {F}^{(3), \beta}_R \approx \frac{\sqrt{\pi\mu}\,e^{-\pi\mu}}{(k_{34}^2-s^2)^{3/4}} e^{-i\pi/4} \exp\left[-i\mu \, \arccosh \frac{k_{34}}{s}\right] \, S(\tau_\bullet''^{(\beta)}) \,,
\end{equation}
where we have used the relation $\Re{F_s(\tau_i)} = \pi\mu$. As for the positive-frequency integral ${F}^{(3), \alpha}_R$, we need to continuously deform the integration contour so that it crosses the saddle point $\tau_\bullet''^{(\alpha)}$ on the second Riemann sheet $\mathbb{C}_{\rm II}$. According to the saddle-point method, the complete integral is proportional to the integrand evaluated at the saddle point. Therefore, the size of the integral is estimated by
\begin{equation}
    |{F}^{(3), \alpha}_R| \sim \left|\exp\left(ik_{34}\tau_\bullet''^{(\alpha)}+ i\int_{-\infty}^{\tau_\bullet''^{(\alpha)}}\d\tau\, \omega_s(\tau)\right)\right| = \exp\left(-\Im \int_{-\infty}^{\tau_\bullet''^{(\alpha)}}\d\tau\, \omega_s(\tau)\right) \,.
\end{equation}
The original integral contour is now deformed to pass through the saddle $\tau_\bullet''^{(\alpha)} \in \mathbb{C}_{\rm II}$. The new contour consists of three portions
\begin{equation}
    \int_{-\infty}^{\tau_\bullet''^{(\alpha)}} = \int_{\textcolor{pyblue}{\C_1}} + \int_{\textcolor{pyred}{\C_2}} + \int_{\textcolor{pygreen}{\C_3}} \,,
\end{equation}
where the various contours are depicted in Fig.~\ref{fig: saddle FR no sound speed no chemical potential}. Both $\C_1$ and $\C_3$ give purely real contributions, which directly drop out of our estimation. The only contribution to the imaginary part is from the semicircle $\C_2$, which can be nicely written as half the residue at the origin
\begin{equation}
    |{F}^{(3), \alpha}_R| \sim \exp\left(-\Im \int_{\C_2}\d\tau \,\omega_s(\tau)\right) = e^{\pi \underset{\tau=0}{\Res} \, \omega_s(\tau)} = e^{-\pi\mu} \,.
\end{equation}
Interestingly, the price to pay for an ``off-shell'' vertex resonance is an exponential factor closely related to the location of the saddle point and the late-time behaviour of the effective frequency. We will see later that this is a general pattern. We can now finish the saddle-point approximation by performing the rest of the Gaussian integral. This yields 
\begin{equation}
\label{eq: saddle F3Ralpha}
    {F}^{(3), \alpha}_R \approx \frac{-\sqrt{\pi\mu}\,e^{-\pi\mu}}{(k_{34}^2-s^2)^{3/4}} e^{-i\pi/4} \exp\left[-i\mu \, \arccosh \frac{k_{34}}{s}\right] \,.
\end{equation}
Combining~\eqref{eq: saddle F3L},~\eqref{eq: saddle F3Rbeta} and~\eqref{eq: saddle F3Ralpha}, and setting $S(\tau_\bullet''^{(\beta)})=1$ in the soft limit, we obtain the full approximate solution for the cosmological collider signal
\begin{eBox}
\begin{equation}
\label{eq: full saddle point approx solution}
    \begin{aligned}
        F_{\rm F} + \text{c.c.} &\approx \frac{2\pi\mu}{\left[(k_{12}^2-s^2)(k_{34}^2-s^2)\right]^{3/4}}
        \left\{ e^{-\pi\mu} \sin \, \mu \left[\arccosh\frac{k_{12}}{s} - \arccosh\frac{k_{34}}{s}\right] \right.\\
        &\left.+ \, e^{-\pi\mu}\cos\, \mu \left[\arccosh\frac{k_{12}}{s} + \arccosh\frac{k_{34}}{s}\right] \right\} \,,
    \end{aligned}
\end{equation}
\end{eBox}
for $k_{12}>k_{34}$. The opposite regime $k_{12}<k_{34}$ is obtained by swapping $k_{12} \leftrightarrow k_{34}$. 
This method gives very accurate results already for $\mu\gtrsim2$, as we will see in Sec.~\ref{subsec: templates}. The small phase shift for the non-local signal---oscillating in the $s^2/k_{12}k_{34}$ direction---can be fixed by considering higher-order corrections in $\mu^{-1}$. 

\paragraph{Discussion.} Let us make a few comments regarding the found approximate solution. 

\begin{itemize}
    \item The first line in~\eqref{eq: full saddle point approx solution} comes from the positive-frequency integral ${F}_R^{(3), \alpha}$ and asymptotes to $\sin[\mu\log(k_{12}/k_{34})]$ in the internal collapsed limit $s\to0$. This term is known as the local cosmological collider signal as it is analytic in $s$. We recall that it captures the propagation of massive particles from one vertex to the other one \cite{Tong:2021wai}. More specifically, we see that the local signal is given by the time elapsed between the saddle point $\tau_\bullet''^{(\alpha)}$ that creates the particle and the saddle $\tau_\bullet'^{(\alpha)}$ that destroys the particle.

    \item The second line in~\eqref{eq: full saddle point approx solution} comes from the negative-frequency integral ${F}_R^{(3), \beta}$ and asymptotes to $\cos[\mu\log(4k_{12}k_{34}/s^2)]$ in the internal collapsed limit $s\to0$. These oscillations are known as the non-local cosmological collider signal due to their non-analyticity in $s$. In contrast to the local signal, this one represents the propagation effect of massive particles from vacuum pair production to their subsequent decay at $\tau_\bullet'^{(\alpha)}$ and $\tau_\bullet''^{(\beta)}$ \cite{Tong:2021wai}.

    \item In this minimal setup, both signals are suppressed by {\it half} the Boltzmann factor $e^{-\pi\mu} = e^{-m/(2T_{\rm dS})}$, where $T_{\rm dS} \equiv \tfrac{H}{2\pi}$ is the de Sitter temperature, but due to different reasons. For the local signal, it is due to the {\it semi-circle} we take to reach the saddle in the unphysical domain, whereas for the non-local signal, it is essentially the Bogoliubov  coefficient. We will see in more general scenarios that the size of these signals can be drastically different. 

    \item The seed integral we started with does not correspond to any physical coupling. However, all physical interaction vertices respecting the shift symmetry of external legs come with higher conformal time weights in the integrand, and their effects can be derived by differentiating the found result with respect to external kinematics. For instance, acting with $k_{12}\partial_{k_{12}}$ on~\eqref{eq: full saddle point approx solution} raises the conformal time power from zero to one. In the large-mass limit, the leading effect of such derivatives is to bring out powers of $\mu$ from the waveform while shifting its phase, 
    \begin{equation}
        \begin{aligned}
            \left(k_{12}\partial_{k_{12}}\right)^{p_1} \left(k_{34}\partial_{k_{34}}\right)^{p_2} (F_{\rm F}+\text{c.c.}) &\approx \left(\frac{\mu k_{12}}{\sqrt{k_{12}^2-s^2}}\right)^{p_1} \left(\frac{\mu k_{34}}{\sqrt{k_{34}^2-s^2}}\right)^{p_2} \\
            &\times (F_{\rm F}+\text{c.c.})_{\text{shifted phase}} \,,
        \end{aligned}
    \end{equation}
    with 
    \begin{align}
	\text{shifted phase}=\left\{
	\begin{aligned}
	\frac{\pi(p_1-p_2)}{2} \,, &\quad \text{local signal}\\
	\frac{\pi(p_1+p_2)}{2} \,, &\quad \text{non-local signal}
	\end{aligned}\right.\,.
	\end{align}
    This feature does not alter the qualitative properties of the cosmological collider signals discussed above. In particular, the waveforms are unchanged modulo discrete phase shifts. 

    \item Readers looking for mathematical rigour may have noticed that our ``steepest'' contour is, strictly speaking, invalid for the integral ${F}_R^{(3)}$. This is because the contour is restricted by the branching point $\tau_c^*$, and is in fact not a path for which the saddle-point approximation is justified. In other words, there exist points along this contour where the size of the integrand is actually larger than the height of the saddle at $\tau_\bullet''^{(\alpha)}$. To remedy this seemingly fatal mathematical flaw, we note that the defining integrand given in terms of the exact mode function is analytic in the upper-half complex $\tau''$-plane without branching points. Moreover, the Bunch-Davies $i\epsilon$ prescription ensures that the integrand vanishes for the upper arc at infinity. Therefore, we can Wick rotate the contour from the very beginning, and then insert the WKB approximation. This way, the saddle can be picked up by the Wick rotated contour directly, eventually giving the same result as in~\eqref{eq: saddle F3Ralpha}. 
\end{itemize}

\subsubsection{Turning on a Non-Trivial Sound Speed}

\begin{figure}[h!]
     \centering
     \begin{tikzpicture}
		[scale = 2]
        \draw[->] (0, 0) -- (-2, 2) coordinate[label=below:$ $] (k1);
	\draw[->] (0, 0) -- (2, 2)
    coordinate[label=below:$ $] (k1);
        \node at (-2.2, 2.2) {$k_{34}/s$};
	\node at (2.2, 2.2) {$k_{12}/s$};

        \draw (-0.7,0.7) +(-135:0.05) -- + (45:0.05); 
        \draw (0.7,0.7) +(-45:0.05) -- + (135:0.05); 
        \node at (-1, 0.6) {$c_s^{-1}$};
	\node at (1, 0.6) {$c_s^{-1}$};

        \draw[pyred, fill = pyred, opacity = 0.1] (0, 0) -- (0.7, 0.7) -- (0, 1.4) -- (-0.7, 0.7) -- (0, 0);
        \draw[pygreen, fill = pygreen, opacity = 0.1] (0.7, 0.7) -- (2, 2) -- (1.3, 2.7) -- (0, 1.4) -- (0.7, 0.7);
        \draw[pygreen, fill = pygreen, opacity = 0.1] (-0.7, 0.7) -- (-2, 2) -- (-1.3, 2.7) -- (0, 1.4) -- (-0.7, 0.7);
        \draw[pyblue, fill = pyblue, opacity = 0.1] (0, 1.4) -- (1.3, 2.7) -- (0, 4) -- (-1.3, 2.7) -- (0, 1.4);

        \draw[<->, thick] (0, 2.5) -- (0, 3.5) node[midway, rotate=90, above] {non-local};
        \draw[<->, thick] (-0.5, 2.2) -- (0.5, 2.2) node[midway, below] {local};
    \end{tikzpicture}
    \caption{Kinematic space of the four-point function in the presence of a reduced sound speed for the external legs $c_s<1$. The local signal direction is horizontal and the non-local signal direction is vertical.}
    \label{fig: kinematic space with sound speed}
\end{figure}
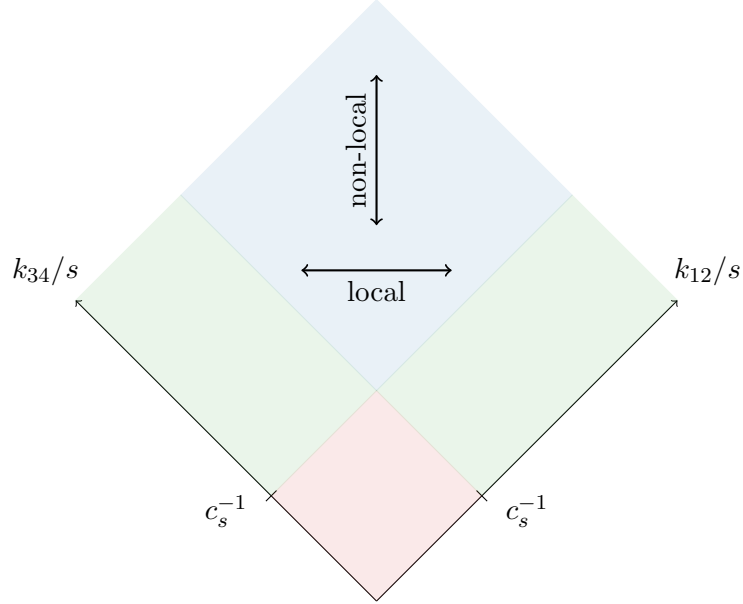

Let us now introduce a non-trivial sound speed. Recall that we have rescaled spacetime coordinates such that the speed of sound of the massive field remains unity while the external fields acquire a sound speed $c_s$, which is the ratio of both speeds. For $c_s>1$, i.e.~the exchanged massive field propagating slower than the external one, the analysis from the previous section applies and one only needs to replace $k_{12, 34}\to c_s k_{12, 34}$. The converse case $c_s<1$, however, needs to be treated with care. Indeed, the bulk cutting rules require that both factorised integrals converge. Yet for $k_{12}>k_{34}$, convergence is achieved only if $k_{12}/s<c_s^{-1}$ as seen in Sec.~\ref{subsec: Via Boundary Kinematic Differential Equation}. In the opposite kinematic regime $k_{34}>k_{12}$, one needs to flip the other theta function, therefore requiring $k_{34}/s<c_s^{-1}$ for convergence. In combination, the convergence of the factorised time integrals for a reduced sound speed carves out a diamond-shape region in kinematic space to which our approximation scheme no longer applies, see the region in \textcolor{pyred}{red} in Fig.~\ref{fig: kinematic space with sound speed}. Moreover, two new ``edge'' regions in \textcolor{pygreen}{green} defined by $k_{34}/s<c_s^{-1}<k_{12}/s$ and $k_{12}/s<c_s^{-1}<k_{34}/s$ emerge where our approximate procedure still applies. 
From now on, we will assume $c_s<1$ and focus on the edge region $k_{34}/s<c_s^{-1}<k_{12}/s$. The integral $F_L^{(3)}$ can again be separated into two pieces
\begin{equation}
    \begin{aligned}
        F^{(3)}_L &\approx \int_{-\infty^+}^0 \frac{\d\tau'}{\sqrt{2\omega_s(\tau')}}\left[e^{i\left(c_sk_{12}\tau'-\int_{\tau_i}^{\tau'} \d\tau\, \omega_s(\tau)\right)} + \beta_s(\tau')e^{i\left(c_sk_{12}\tau'+\int_{\tau_i}^{\tau'} \d\tau\, \omega_s(\tau)\right)}\right] \\
        &\equiv F^{(3), \alpha}_L + F^{(3), \beta}_L \,, 
    \end{aligned}
\end{equation}
Since the massive field remains untouched by the external reduced sound speed, the negative-frequency integral $F^{(3), \beta}_L$ remains subdominant and can be neglected. The saddle point for the positive-frequency integral $F^{(3), \alpha}_L$ is found to be
\begin{equation}
    \tau_\bullet'^{(\alpha)} = \frac{-\mu}{\sqrt{c_s^2k_{12}^2-s^2}} \in \mathbb{C}_{\rm I} \,.
\end{equation}
Because $k_{12}/s>c_s^{-1}$, the saddle $\tau_\bullet'^{(\alpha)}$ lies in the physical region. Therefore, we can directly apply the result~\eqref{eq: saddle F3L} after replacing $k_{12}\to c_sk_{12}$. Similarly, when $k_{34}>c_s^{-1}$, the second integral ${F}_R^{(3)}$ is also well approximated by~\eqref{eq: saddle F3Rbeta} and~\eqref{eq: saddle F3Ralpha} with $k_{34}\to c_s k_{34}$. However, in the edge region $k_{34}/s<c_s^{-1}$, the saddle points for the integral 
\begin{equation}
    \begin{aligned}
        {F}^{(3)}_R &\approx \int_{-\infty^+}^0 \frac{\d\tau''}{\sqrt{2\omega_s(\tau'')}}\left[e^{i\left(c_sk_{34}\tau''+\int_{\tau_i}^{\tau''} \d\tau\, \omega_s(\tau)\right)} + \beta_s^*(\tau'')e^{i\left(c_sk_{34}\tau''-\int_{\tau_i}^{\tau''} \d\tau\, \omega_s(\tau)\right)}\right] \\
        &\equiv {F}^{(3), \alpha}_R + {F}^{(3), \beta}_R \,, 
    \end{aligned}
\end{equation}
lie along the imaginary axis
\begin{equation}
    \tau_\bullet''^{(\alpha)} = \frac{+i\mu}{\sqrt{s^2 - c_s^2k_{34}^2}} \in \mathbb{C}_{\rm II}\,, \quad \tau_\bullet''^{(\beta)} = \frac{+i\mu}{\sqrt{s^2 - c_s^2k_{34}^2}} \in \mathbb{C}_{\rm I} \,.
\end{equation}
Now, $\tau_\bullet''^{(\beta)}$ is moved away from the physical regime. Using the rule of thumb, we expect ${F}^{(3), \beta}_R$ to be suppressed by $\O(e^{-(\pi+\#)\mu})$, making it negligible compared to ${F}^{(3), \alpha}_R\sim\O(e^{-\#\mu})$. We therefore drop the negative-frequency integral and focus on ${F}^{(3), \alpha}_R$. Following the saddle point approximation method, we deform the integration contour to pass through $\tau_\bullet''^{(\alpha)}$. The size of the integral is given by
\begin{equation}
    \begin{aligned}
        |{F}^{(3), \alpha}_R| &\sim \left|\exp\left(ic_sk_{34}\tau_\bullet''^{(\alpha)}+ i\int_{-\infty}^{\tau_\bullet''^{(\alpha)}}\d\tau\, \omega_s(\tau)\right)\right| \\
        &= \exp\left(-c_s k_{34} \,\Im \tau_\bullet''^{(\alpha)}\right) \exp\left(-\Im \int_{-\infty}^{\tau_\bullet''^{(\alpha)}} \d\tau\,\omega_s(\tau)\right) \,.
    \end{aligned}
\end{equation}
To reach the imaginary saddle point on the second Riemann sheet, we need to rotate around the pole at the origin only a quarter of circle, plus an imaginary contour from the branching point to the saddle point. The new contour consists now of four portions
\begin{equation}
    \int_{-\infty}^{\tau_\bullet''^{(\alpha)}} = \int_{\textcolor{pyblue}{\C_1}} + \int_{\textcolor{pyred}{\C_2}} + \int_{\textcolor{pygreen}{\C_3}} + \int_{\textcolor{pypurple}{\C_4}} \,,
\end{equation}
where the various contours are depicted in Fig.~\ref{fig: saddle FR sound speed no chemical potential}. Picking up a quarter of residue at the origin, we obtain
\begin{figure}[h!]
     \centering
     \begin{tikzpicture}
		[scale = 2]
        \node at (-1, 1) {\textcolor{pyblue}{$\mathbb{C}_{\rm I}$}};
        \node at (1, 1) {\textcolor{pyred}{$\mathbb{C}_{\rm II}$}};

        \draw[pyred, fill = pyred, opacity = 0.1] (0, 1.5) -- (0, -1.5) -- (1.5, -1.5) -- (1.5, 1.5) -- (0, 1.5);
        \draw[pyblue, fill = pyblue, opacity = 0.1] (0, 1.5) -- (0, -1.5) -- (-1.5, -1.5) -- (-1.5, 1.5) -- (0, 1.5);

        \draw[->] (0, -1.6) -- (0, 1.6) coordinate[label=below:$ $] (k1);
	\draw[->] (-1.6, 0) -- (1.6, 0) coordinate[label=below:$ $] (k1);
        \node at (0, 1.7) {$\Im\tau''$};
	\node at (1.9, 0) {$\Re\tau''$};

        \draw[-, double, decorate, decoration={segment length=6, amplitude=1.5}, black, line width=0.2mm] (0, 0.5) -- (0, 1.5);
        \draw[-, double, decorate, decoration={segment length=6, amplitude=1.5}, black, line width=0.2mm] (0, -0.5) -- (0, -1.5);
        \filldraw[black] (0, 0.5) circle (.03cm) node[left] {$\tau_c^*$};
        \filldraw[black] (0, -0.5) circle (.03cm) node[left] {$\tau_c$};

        \path[pyblue, draw, line width = 1.2pt, postaction = decorate, decoration={markings,
			mark=at position 0.5 with {\arrow[line width=1pt]{>}}}] (-1.5, 0) -- (-0.2, 0);
        \node at (-0.75, 0.2) {\textcolor{pyblue}{$\C_1$}};

        \path[pygreen, draw, line width = 1.2pt, postaction = decorate, decoration={markings,
			mark=at position 0.5 with {\arrow[line width=1pt]{>}}}] (0, 0.2) -- (0.1, 0.2) -- (0.1, 0.5);
        \node at (0.4, 0.2) {\textcolor{pygreen}{$\C_3$}};

        \path[pypurple, draw, line width = 1.2pt, postaction = decorate, decoration={markings,
			mark=at position 0.5 with {\arrow[line width=1pt]{>}}}] (0.1, 0.5) -- (0.1, 1);
        \node at (0.4, 0.7) {\textcolor{pypurple}{$\C_4$}};

        \path[draw, line width = 1.2pt, pyred, postaction = decorate, decoration={markings,
			mark=at position 0.7 with {\arrow[line width=1pt]{<}}}] (0, 0.2)  arc (90:180:0.2);
        \node at (-0.1, -0.2) {\textcolor{pyred}{$\C_2$}};
            
        \filldraw[black] (0.1, 1) circle (.03cm) node[right] {$\tau_\bullet^{''(\alpha)}$};;
        \filldraw[black] (0, 0) circle (.03cm); 
    \end{tikzpicture}
    \caption{Analytic structure of the effective frequency $\omega_s$ in the complex $\tau''$ plane to compute ${F}^{(3)}_R$ in the case of a reduced sound speed $c_s<1$ with $k_{34}/s<c_s^{-1}$.}
    \label{fig: saddle FR sound speed no chemical potential}
\end{figure}
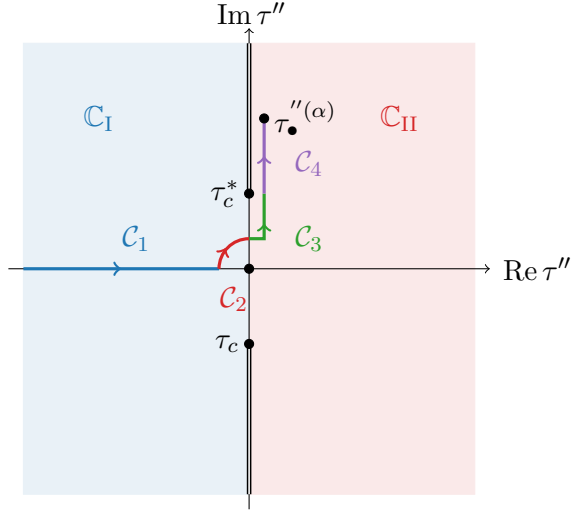
\begin{equation}
    \begin{aligned}
        |{F}_R^{(3)}| &\sim \exp\left(-\Im \int_{\C_2} \d\tau \,\omega_s(\tau)\right) \exp\left(-\Im \int_{\C_4} \d\tau \,\omega_s(\tau)\right) \exp\left(-c_s k_{34} \,\Im \tau_\bullet''^{(\alpha)}\right) \\
        &= e^{\tfrac{\pi}{2} \underset{\tau=0}{\Res} \, \omega_s(\tau)} \exp\left(\int_{\mu/s}^{\mu/\sqrt{s^2-c_s^2k_{34}^2}}\sqrt{s^2 - \frac{\mu^2}{x^2}}\d x - \frac{c_sk_{34}\mu}{\sqrt{s^2-c_s^2k_{34}^2}}\right)\\
        &= e^{-\pi\mu/2} \exp\left(-\mu\arcsin\frac{c_sk_{34}}{s}\right) \,.
    \end{aligned}
\end{equation}
Applying the full saddle point approximation yields
\begin{equation}
    {F}_R^{(3)} = \frac{-i\sqrt{\pi\mu}\,e^{-\pi\mu/2}}{(s^2 - c_s^2k_{34}^2)^{3/4}}\exp\left[-\mu\arcsin\frac{c_sk_{34}}{s}\right] \,.
\end{equation}
Combining all relevant expressions, we obtain the full approximate solution for the cosmological collider signal
\begin{eBox}
\begin{equation}
\label{eq: full saddle point approx solution with cs}
    \begin{aligned}
        F_{\rm F} + \text{c.c.} &\approx \frac{2\pi\mu}{[(c_s^2k_{12}^2-s^2)(s^2-c_s^2k_{34}^2)]^{3/4}} \\
        &\times e^{-\pi\mu/2} \exp\left[-\mu\arcsin\frac{c_sk_{34}}{s}\right] \cos\left[\mu\, \arccosh\frac{c_sk_{12}}{s} + \frac{3\pi}{4}\right]\,.
    \end{aligned}
\end{equation}
\end{eBox}
Again, the signal in the edge region $k_{12}/s<c_s^{-1}<k_{34}/s$ is obtained by swapping the kinematic variables $k_{12}\leftrightarrow k_{34}$.

\paragraph{Discussion.} A few comments are in order.

\begin{itemize}
    \item In contrast to the conventional cosmological collider signals, the one with a reduced sound speed is suppressed by only a \textit{quarter} of the usual Boltzmann factor $e^{-\pi\mu/2}=e^{-m/(4T_{\rm dS})}$ due to the \textit{quarter} circle we take to reach the saddle. Such amplified oscillations have long been known in the literature but without much explanation nor interpretation~\cite{Lee:2016vti, Jazayeri:2022kjy}.

    \item The behaviour of the cosmological collider signal is intimately related to the location of the saddle $\tau_\bullet''^{(\alpha)}$. In the edge region of kinematic space, $\tau_\bullet'^{(\alpha)}$ is real, and it shifts in the real direction when we vary the ratio $k_{12}/s$, giving rise to oscillations. On the other hand, $\tau_\bullet''^{(\alpha)}$ is purely imaginary. Hence when we vary the ratio $k_{34}/s$, the saddle point shifts in the imaginary direction, giving rise to smooth exponential attenuation instead of oscillations. In addition, the arcsin function nicely produces a smooth interpolation between $e^{-\pi\mu/2}$ (reduced sound speed) and $e^{-\pi\mu}$ (conventional collider), as $c_s k_{34}/s$ increases from zero to one.

    \item Interestingly, we notice that~\eqref{eq: full saddle point approx solution with cs} can be derived from~\eqref{eq: full saddle point approx solution} via direct analytic continuation. To see this, note that the $\arccosh(z)$ function has a branching point at $z\equiv c_s k_{34}/s = 1$. To consistently continue beyond this singularity, we recall that $k_{34}=|k_{34}|-i\epsilon$ is defined with a negative imaginary part to ensure the convergence of ${F}_R^{(3)}$. Therefore, the correct transformation of the $\arccosh(z)$ function is
    \begin{equation}
        \arccosh(z)\to -i\arccos(z) \,, \quad 0<\Re z<1 \,, \quad \Im z < 0 \,.
    \end{equation}
    The power law prefactor transforms as
    \begin{equation}
    	(c_s^2 k_{34}^2-s^2)^{3/4}\to e^{-3i\pi/4}(s^2-c_s^2 k_{34}^2)^{3/4} \,.
    \end{equation}
    Expansion of the trigonometric functions yields terms of higher order in $\mathcal{O}(e^{-\#\mu})$ which can be dropped. This allows us to directly land on the correct result~\eqref{eq: full saddle point approx solution with cs} without repeating the entire saddle point analysis. It is intriguing why this should be the case since approximate solutions are not expected to enjoy precise analytic continuations in general. We leave this pleasant yet seemingly coincidental phenomenon for future work.

    \item Similar to the case with unit sound speed, realistic correlators with more derivatives in the interaction vertices can be obtained by differentiating the seed integral with respect to the external kinematics. In the large-$\mu$ asymptotic limit, we can approximate
    \begin{equation}
        \begin{aligned}
            \left(k_{12}\partial_{k_{12}}\right)^{p_1} \left(k_{34}\partial_{k_{34}}\right)^{p_2} (F_{\rm F}+\text{c.c.}) &\approx \left(\frac{c_s\mu k_{12}}{\sqrt{c_s^2k_{12}^2-s^2}}\right)^{p_1} \left(\frac{-c_s\mu k_{34}}{\sqrt{s^2-c_s^2k_{34}^2}}\right)^{p_2} \\
            &\times (F_{\rm F}+\text{c.c.})_{\text{shifted phase}} \,.
        \end{aligned}
    \end{equation}
    Again, the waveform changes at most by a discrete phase shift,
    \begin{equation}
	\text{shifted phase}=\frac{\pi p_1}{2} \,.
    \end{equation}
    Consequently, the phase of the low-speed collider signal is fixed at $\pm\pi/4,\pm 3\pi/4$ at leading order in $\mu^{-1}$, which is a unique prediction of the approximated signal.
\end{itemize}

\subsubsection{Further Turning on the Chemical Potential}

We now move to the most general case where we further turn on a non-zero chemical potential $\kappa$. Since the WKB approximation \eqref{eq: WKB approx mode function} is valid only when $|\beta_s(\tau)|\ll 1$, we will only consider the parameter regime with weak particle production, i.e.~$\kappa<\mu$. This avoids large backreaction from $\sigma$ loops \cite{Tong:2022cdz}. Without loss of generality, we set $k_{12}>k_{34}$ and concentrate on the enhanced helicity $\lambda = -1$. The result for the $\lambda=+1$ mode can be found via a simple flip of the chemical potential $\kappa\to -\kappa$. As discussed in the previous subsection, we also require $k_{12}/s>c_s^{-1}$ to ensure the validity of the bulk cutting rule. The effect of the chemical potential is to add a linear term in the massive effective frequency
\begin{equation}
    \omega_s^2(\tau) = s^2 + \frac{2s\kappa}{\tau} + \frac{\mu^2}{\tau^2} \,.
\end{equation}
This additional linear term does not alter the leading late-time pole at $\tau=0$, but it shifts the turning points away from the imaginary axis,
\begin{equation}
    \tau_c = \frac{1}{s}\left(-\kappa - i \sqrt{\mu^2-\kappa}\right) \,.
\end{equation}
The branch cut at $\omega_s^2(\tau)<0$ joins $\tau_c, \tau_c^*$ and $0$, separating the complex plane into different Riemann sheets. The effective frequency is therefore defined as double-valued function
\begin{equation}
    \begin{aligned}
        \omega_s(\tau) &= +\sqrt{s^2+\frac{2s\kappa}{\tau}+\frac{\mu^2}{\tau^2}}\,, \,\,\, \tau \in \mathbb{C}_{\rm I}\,, \\
        \omega_s(\tau) &= -\sqrt{s^2+\frac{2s\kappa}{\tau}+\frac{\mu^2}{\tau^2}}\,, \,\,\, \tau \in \mathbb{C}_{\rm II}\,.
    \end{aligned}
\end{equation}
The singulant function admits a closed-form expression
\begin{equation}
    \begin{aligned}
        F_s(\tau) &= -2i\left[\tau \omega_s(\tau) + \mu\,\arctanh\left(\frac{s\tau\kappa+\mu^2}{-\mu\tau\omega_s(\tau)}\right) \right.\\
        &\left.-\kappa\,\arctanh\left(\frac{s\tau+\kappa}{-\tau\omega_s(\tau)}\right)\right] + \pi(\mu-\kappa)\,.
    \end{aligned}
\end{equation}
The chemical potential assists spontaneous particle production by alleviating the effective mass in the Boltzmann factor $|\beta_s|^2=e^{-2\Re F_s(\tau_i)}=e^{-2\pi(\mu-\kappa)}$. Following the same procedure as for the previous cases, we start with the left time integral
\begin{equation}
    \begin{aligned}
        F^{(3)}_L &\approx \int_{-\infty^+}^0 \frac{\d\tau'}{\sqrt{2\omega_s(\tau')}}\left[e^{i\left(k_{12}\tau'-\int_{\tau_i}^{\tau'} \d\tau\, \omega_s(\tau)\right)} + \beta_s(\tau')e^{i\left(k_{12}\tau'+\int_{\tau_i}^{\tau'} \d\tau\, \omega_s(\tau)\right)}\right] \\
        &\equiv F^{(3), \alpha}_L + F^{(3), \beta}_L \,.
    \end{aligned}
\end{equation}
Solving for the saddle points, we find
\begin{equation}
    \tau_\bullet'^{(\alpha)} = \frac{s(\kappa - \mu \Delta_L)}{c_s^2k_{12}^2-s^2} \in \mathbb{C}_{\rm I} \,, \quad
    \tau_\bullet'^{(\beta)} = \frac{s(\kappa + \mu \Delta_L)}{c_s^2k_{12}^2-s^2} \in \mathbb{C}_{\rm II} \,,
\end{equation}
with $\Delta_L = \sqrt{c_s^2k_{12}^2/s^2-(1-\kappa^2/\mu^2)}$. Again, the integral $F^{(3), \alpha}_L$ has a physical saddle point, and thereby dominates over $F^{(3), \beta}_L$. Finishing the Gaussian integral around $\tau_\bullet'^{(\alpha)}$, we obtain
\begin{equation}
    \begin{aligned}
        F_L^{(3)} &\approx \frac{\sqrt{\pi}\mu^{3/2}}{s^{3/2}(\kappa+\mu\Delta_L)\sqrt{\Delta_L}} e^{-i\pi/4-i\uptheta} \exp\left[-i\mu\left(\arccosh\frac{c_sk_{12}/s}{\sqrt{1-\kappa^2/\mu^2}}\right) \right.\\
        &\left.-i\kappa\left(\arccoth\frac{c_sk_{12}}{s} - \arctanh\frac{c_s k_{12} \kappa}{s\mu\Delta_L}\right)\right] \,,
    \end{aligned}
\end{equation}
where $\uptheta$ is a kinematic-independent phase, given by
\begin{equation}
    \uptheta(\mu, \kappa) \equiv \frac{\mu+\kappa}{2} \log(\mu+\kappa) - \frac{\mu-\kappa}{2} \log(\mu-\kappa) - \kappa \,.
\end{equation}
As usual, the integral ${F}_R^{(3)}$ is more complicated. We define
\begin{equation}
    \begin{aligned}
        {F}^{(3)}_R &\approx \int_{-\infty^+}^0 \frac{\d\tau''}{\sqrt{2\omega_s(\tau'')}}\left[e^{i\left(c_sk_{34}\tau''+\int_{\tau_i}^{\tau''} \d\tau\, \omega_s(\tau)\right)} + \beta_s^*(\tau'')e^{i\left(c_sk_{34}\tau''-\int_{\tau_i}^{\tau''} \d\tau\, \omega_s(\tau)\right)}\right] \\
        &\equiv {F}^{(3), \alpha}_R + {F}^{(3), \beta}_R \,.
    \end{aligned}
\end{equation}
The location of the saddle points are
\begin{equation}
    \tau_\bullet''^{(\alpha)} = \frac{s(\kappa + \mu \Delta_R)}{c_s^2k_{34}^2-s^2} \in \mathbb{C}_{\rm II} \,, \quad
    \tau_\bullet''^{(\beta)} = \frac{s(\kappa - \mu \Delta_R)}{c_s^2k_{34}^2-s^2} \in \mathbb{C}_{\rm I} \,,
\end{equation}
with $\Delta_R = \sqrt{c_s^2k_{34}^2/s^2-(1-\kappa^2/\mu^2)}$. For $k_{34}/s>c_s^{-1}\sqrt{1-\kappa^2/\mu^2}$, we have a real $\Delta_R$ and the saddle points lie on the real axis. For $k_{34}/s<c_s^{-1}\sqrt{1-\kappa^2/\mu^2}$, $\Delta_R$ is purely imaginary and the saddle points can move around in the complex plane. The point $k_{34}/s=c_s^{-1}$ also distinguishes the position of $\tau_\bullet''^{(\alpha)}$ relative to infinity. As the qualitative behaviour of cosmological collider signals dramatically depends on the location of the saddle points, we separate the discussion according to kinematic regions.

\paragraph{Interior region $k_{34}/s>c_s^{-1}$.} This region is depicted in \textcolor{pyblue}{blue} in Fig.~\ref{fig: kinematic space with sound speed} (recall that $k_{12}>k_{34}$). In this region, $\tau_\bullet''^{(\alpha)}$ is along the positive real axis on the second Riemann sheet $\mathbb{C}_{\rm II}$, whereas $\tau_\bullet''^{(\beta)}$ is along the negative real axis on $\mathbb{C}_{\rm I}$. The integral ${F}_R^{(3), \beta}$ picks up the saddle in the physical region, which leads to
\begin{equation}
    \begin{aligned}
        {F}_R^{(3), \beta} &\approx \frac{\sqrt{\pi}\mu^{3/2}e^{-\pi(\mu-\kappa)}}{s^{3/2}(\kappa+\mu\Delta_R)\sqrt{\Delta_R}} e^{i\pi/4+i\uptheta} \exp\left[-i\mu\left(\arccosh\frac{c_sk_{34}/s}{\sqrt{1-\kappa^2/\mu^2}}\right) \right.\\
        &\left.-i\kappa\left(\arccoth\frac{c_sk_{34}}{s} - \arctanh\frac{c_s k_{34} \kappa}{s\mu\Delta_R}\right)\right] \,.
    \end{aligned}
\end{equation}
In order to apply the saddle-point approximation for the integral ${F}_R^{(3), \alpha}$, we deform the initial integration contour such that it dives under the branch cut, passing through $\tau_\bullet''^{(\alpha)}$ in the steepest direction. As represented in Fig.~\ref{fig: saddle FR sound speed chemical potential}, we again need to rotate around the origin, therefore picking up the same suppression power
\begin{equation}
    |{F}_R^{(3), \alpha}| \sim \exp\left(-\Im\int_{\C_2} \d\tau\, \omega_s(\tau)\right) = e^{\pi \underset{\tau=0}{\Res} \, \omega_s(\tau)} = e^{-\pi\mu} \,.
\end{equation}
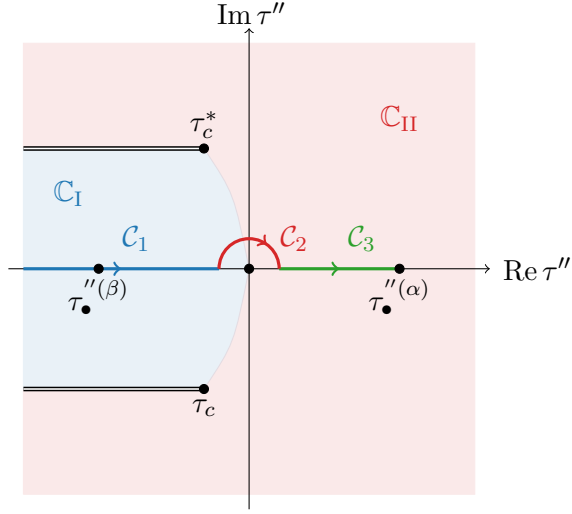
\begin{figure}[t!]
     \centering
     \begin{tikzpicture}
		[scale = 2]
        \node at (-1.2, 0.5) {\textcolor{pyblue}{$\mathbb{C}_{\rm I}$}};
        \node at (1, 1) {\textcolor{pyred}{$\mathbb{C}_{\rm II}$}};

        \draw[pyred, fill=pyred, opacity=0.1] 
        (-1.5, -1.5) -- (1.5, -1.5) -- (1.5, 1.5) -- (-1.5, 1.5) 
        -- (-1.5, 0.8) -- (-0.3, 0.8) 
        .. controls (-0.1, 0.5) .. (0, 0) 
        .. controls (-0.1, -0.5) .. (-0.3, -0.8) 
        -- (-1.5, -0.8) -- (-1.5, -1.5);
        \draw[pyblue, fill=pyblue, opacity=0.1] 
        (-1.5, 0.8) -- (-0.3, 0.8) .. controls (-0.1, 0.5) .. (0, 0) 
        .. controls (-0.1, -0.5) .. (-0.3, -0.8) 
        -- (-1.5, -0.8) -- (-1.5, 0.8);

        \draw[->] (0, -1.6) -- (0, 1.6) coordinate[label=below:$ $] (k1);
	\draw[->] (-1.6, 0) -- (1.6, 0) coordinate[label=below:$ $] (k1);
        \node at (0, 1.7) {$\Im\tau''$};
	\node at (1.9, 0) {$\Re\tau''$};

        \draw[-, double, decorate, decoration={segment length=6, amplitude=1.5}, black, line width=0.2mm] (-0.3, 0.8) -- (-1.5, 0.8);
        \draw[-, double, decorate, decoration={segment length=6, amplitude=1.5}, black, line width=0.2mm] (-0.3, -0.8) -- (-1.5, -0.8);
        \filldraw[black] (-0.3, 0.8) circle (.03cm) node[above] {$\tau_c^*$};
        \filldraw[black] (-0.3, -0.8) circle (.03cm) node[below] {$\tau_c$};

        \path[pyblue, draw, line width = 1.2pt, postaction = decorate, decoration={markings,
			mark=at position 0.5 with {\arrow[line width=1pt]{>}}}] (-1.5, 0) -- (-0.2, 0);
        \node at (-0.75, 0.2) {\textcolor{pyblue}{$\C_1$}};

        \path[pygreen, draw, line width = 1.2pt, postaction = decorate, decoration={markings,
			mark=at position 0.5 with {\arrow[line width=1pt]{>}}}] (0.2, 0) -- (1, 0);
        \node at (0.75, 0.2) {\textcolor{pygreen}{$\C_3$}};

        \path[draw, line width = 1.2pt, pyred, postaction = decorate, decoration={markings,
			mark=at position 0.7 with {\arrow[line width=1pt]{>}}}] (-0.2, 0)  arc (180:0:0.2);
        \node at (0.3, 0.2) {\textcolor{pyred}{$\C_2$}};
            
        \filldraw[black] (-1, 0) circle (.03cm) node[below] {$\tau_\bullet^{''(\beta)}$};
        \filldraw[black] (1, 0) circle (.03cm) node[below] {$\tau_\bullet^{''(\alpha)}$};;
        \filldraw[black] (0, 0) circle (.03cm); 
    \end{tikzpicture}
    \caption{Analytic structure of the effective frequency $\omega_s$ in the complex $\tau''$ plane to compute ${F}^{(3)}_R$ in the case of a non-unity sound speed and non-zero chemical potential in the interior region.}
    \label{fig: saddle FR sound speed chemical potential}
\end{figure}
Finishing the Gaussian integral, we obtain
\begin{equation}
    \begin{aligned}
        {F}_R^{(3), \alpha} &\approx \frac{-\sqrt{\pi}\mu^{3/2}e^{-\pi\mu}}{s^{3/2}(\kappa-\mu\Delta_R)\sqrt{\Delta_R}} e^{i\pi/4+i\uptheta} \exp\left[i\mu\left(\arccosh\frac{c_sk_{34}/s}{\sqrt{1-\kappa^2/\mu^2}}\right) \right.\\
        &\left.-i\kappa\left(\arccoth\frac{c_sk_{34}}{s} + \arctanh\frac{c_s k_{34} \kappa}{s\mu\Delta_R}\right)\right] \,.
    \end{aligned}
\end{equation}
Combining all found results, we obtain the leading approximation of the local and non-local signals in the interior region $k_{12}/s>k_{34}/s>c_s^{-1}$,
\begin{eBox}
\begin{equation}
\label{eq: full saddle point approx solution with cs with kappa}
    \begin{aligned}
        &F_{\rm F} + \text{c.c.} \approx \frac{2\pi\mu^3}{s^3(\kappa+\mu\Delta_L)\sqrt{\Delta_L\Delta_R}} \\
        &\times\left\{ \frac{e^{-\pi\mu}}{\mu\Delta_R-\kappa} \sin\left[\mu\left(\arccosh\frac{c_sk_{12}/s}{\sqrt{1-\kappa^2/\mu^2}} - \arccosh\frac{c_sk_{34}/s}{\sqrt{1-\kappa^2/\mu^2}}\right)+\Xi\right] \right.\\
        &\left.\hspace*{0.38cm}+\frac{e^{-\pi(\mu-\kappa)}}{\mu\Delta_R+\kappa}\cos\left[\mu\left(\arccosh\frac{c_sk_{12}/s}{\sqrt{1-\kappa^2/\mu^2}} + \arccosh\frac{c_sk_{34}/s}{\sqrt{1-\kappa^2/\mu^2}}\right)+\Upsilon\right]\right\}\,,
    \end{aligned}
\end{equation}
\end{eBox}
where the dephasing functions are given by
\begin{equation}
    \begin{aligned}
        \Xi &\equiv \kappa\left(\arccoth\frac{c_s k_{12}}{s}+\arccoth\frac{c_s k_{34}}{s}-\arctanh\frac{c_s k_{12} \kappa}{s \mu \Delta_L}+\arctanh\frac{c_s k_{34} \kappa}{s \mu \Delta_R}\right) \,, \\
        \Upsilon &\equiv \Xi-2\kappa\,\arctanh\frac{c_s k_{34} \kappa}{s \mu \Delta_R} \,.
    \end{aligned}
\end{equation}
Let us now comment on the result.

\begin{itemize}
    \item In the interior region, the cosmological collider signals here behave qualitatively the same way as those in the case $c_s=1$ and $\kappa=0$. The first line in~\eqref{eq: full saddle point approx solution with cs with kappa} represents the local signal. Its size is dictated only by the mass $\O(e^{-\pi\mu})$ coming from the semicircle around the origin. In the soft limit, the dephasing function approaches zero $\Xi\to0$ and the waveform approaches $\sin \mu \log(k_{12}/k_{34})$. In other words, the phase of the local cosmological collider signal is {\it universally} zero, independent of any parameters of the theory.

    \item The second line in~\eqref{eq: full saddle point approx solution with cs with kappa} gives the non-local signal, whose size is amplified by the chemical potential $\O(e^{-\pi(\mu-\kappa)})$. The corresponding dephasing function asymptotes to a constant that does depend on the parameters of the theory $\Upsilon \to -2\kappa \, \arctanh\kappa/\mu$. In general, the presence of the dephasing function implies that the waveform can significantly deviate from a simple logarithmically oscillating pattern. 

    \item As in the previous cases, results from realistic interaction vertices are straightforwardly obtained by taking derivatives with respect to external kinematics. The waveforms stay invariant modulo discrete phase shifts.
\end{itemize}

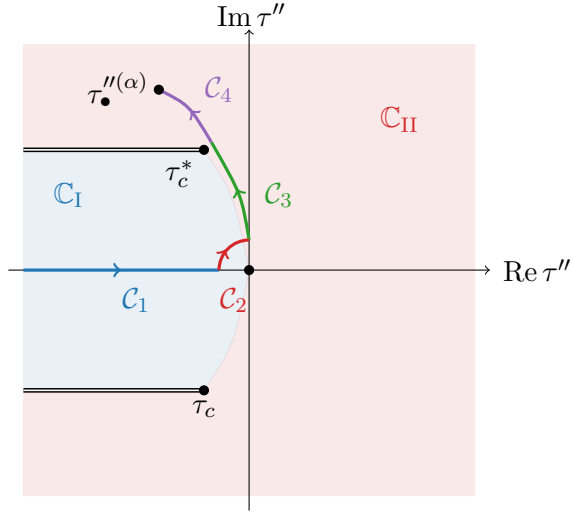
\begin{figure}[h!]
     \centering
     \begin{tikzpicture}
		[scale = 2]
        \node at (-1.2, 0.5) {\textcolor{pyblue}{$\mathbb{C}_{\rm I}$}};
        \node at (1, 1) {\textcolor{pyred}{$\mathbb{C}_{\rm II}$}};

        \draw[pyred, fill=pyred, opacity=0.1] 
        (-1.5, -1.5) -- (1.5, -1.5) -- (1.5, 1.5) -- (-1.5, 1.5) 
        -- (-1.5, 0.8) -- (-0.3, 0.8) 
        .. controls (-0.1, 0.5) .. (0, 0) 
        .. controls (-0.1, -0.5) .. (-0.3, -0.8) 
        -- (-1.5, -0.8) -- (-1.5, -1.5);
        \draw[pyblue, fill=pyblue, opacity=0.1] 
        (-1.5, 0.8) -- (-0.3, 0.8) .. controls (-0.1, 0.5) .. (0, 0) 
        .. controls (-0.1, -0.5) .. (-0.3, -0.8) 
        -- (-1.5, -0.8) -- (-1.5, 0.8);

        \draw[->] (0, -1.6) -- (0, 1.6) coordinate[label=below:$ $] (k1);
	\draw[->] (-1.6, 0) -- (1.6, 0) coordinate[label=below:$ $] (k1);
        \node at (0, 1.7) {$\Im \tau''$};
	\node at (1.9, 0) {$\Re \tau''$};

        \draw[-, double, decorate, decoration={segment length=6, amplitude=1.5}, black, line width=0.2mm] (-0.3, 0.8) -- (-1.5, 0.8);
        \draw[-, double, decorate, decoration={segment length=6, amplitude=1.5}, black, line width=0.2mm] (-0.3, -0.8) -- (-1.5, -0.8);
        \filldraw[black] (-0.3, 0.8) circle (.03cm) node[below left] {$\tau_c^*$};
        \filldraw[black] (-0.3, -0.8) circle (.03cm) node[below] {$\tau_c$};

        \path[pyblue, draw, line width = 1.2pt, postaction = decorate, decoration={markings,
			mark=at position 0.5 with {\arrow[line width=1pt]{>}}}] (-1.5, 0) -- (-0.2, 0);
        \node at (-0.75, -0.2) {\textcolor{pyblue}{$\C_1$}};

        \path[pygreen, draw, line width = 1.2pt, postaction = decorate, decoration={markings,
			mark=at position 0.5 with {\arrow[line width=1pt]{>}}}] (0, 0.2) .. controls (-0.05, 0.5) .. (-0.25, 0.85);
        \node at (0.2, 0.5) {\textcolor{pygreen}{$\C_3$}};

        \path[pypurple, draw, line width = 1.2pt, postaction = decorate, decoration={markings,
			mark=at position 0.5 with {\arrow[line width=1pt]{>}}}] (-0.25, 0.85) .. controls (-0.4, 1.1) .. (-0.6, 1.2);
        \node at (-0.2, 1.2) {\textcolor{pypurple}{$\C_4$}};

        \path[draw, line width = 1.2pt, pyred, postaction = decorate, decoration={markings,
			mark=at position 0.7 with {\arrow[line width=1pt]{<}}}] (0, 0.2)  arc (90:180:0.2);
        \node at (-0.1, -0.2) {\textcolor{pyred}{$\C_2$}};
            
        \filldraw[black] (-0.6, 1.2) circle (.03cm) node[left] {$\tau_\bullet^{\prime\prime(\alpha)}$};;
        \filldraw[black] (0, 0) circle (.03cm); 
    \end{tikzpicture}
    \caption{Analytic structure of the effective frequency $\omega_s$ in the complex $\tau''$ plane to compute ${F}^{(3)}_R$ in the case of a non-unity sound speed and non-zero chemical potential in the edge region. }
    \label{fig: saddle FR sound speed and chemical potential}
\end{figure}

\paragraph{Edge region $k_{34}/s<c_s^{-1}\sqrt{1-\kappa^2/\mu^2}$.} This region is depicted in \textcolor{pygreen}{green} in Fig.~\ref{fig: kinematic space with sound speed}. Here, both saddle points are complex. Using the rule of thumb, the saddle at $\tau_\bullet''^{(\beta)}$ can then be dropped due to its additional suppression. We then focus on the saddle located at
\begin{equation}
    \tau_\bullet''^{(\alpha)} = \frac{s(-\kappa+i\mu|\Delta_R|)}{s^2-c_s^2k_{34}^2} \in \mathbb{C}_{\rm II}\,.
\end{equation}
To reach this saddle, we again need to dive under the branch cut to enter the second Riemann sheet $\mathbb{C}_{\rm II}$. The size of the integral is estimated from the exponent
\begin{equation}
    \begin{aligned}
        |{F}^{(3), \alpha}_R| &\sim \left|\exp\left(ic_sk_{34}\tau_\bullet''^{(\alpha)}+ i\int_{-\infty}^{\tau_\bullet''^{(\alpha)}}\d\tau\, \omega_s(\tau)\right)\right| \\
        &= \exp\left(-c_s k_{34} \,\Im \tau_\bullet''^{(\alpha)}\right) \exp\left(-\Im \int_{-\infty}^{\tau_\bullet''^{(\alpha)}} \d\tau \omega_s(\tau)\right) \,.
    \end{aligned}
\end{equation}
As depicted in Fig.~\ref{fig: saddle FR sound speed and chemical potential}, the path to reach the saddle consists of four segments
\begin{equation}
    \int_{-\infty}^{\tau_\bullet''^{(\alpha)}} = \int_{\textcolor{pyblue}{\C_1}} + \int_{\textcolor{pyred}{\C_2}} + \int_{\textcolor{pygreen}{\C_3}} + \int_{\textcolor{pypurple}{\C_4}} \,.
\end{equation}
The contour $\C_1$ is purely real and drops out of the imaginary part. The contour $\C_4$ part is negligible if $k_{34}/s \ll c_s^{-1}$ because both the integrand and the length of the integration path tend to zero in this regime. We are left with an integral along a quarter circle around the origin, which gives
\begin{equation}
    \Im \int_{\C_2}\d\tau \omega_s(\tau) = -\Im\left(\frac{i\pi}{2}\,\underset{\tau=0}{\Res} \, \omega_s(\tau)\right) = \frac{\pi\mu}{2}\,,
\end{equation}
and a contour around the branch cut on $\mathbb{C}_{\rm II}$, whose imaginary part can be computed as
\begin{align}
    \Im\int_{\mathcal{C}_3} \d\tau\, \omega_s(\tau)
    &=\frac{1}{4}\Im\int_{\mathcal{C}_{\rm cut}\subset \mathbb{C}_{\rm II}} \d\tau\, \omega_s(\tau)=\frac{1}{4}\Im\int_{\mathcal{C}_\infty\subset \mathbb{C}_{\rm II}} \d\tau\, \omega_s(\tau)\nonumber
    \\
    &=\Im\left(\frac{-\pi i}{2}\underset{\tau=\infty}{\Res} \omega_s(\tau)\right)=-\frac{\pi {\kappa}}{2}~.
\end{align}
Here in the first step we have used the fact that $\omega_s^*(\tau)=\omega_s(\tau)$ is a real function and that along the two sides of the branch cut, $\omega_s(\tau)$ differs by a minus sign. In the second step, we have deformed the contour around the branch cut $\mathcal{C}_{\rm cut}$ to a large circle at infinity $\mathcal{C}_\infty$ on Sheet II, which is then evaluated as a residue at infinity.
We thus obtain a neat size formula $|{F}^{(3), \alpha}_R|\sim e^{-\pi(\mu-\kappa)/2}$. We note in passing that it is remarkable to behold the intimate connection between the cosmological collider signal strength, which reflects the amount of kinematic non-analyticities in boundary correlators, and the residue of the dispersion relation at early and late times, which reflects the amount of temporal non-analyticities of the bulk Lagrangian. Completing the Gaussian integral, we then have
\begin{equation}
    \begin{aligned}
        F_R^{(\alpha)} &\approx \frac{\sqrt{\pi}\mu^{3/2}e^{-\pi(\mu-\kappa)/2}}{s^{3/2}(\kappa+i\mu|\Delta_R|)\sqrt{|\Delta_R|}} e^{i\uptheta} \exp\left[-\mu\,\arcsin\frac{c_sk_{34}/s}{\sqrt{1-\kappa^2/\mu^2}} \right.\\
        &\left.+\kappa \arctan\frac{c_s k_{34}\kappa}{s\mu|\Delta_R|} - i\kappa \arctanh\frac{c_s k_{34}}{s}\right]\,.
    \end{aligned}
\end{equation}
Combining all found results, we obtain the leading approximation of the signals in the edge region
\begin{eBox}
\begin{equation}
\label{eq: full saddle point approx solution with cs with kappa edge}
    \begin{aligned}
        &F_{\mathrm{F}}+\text{c.c.} \approx \frac{2\pi\mu^3 e^{-\pi(\mu-\kappa)/2}}{s^3(\kappa+\mu\Delta_L)[(\kappa^2+\mu^2|\Delta_R|^2)\Delta_L|\Delta_R|]^{1/2}} \\
        &\times \exp\left[-\mu \arcsin\frac{c_s k_{34}/s}{\sqrt{1-\kappa^2/\mu^2}}+\kappa\arctan\frac{c_s k_{34}\kappa}{s\mu|\Delta_R|}\right]\\
		&\times\cos\left[\mu \, \arccosh\frac{c_s k_{12}/s}{\sqrt{1-\kappa^2/\mu^2}}+\kappa\arctanh\frac{c_s k_{34}}{s}+\frac{3\pi}{4}+\Omega\right]\,,
    \end{aligned}
\end{equation}
\end{eBox}
where the dephasing function reads
\begin{equation}
    \Omega \equiv \kappa\left(\arccoth\frac{c_s k_{12}}{s}-\arctanh\frac{c_s k_{12}\kappa}{s\mu\Delta_L}\right)-\arctan\frac{\kappa}{\mu|\Delta_R|} \,.
\end{equation}

\begin{itemize}
    \item In the edge region of the kinematics space, the cosmological collider signal has important chemical-potential-dependent corrections. Apart from the overall enhancement, the corrected attenuation factor renders the signal amplitude to smoothly interpolate between $e^{-\pi(\mu-\kappa)/2}$ and $e^{-\pi(\mu-\kappa)}$ when $c_s k_{34}/s$ varies from $0$ to $\sqrt{1-\kappa^2/\mu^2}$. This corresponds to the shift of the saddle $\tau_\bullet''^{(\alpha)}$ in the imaginary direction.

    \item The waveform is also corrected by the dephasing function $\Omega$, which tends to a constant in the hierarchical soft limit $c_s k_{12}/s\to \infty$, $c_s k_{34}/s\to 0$,
    \begin{equation}
	   \lim_{\substack{c_s k_{12}/s\to \infty\\c_s k_{34}/s\to 0}}\Omega =-\kappa\arctanh \frac{\kappa}{\mu}-\arctan\frac{\kappa}{\sqrt{1-\kappa^2/\mu^2}} \,.
    \end{equation}
    Interestingly, in addition to the dephasing function, the waveform acquires a term proportional to the chemical potential (the second term in the second line of~\eqref{eq: full saddle point approx solution with cs with kappa edge}). With $k_{12}$ held fixed, there is an intermediate stage where $k_{34}/s\ll c_s^{-1}$ and the waveform is approximated by
    \begin{equation}
	F_{\rm F} + \text{c.c.} \supset \exp\left(-c_s\sqrt{\mu^2-\kappa^2}\,\frac{k_{34}}{s} + \cdots\right)\cos\left(c_s\kappa\,\frac{k_{34}}{s} + \cdots\right) \,.
    \end{equation}
    Consequently, a {\it new type of cosmological collider signal} arises. Drastically distinct from all other more standard signals above, this new signal oscillates linearly with respect to the momentum ratio $k_{34}/s$ with a frequency $c_s \kappa$ determined by the sound speed and the chemical potential. Meanwhile, the amplitude of the oscillations decays exponentially at a rate $c_s \sqrt{\mu^2-\kappa^2}$ which is often comparable to oscillation frequency. Due to its large decay rate, we name this new signal the \textit{transient cosmological collider signal}. Such a transient signal was found previously in the parity-odd sector of non-local EFT description of the same system~\cite{Jazayeri:2023kji}, but without apparent physical interpretation. Here we see clearly that the origin of the transient signal is the curvilinear motion of the saddle point $\tau_\bullet''^{(\alpha)}$ on the complex plane. When the ratio $c_s k_{34}/s$ increases from nearly zero, the saddle point $\tau_\bullet''^{(\alpha)}$ moves away from $\tau_c^*$ and shifts simultaneously in the real and imaginary direction, leading to the damped oscillations in the transient signal. The oscillations stop exponentially decaying when $\tau_\bullet''^{(\alpha)}$ hits the negative real axis on the second Riemann sheet $\mathbb{C}_{\rm II}$, which corresponds to the boundary of the edge region where $k_{34}/s=c_s^{-1}\sqrt{1-\kappa^2/\mu^2}$ and our approximation fails.

    \item Similar to the previous case, the found result \eqref{eq: full saddle point approx solution with cs with kappa edge} can be derived from direct analytic continuation from that in the interior region \eqref{eq: full saddle point approx solution with cs with kappa}, upon utilising the following continuation rules,
    \begin{equation}
        \begin{aligned}
		\arccosh\frac{c_s k_{34}/s}{\sqrt{1-\kappa^2/\mu^2}}&\to -i\arccos\frac{c_s k_{34}/s}{\sqrt{1-\kappa^2/\mu^2}} \,,\\
		\Delta_R&\to -i|\Delta_R| \,,\\
		\arccoth\frac{c_s k_{34}}{s}&\to \arctanh\frac{c_s k_{34}}{s}+\frac{i\pi}{2} \,.
        \end{aligned}
    \end{equation}
\end{itemize}

\paragraph{Joint region $c_s^{-1}\sqrt{1-\kappa^2/\mu^2} < k_{34}/s < c_s^{-1}$.} Entering the join region from above, $k_{34}>s/c_s\to k_{34}<s/c_s$, the location of $\tau_\bullet''^{(\alpha)}$ switches from $+\infty$ to $-\infty$, while the location of $\tau_\bullet''^{(\beta)}$ changes continuously. In principle, we can run the same reasoning as before and perform the same saddle-point analysis. To avoid technical repetitions, instead of computing the approximate results from scratch, we shall take advantage of the analyticity property we discovered above. The continued results are then checked against the exact results. We observe that the dephasing functions contain singularities at $k_{34}/s=c_s^{-1}$ (where $\Delta_R=\tilde\kappa/\mu$) for different terms. In $\Upsilon$, the singularities of two trigonometrical functions cancel out, leaving a straightforward analytic continuation rule
\begin{equation}
    \Upsilon~:~\left\{ \begin{aligned}
        \arccoth\frac{c_s k_{34}}{s}&\to \arctanh\frac{c_s k_{34}}{s}+\frac{i\pi}{2}\\
        \arctanh\frac{c_s k_{34}\kappa}{s\mu\Delta_R}&\to \arccoth\frac{c_s k_{34}\kappa}{s\mu\Delta_R}+\frac{i\pi}{2}
    \end{aligned} \,.\right.
\end{equation}
In $\Xi$, the singularities add up, suggesting that the consistent continuation rule is
\begin{equation}
    \Xi~:~\left\{ \begin{aligned}
        \arccoth\frac{c_s k_{34}}{s}&\to \arctanh\frac{c_s k_{34}}{s}+\frac{i\pi}{2}\\
        \arctanh\frac{c_s k_{34}\kappa}{s\mu\Delta_R}&\to \arccoth\frac{c_s k_{34}\kappa}{s\mu\Delta_R}-\frac{i\pi}{2}
    \end{aligned} \,.\right.
\end{equation}
All other terms remain unchanged.

\subsection{Refined Cosmological Collider Templates}
\label{subsec: templates}

Unravelling the rich bulk physics hidden in the boundary correlators requires searching for cosmological collider signals as intricate kinematic dependences in the non-Gaussian statistics of cosmological perturbations. Due to the high dimensionality of the CMB/LSS datasets, reconstructing the correlators in the full kinematic domain using an unbiased estimator is extremely challenging. Instead, one is compelled to use biased estimators which are derived or motivated from specific theoretical models. In other words, this amounts to computing the inner product between a shape template and the data to check for the detection significance, which in turn requires an efficient sampling of the high-dimensional kinematics space as well as parameter space. As seen in Sec.~\ref{sec: Bootstrapping Boost-Breaking Cosmological Correlators}, exact solutions of the cosmological collider signals often involve sums of products of hypergeometric functions, leading to significant computational complexity. This motivates us to propose {\rm approximate} shape templates that substantially reduce computational cost while maintaining accuracy across most of the relevant kinematic and parameter spaces.

\vskip 4pt
The saddle point method has proved to be a powerful approximation scheme that allowed us to link kinematic dependence of the boundary correlators to the resonance history in the inflationary bulk. As a useful by-product, it has provided clean boost-breaking templates expressed in terms of elementary functions, yet more accurate than the ones previously used. Here, we propose new refined templates for cosmological collider signals that match exact calculations both in terms of amplitude and function dependence of the waveform. Importantly, as probing ultra-soft limits of cosmological correlators requires a large hierarchy of scales and is limited by cosmic variance, these new templates are valid in extended and intermediate regions of the parameters space (e.g.~covering mildly reduced sound speed cases) and of the kinematic configurations (e.g.~covering mildly soft limits). 

\paragraph{Trispectrum shape templates.} Following standard conventions, we define the dimensionless trispectrum shape function by
\begin{equation}
    \braket{\zeta_{\k_1} \zeta_{\k_2} \zeta_{\k_3} \zeta_{\k_4}} \equiv (2\pi)^9 \delta^{3}(\k_1 + \k_2 + \k_3 + \k_4) \frac{(k_T/4)^3 \P_\zeta^3}{(k_1k_2k_3k_4)^3} \, \T(\k_1, \k_2, \k_3, \k_4)\,,
\end{equation}
where $k_T \equiv \sum_i k_i$ and $\P_\zeta \equiv \tfrac{k^3}{2\pi^2}P_\zeta \approx 2.1\times 10^{-9}$ is the dimensionless power spectrum of the curvature perturbation $\zeta$. The function $\T$ is a dimensionless scalar function of momenta that is invariant under the (same) rescaling of all momenta $\k_i \to \alpha \k_i$. The dimensionless trispectrum $\T$ can be further decomposed into a polarisation factor $\Pi$ and a waveform function $\F$ 
\begin{equation}
\label{eq: trispectrum waveform def}
    \mathcal{T}({\bm k}_1,{\bm k}_2,{\bm k}_3,{\bm k}_4) = \left(\frac{s^2}{k_{12}k_{34}}\right)^{3/2}\sum_{\lambda}\,\Pi_{S, \lambda}({\hat{\bm  k}}_1,{\hat{\bm k}}_2,{\hat{\bm k}}_3,{\hat{\bm k}}_4,{\hat{\bm s}}) \, \F^{\lambda} (k_{12},k_{34},s) + {\rm 2 \,\, perms}\,,
\end{equation}
where $\hat{\bm k}_i\equiv \bm k_i/k_i$, $\hat{\bm s}\equiv \bm s/s$ are unit vectors, and $\bm{s} \equiv \k_1+\k_2$. The two permutations correspond to the $t$- and $u$-channels, with ${\bm t} \equiv \k_1+\k_3$ and ${\bm u} \equiv \k_1+\k_4$. In the following, we only consider the dominant helicity component of $\F^\lambda$. The polarisation factor $\Pi_{S, \lambda}$ depends on the (integer) spin of the exchanged field $S$, the helicity $-S \leq\lambda\leq S$ and the detailed form of the interaction vertices. In the soft limit, the polarisation factor is well approximated by \cite{Arkani-Hamed:2015bza}
\begin{equation}
    \Pi_{S, \lambda}({\hat{\bm  k}}_1,{\hat{\bm k}}_2,{\hat{\bm k}}_3,{\hat{\bm k}}_4,{\hat{\bm s}}) \approx e^{i\lambda\psi}\, P_S^\lambda({\hat{\bm  k}}_1\cdot {\hat{\bm  s}})P_S^{-\lambda}({\hat{\bm  k}}_3\cdot {\hat{\bm  s}}) \,,
\end{equation}
where $P_S^\lambda$ is the associated Legendre polynomial, and $\psi$ is the dihedral angle between the planes spanned by ${\bm k}_1,{\bm k}_2$ and ${\bm k}_3,{\bm k}_4$, such that $\cos\psi \equiv \hat{\bm{n}}_{12} \cdot \hat{\bm{n}}_{34}$ with $\hat{\bm{n}}_{12} \equiv \hat{\k}_1 \times \hat{\k}_2$ (resp.~$\hat{\bm{n}}_{34} \equiv \hat{\k}_3 \times \hat{\k}_4$) being the normal unit vector to the plane generated by the vectors $\k_1, \k_2$ (resp.~$\k_3, \k_4$). Explicitly, the first few terms are $\Pi_{0, 0} = 1, \Pi_{1, 0} = (\hat{\k}_1\cdot\hat{\bm{s}})\, (\hat{\k}_3\cdot\hat{\bm{s}}),$ and $\Pi_{1, \pm1} = -\tfrac{1}{2}e^{\pm i \psi} \sqrt{[1-(\hat{\k}_1\cdot\hat{\bm{s}})^2][1-(\hat{\k}_3\cdot\hat{\bm{s}})^2]}$. The $s$-channel dimensionless waveform function $\F(k_{12}, k_{34}, s)$ can be decomposed into a {\it local part} that exhibits an oscillatory behaviour in the logarithm of the momentum ratio $k_{12}/k_{34}$ (assuming $k_{12}>k_{34}$) and a {\it non-local part} oscillating in the logarithm of the momentum ratio $k_{12}k_{34}/s^2$:\footnote{The $s$-channel dimensionless waveform function only depends on $(k_{12}, k_{34}, s)$ as we consider the simple seed correlator~\eqref{eq: helical seed correlator}. The exchange trispectrum corresponding to specific cubic interactions, with a more complicated kinematics dependence, can then be deduced by the application of suitable differential operators, see, e.g.~, Eq.~\eqref{eq:example-trispectrum-specific-interactions}.}
\begin{equation}
    \F(k_{12}, k_{34}, s) = \F^{\rm L}(k_{12}, k_{34}, s) + \F^{\rm NL}(k_{12}, k_{34}, s) \,.
\end{equation}
The waveforms $\F^{\rm L}$ and $\F^{\rm NL}$ are parametrised by three constants: (i) a dimensionless mass parameter $\mu$ of the exchanged massive field, (ii) a speed of sound\footnote{Recall that $c_s$ denotes the ratio between the sound speed of inflaton fluctuations and that of the exchanged massive field, and therefore can exceed unity.} $0<c_s<\infty$, and (iii) a dimensionless chemical potential $|\kappa|<\mu$. In the $s$-channel, all cosmological collider signals lie in the soft limit $s\to0$. Without loss of generality, we set $k_{12}>k_{34}$ and the opposite case is obtained by the trivial swap $k_{12}\leftrightarrow k_{34}$. With the presence of non-trivial sound speed $c_s\neq1$ and a non-zero chemical potential $\kappa \neq 0$, which signals the breaking of de Sitter boosts, the kinematics space can be divided into an {\it interior region} $k_{34}/s\geq c_s^{-1}$ and an {\it edge region} $k_{34}/s\leq c_s^{-1}$, see Fig.~\ref{fig: kinematic space with sound speed}. The template then takes the form of a piecewise function of $k_{12}$, $k_{34}$ and $s$. Based on the analysis of Sec.~\ref{subsec: Application to Tree-Level Exchange Correlators}, we suggest the following simple local waveform template
\begin{eBox}
\begin{equation}
\label{eq: boost-breaking local template}
    \begin{aligned} 
        \F^{\rm L}(k_{12}, k_{34}, s) = e^{-\pi\mu}\sin&\left[\mu\left( \arccosh\frac{\gamma k_{12}}{s} - \arccosh\frac{\gamma k_{34}}{s}\right)\right.\\
        &\left.+\,\kappa\left(\arccoth\frac{c_s k_{12}}{s} + \arccoth\frac{c_s k_{34}}{s}\right) + \delta_{\rm L}\right] \quad k_{34}/s\geq c_s^{-1}\,,
    \end{aligned}
\end{equation}
\end{eBox}
where $\gamma \equiv c_s /\sqrt{1-\kappa^2/\mu^2}$ is a dimensionless boost-breaking parameter, and $\delta$ is a constant phase. 
Similarly, a simple non-local waveform template is given by 
\begin{eBox}
\begin{equation}
\label{eq: boost-breaking non-local template}
\F^{\rm NL}(k_{12}, k_{34}, s) =
    \begin{cases}
        \begin{aligned}
            e^{-\pi(\mu-\kappa)}\cos&\left[\mu\left( \arccosh\frac{\gamma k_{12}}{s} + \arccosh\frac{\gamma k_{34}}{s}\right)\right.\\
            &\left.+\,\kappa\left(\arccoth\frac{c_s k_{12}}{s} + \arccoth\frac{c_s k_{34}}{s}\right) + \delta_{\rm NL}\right]
        \end{aligned}
            & k_{34}/s\geq c_s^{-1} \,,\\
        \begin{aligned}
            e^{-\pi(\mu-\kappa)/2}&\exp\left[-\mu \, \arcsin\frac{\gamma k_{34}}{s}+\kappa\arctan\frac{c_s k_{34}\kappa}{s\mu|\Delta_R|}\right] \\
            &\times\cos\left[\mu \arccosh\frac{\gamma k_{12}}{s} + \kappa \arctanh\frac{c_sk_{34}}{s}+\delta_{\rm NL}\right]
        \end{aligned}
            & k_{34}/s\leq c_s^{-1} \,.
    \end{cases}
\end{equation}
\end{eBox}
with $\Delta_R = \sqrt{c_s^2k_{34}^2/s^2-(1-\kappa^2/\mu^2)}$. Let us now comment this template, and show that it encodes all the boost-breaking physics.

\begin{itemize}
    \item {\it de Sitter limit}: In the canonical de Sitter invariant case $\kappa=0$ and $c_s=1$, we have $\gamma = 1$ and the edge region $(k_{34}/s\leq c_s^{-1})$ disappears due to the triangle inequality. Therefore, the local and non-local signals reduce to 
    \begin{equation}
    \label{eq: trispectrum canonical template}
    \begin{aligned}
        \F^{\rm L}(k_{12}, k_{34}, s) &= e^{-\pi\mu}\sin\left[\mu\left( \arccosh\frac{k_{12}}{s} - \arccosh\frac{k_{34}}{s}\right) +\delta_{\rm L}\right]\,, \\
        \F^{\rm NL}(k_{12}, k_{34}, s) &= e^{-\pi\mu}\cos\left[\mu\left( \arccosh\frac{k_{12}}{s} + \arccosh\frac{k_{34}}{s}+\delta_{\rm NL}\right) \right]\,.
    \end{aligned}
    \end{equation}
    Notice in particular that both the local and the non-local signals have the same Boltzmann suppressed amplitude. In the soft limit $s\to0$, one recovers the conventional waveforms
    \begin{equation}
    \label{eq: trispectrum conventional template}
        \F^{\rm L} \approx e^{-\pi\mu} \sin[\mu \log(k_{12}/k_{34})+\delta_{\rm L}]\,, \quad \F^{\rm NL} \approx e^{-\pi\mu}\cos[\mu \log(k_{12}k_{34}/s^2)+\delta_{\rm NL}]\,.
    \end{equation}
    In Fig.~\ref{fig: Trispectrum template canonical}, we show both the local and non-local signals in the dimensionless trispectrum waveform function $\F$ in the $s$-channel, varying the mass parameter $\mu$. Remarkably, the refined waveform~\eqref{eq: trispectrum canonical template} matches well the exact result, even close to regular kinematic configurations $k_{12}/k_{34}\approx 1$ and $k_{12}k_{34}/s^2\approx 1$, even for $\mu =\O(1)$. Meanwhile, the conventional template exhibits an incorrect phase for $\mu \lesssim 3$, and fails to reproduce the amplitude of the local signal as the mass parameter increases. Notice that for the local signal, the conventional template does not accurately reproduce the exact signal in mildly soft kinematic configurations, where the signal-to-noise ratio would be the greater.

    \begin{figure}[h!]
    \centering
    \begin{subfigure}{.5\textwidth}
        \centering
        \includegraphics[width=1\linewidth]{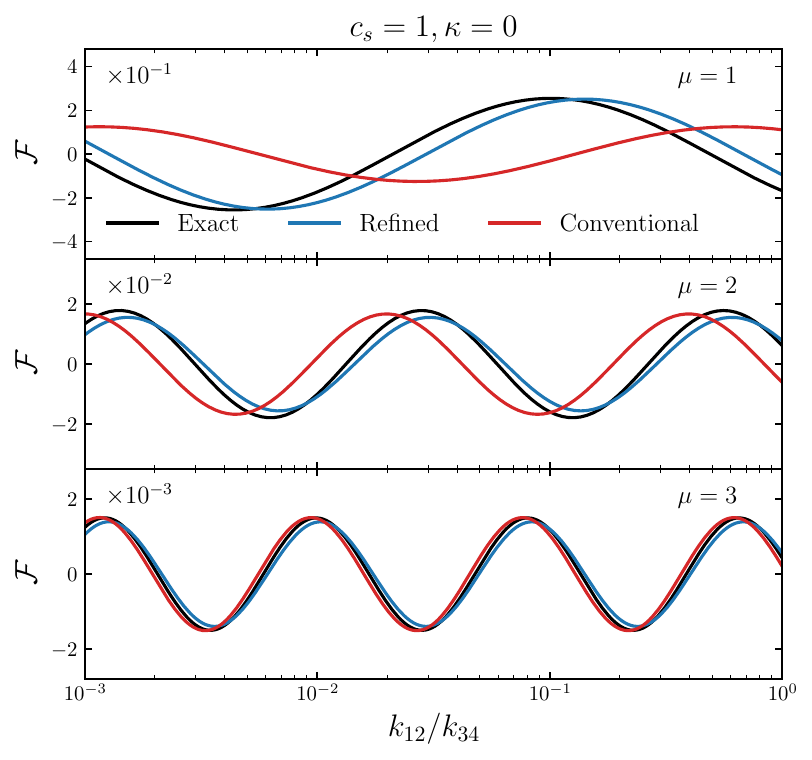}
    \end{subfigure}%
    \begin{subfigure}{.5\textwidth}
        \centering
        \includegraphics[width=1\linewidth]{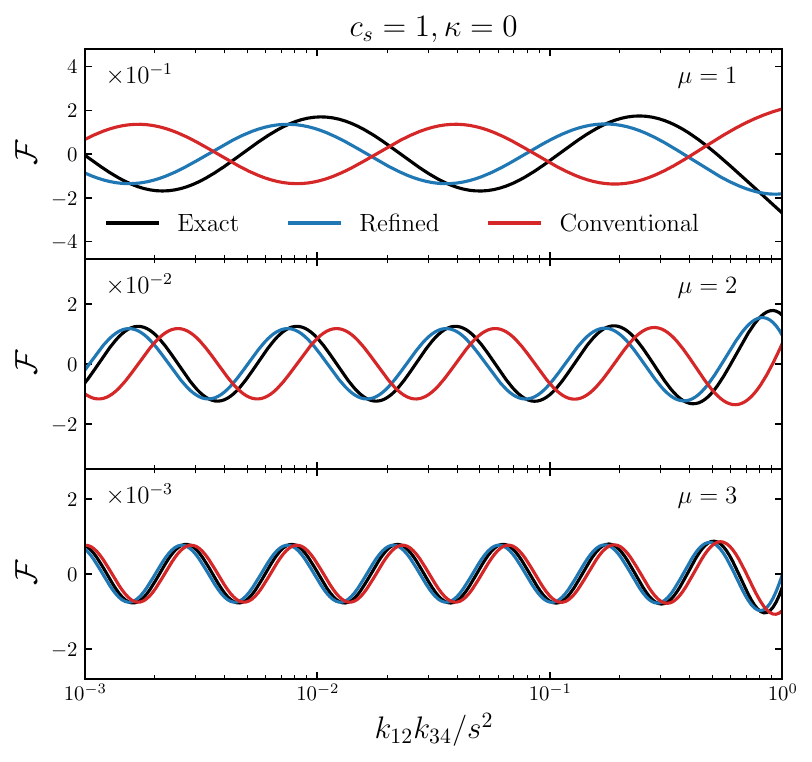}
    \end{subfigure}
   \caption{Dimensionless de Sitter invariant trispectrum waveform function $\F(k_{12}, k_{34}, s)$ in the $s$-channel, as defined in~\eqref{eq: trispectrum waveform def}, as function of the momentum ratio $k_{12}/k_{34}$ for the local signal (\textit{left panel}), and as function of the momentum ratio $k_{12}k_{34}/s^2$ for the non-local signal (\textit{right panel}), varying the mass parameter $\mu=1, 2, 3$. The exact result (\textcolor{black}{black}) is expressed in terms of product of hypergeometric functions, see Sec.~\ref{sec: Bootstrapping Boost-Breaking Cosmological Correlators}, the refined template (\textcolor{pyblue}{blue}) is given by~\eqref{eq: trispectrum canonical template}, and the conventional template (\textcolor{pyred}{red}) is given in~\eqref{eq: trispectrum conventional template}. We have taken $\delta=0$.}
  \label{fig: Trispectrum template canonical}
\end{figure}

    \item \textit{Reduced sound speed limit}: The notable feature of a reduced sound speed as the only boost-breaking effect is to modify the amplitude of the non-local signal which reads
    \begin{equation}
        \F^{\rm NL}(k_{12}, k_{34}, s) =
    \begin{cases}
        e^{-\pi\mu/2}\cos\left[\mu\left( \arccosh\frac{c_s k_{12}}{s} + \arccosh\frac{c_s k_{34}}{s}\right)\right]
        & k_{34}/s\geq c_s^{-1} \,,\\
        e^{-\pi\mu/2-\mu \, \arcsin c_s k_{34}/s} \cos\left[\mu \arccosh\frac{c_s k_{12}}{s} +\delta_{\rm NL}\right]
        & k_{34}/s\leq c_s^{-1} \,.
    \end{cases}
    \end{equation}
    In the ultra-reduced sound speed case $c_s\to0$ deep in the edge region, this template reduces to the conventional one
    \begin{equation}
        \F^{\rm NL} \approx e^{-\pi\mu/2} \cos[\mu\log(s/c_s k_{12})+\delta_{\rm NL}]\,,
    \end{equation}
    which describes an amplified cosmological collider signal~\cite{Lee:2016vti,Jazayeri:2023xcj}. Notice that the kinematic-dependent amplitude factor $\exp[-\mu \arcsin(c_s k_{34}/s)]$ interpolates from $1$ to $e^{-\pi\mu/2}$ as we scan the edge region $0<c_s k_{34}/s<1$. As depicted in Fig.~\ref{fig: Trispectrum template non-canonical}, where we show the dimensionless trispectrum exact, approximate, and conventional waveform, this dependence is crucial to accurately capture the amplitude of the cosmological collider signal. Indeed, we notice that the amplitude of the cosmological collider signal given by the conventional template is inaccurate, especially close to the interior region of the kinematic configurations.
    \item {\it Helical template}: Further turning on the helical chemical potential yields the full boost-breaking trispectrum templates~\eqref{eq: boost-breaking local template} and~\eqref{eq: boost-breaking non-local template}. Notably, the non-local signal amplitude is boosted by an enhanced Boltzmann factor $e^{-\pi(\mu-\kappa)/2}$, with again the additional kinematic dependence smoothly interpolating between the conventional cosmological amplitude and that with a reduced sound speed.
    \begin{figure}[h!]
    \centering
    \begin{subfigure}{.5\textwidth}
        \centering
        \includegraphics[width=1\linewidth]{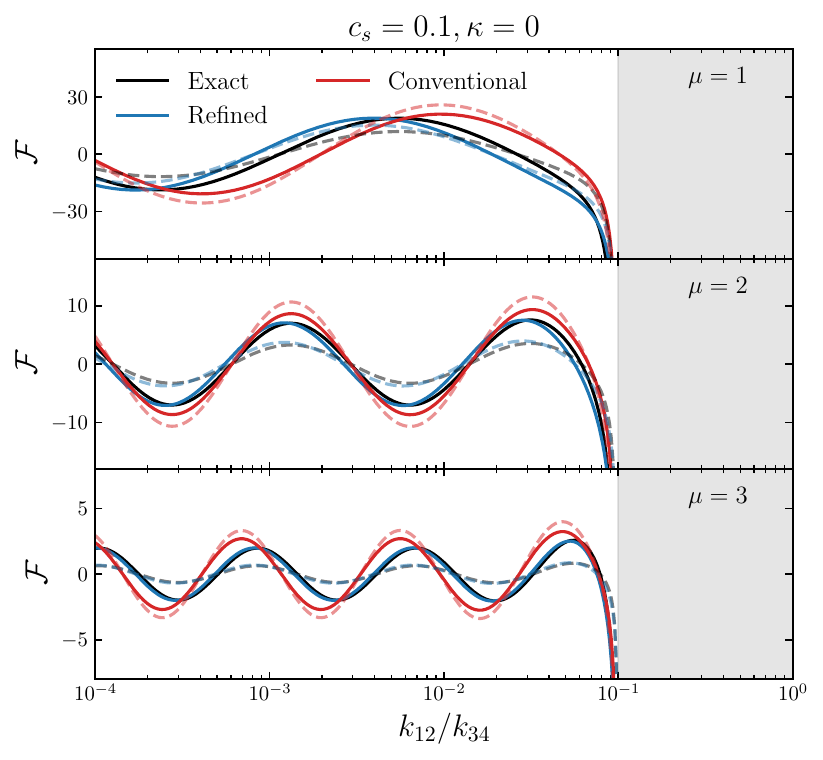}
    \end{subfigure}%
    \begin{subfigure}{.5\textwidth}
        \centering
        \includegraphics[width=1\linewidth]{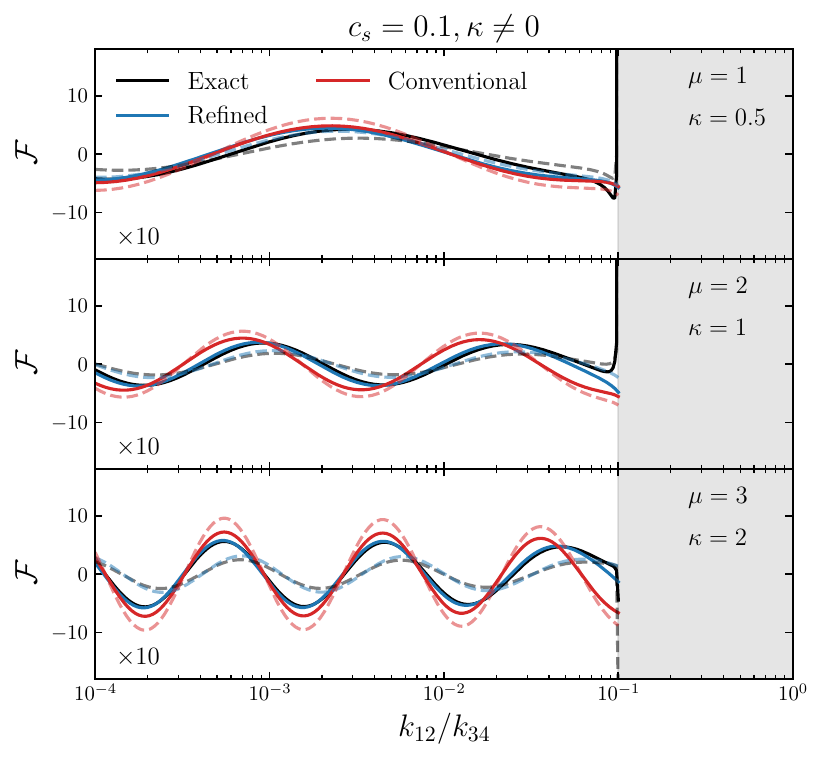}
    \end{subfigure}
   \caption{Dimensionless trispectrum waveform function $\F_s(k_{12}, k_{34}, s)$ in the $s$-channel, as defined in~\eqref{eq: trispectrum waveform def}, as function of the momentum ratio $k_{12}/k_{34}$ in the case of a reduced sound speed $c_s=0.1$ in the edge region (\textit{left panel}), and further turning on the chemical potential $\kappa\neq0$ (\textit{right panel}), varying the mass parameter $\mu=1, 2, 3$. The solid lines correspond to $k_{34}/s=1$ and the dashed line corresponds to $k_{34}/s=5$, closer to the interior region. We refer to Fig.~\ref{fig: Trispectrum template canonical} for details about the exact, approximate, and conventional templates. We have taken $\delta=3\pi/4$. The shaded \textcolor{gray}{grey} region corresponds to the diamond region in \textcolor{pyred}{red} shown in Fig.~\ref{fig: kinematic space with sound speed}.}
  \label{fig: Trispectrum template non-canonical}
\end{figure}
\end{itemize}

\paragraph{Bispectrum shape template.} The bispectrum can be obtained from the trispectrum by taking the soft limit $k_4\to0$. In practice, this amounts to set $k_{34}\to k_3$ and $s\to k_3$ by momentum conservation. The local and non-local signals become indistinguishable, and we refer to the resulting signal collectively as the cosmological collider signal. Following standard conventions, we define the dimensionless bispectrum shape function $S$ such that 
\begin{equation}
    \braket{\zeta_{\k_1} \zeta_{\k_2} \zeta_{\k_3}} \equiv (2\pi)^7 \delta^{(3)}(\k_1 + \k_2 + \k_3) \frac{\P_\zeta^2}{(k_1k_2k_3)^2} \, S(k_1, k_2, k_3)\,.
\end{equation}
Based on the previous dimensionless trispectrum, we propose the following boost-breaking bispectrum shape template
\begin{eBox}
\begin{equation}
\label{eq: bispectrum shape boost-breaking template}
    \begin{aligned}
        S^{\rm collider}(k_1, k_2, k_3) &= \B(\{k_i\}) \, P_S(\hat{\k}_1 \cdot \hat{\k}_3)\,e^{-\pi\mu/2-\mu \arcsin c_s} \\
        &\times \cos\left[\mu \arccosh\left(\frac{c_s k_{12}}{k_3}\right) + \delta\right] + {\rm perms.}\,,
    \end{aligned}
\end{equation}
\end{eBox}
where $\delta$ is $\mu$-dependent constant phase, and $\B(\{k_i\})$ is a form factor that depends on the considered cubic interaction, and in general has a polynomial dependence on the mass parameter $\mu$. As an example, the form factor from the $\dot{\varphi}^2\sigma-\dot{\varphi} \sigma$ scalar exchange reads $\B(\{k_i\}) = \mu^{3/2} k_1k_2/(k_{12})^2 [k_3/(k_{12})]^{1/2}$. Other precise momentum dependences can be found in~\cite{Noumi:2012vr}. Note that in the ultra-soft limit $k_3\to0$, one recovers the standard functional dependence $S^{\rm collider} \sim \cos[\mu \log(\tfrac{k_{12}}{k_3}) + \delta]$. The template~\eqref{eq: bispectrum shape boost-breaking template} is more accurate in the intermediate parameter range $\mu=\O(1)$ and $c_s \lesssim 1$, and also matches well the cosmological collider signal in mildly soft kinematic configurations. Notice that the bispectrum requires a linear mixing between the external and the exchanged massive fields, such as $\partial_{i_1 \ldots i_S} \varphi \sigma_{i_1 \ldots i_S}$. Therefore, only the longitudinal mode of the spinning particle can contribute to the bispectrum, while other components with $\lambda\neq 0$ are projected out by the linear mixing. This leads to the universal angular dependence $P_S(\hat{\k}_1 \cdot \hat{\k}_3)$ written in~\eqref{eq: bispectrum shape boost-breaking template} and no chemical potential. This refined cosmological collider template is both straightforward to implement and practically valuable. The fact that it consists solely of elementary functions significantly reduces computational costs. At the same time, it serves as an excellent approximation to the exact result in mildly soft kinematic regimes, where the signal-to-noise ratio is highest. Finally, it provides a smooth interpolation between the de Sitter invariant and strongly boost-breaking limits.

\section{Conclusions}
\label{sec: Conclusions}

In this paper, we have computed the most general tree-level boost-breaking seed correlator and have classified signals of cosmological collider physics based on our newly developed saddle-point method to evaluate bulk time integrals. The resulting analytical shapes offer valuable templates for cosmological collider studies. These template waveforms are particularly relevant for phenomenologically motivated scenarios with potential observational signatures in current and upcoming cosmological surveys.

\vskip 4pt
More specifically, we have focused on cosmological correlators sourced by the exchange of (spinning) particles and external inflatons with reduced speed of sound. This technically  extends the kinematically allowed configurations where standard series solutions fail to converge. 
To compute the correlators, we employ both the well-established bootstrap approach and a newly developed spectral method, each offering complementary advantages in terms of convergence and computational efficiency.
We then provide a detailed and handy recipe to extract and classify various cosmological collider signals based on expanding the internal mode using a WKB approximation and evaluating the time integrals with the saddle-point method. Notably, the saddle points in the complex time plane map directly onto the dynamics of massive fields, providing clear physical understanding of various bulk processes. We then showcase this method by studying in detail de Sitter-invariant and boost-breaking scenarios. The study on the motion of the saddle point in the complex plane driven by external kinematics reveals a deep connection between the amplitude of boundary kinematical discontinuities and the residue of bulk time singularities, giving a natural explanation to the fractional Boltzmann suppression of cosmological collider signals. By virtue of the saddle-point analysis, we also find novel new signals---e.g.~oscillating linearly in the momentum ratio with a frequency set by the product of the speed of sound and the chemical potential appearing in the edge regions---, as well as a refined waveform template for the standard cosmological collider signals. Importantly, after all boost-breaking effects are captured at tree-level, the resulting signals can be sizeable and potentially lie within reach of current or near-future cosmological observations.

\vskip 4pt
Eventually, our work opens several interesting avenues for future investigation:

\begin{itemize}
    \item {\bf Linking kinematic and time singularities.}  In Sec.~\ref{sec: Classifying non-Analyticities}, we noticed an intriguing connection between the amplitudes of the boundary kinematic singularities and poles in the bulk time. In all cases encountered so far, the discontinuity along the branch cut of kinematic variables are dictated by the exponential of the residue of the dispersion relation at time-zero or infinity. We believe such a connection is no mere coincidence, but reflects a deep correspondence between the boundary and the bulk, which certainly deserves further exploration in the future.
    \item {\bf Towards saddle points in loops.} The saddle-point analysis for cosmological collider signals we adopted in this paper is exclusively focused on tree diagrams. At loop level, one needs to  incorporate momentum integrals in addition to time integrals. Given that the cutting rule still applies in loop diagrams \cite{Qin:2023bjk,Ema:2024hkj}, it is conceivable that a similar saddle-point analysis can be performed there, yielding simplified shape templates for cosmological collider signals at loop level. We leave this interesting possibility for future work.
    \item {\bf Breaking scale invariance.} In this work, we computed the most general boost-breaking correlator using different of analytical and approximation methods. An important and complementary direction for future research involves breaking of scale invariance (also known as the dS dilation), which is commonly violated in many inflationary models with features. Such breaking of scale invariance often leads to enhanced cosmological collider signals \cite{Chen:2022vzh,Werth:2023pfl,Pinol:2023oux,Pajer:2024ckd,Wang:2025qww}. It is therefore of significant interest to develop a general analytical framework for studying these symmetry-breaking scenarios, as well as to extend our approximation methods to gain deeper physical insights into such cases.
    \item {\bf Searching with refined templates.} Our approximation scheme produces many refined templates for cosmological collider signals that are ready to be searched for in present and upcoming data. In particular, theories with broken de Sitter boosts naturally allow for a large amplitude of the trispectrum, over which we now have precise and efficient analytical control. It is thus interesting to put our templates to the test.
\end{itemize}

\paragraph{Acknowledgements.} We thank Shuntaro Aoki, Xingang Chen, Thomas Colas, Sadra Jazayeri, Fumiya Sano, David Stefanyszyn, Yi Wang, Dong-Gang Wang, Zhong-Zhi Xianyu, Masahide Yamaguchi, and Bowei Zhang for helpful discussions. Z.Q. is supported by NSFC under Grant No. 12275146, the National Key R\&D Program of China (2021YFC2203100), and the Dushi Program of Tsinghua University. During the course of this work, S.R-P. and D.W. were supported by the European Research Council under the European Union’s Horizon 2020 research and innovation programme (grant agreement No 758792, Starting Grant project GEODESI). X.T. is supported by STFC consolidated grants ST/T000694/1 and ST/X000664/1. Y.Z. is supported by the IBS under the project code, IBS-R018-D3.
This article is distributed under the Creative Commons Attribution International Licence (\href{https://creativecommons.org/licenses/by/4.0/}{CC-BY 4.0}).


\newpage

\appendix

\section{Mathematical Interlude}
\label{app: Mathematical Interlude}

Classifying non-analyticities of cosmological correlators employs tools from complex analysis. To ensure this paper remains self-contained and accessible, this appendix offers a pedagogical and elementary introduction to the theory of Riemann surfaces and the Saddle point approximation. Emphasis is placed on clarity and intuition rather than formal mathematical rigour.

\subsection{Riemann Surfaces from Multi-Valued Functions}

The study of differentiable functions of a single complex variable naturally unveils the concept of multi-valued functions. This, in turn, introduces an elegant class of mathematical objects known as Riemann surfaces, which beautifully connects analysis to geometry.\footnote{For a general introduction to Riemann surfaces, see e.g.~\cite{RiemannSurfaces}.} The intuitive idea behind Riemann surfaces is to take multiple branches of a complex multi-valued function and glue these together to obtain a continuous single-valued function. The main question we will try answer is: ``What do we really mean by multi-valued functions?".

\paragraph{Complex logarithm.} Let us take the canonical example of the complex logarithm function to introduce the concept of multi-valued functions and the corresponding Riemann surface. We try to extend the real logarithm $x \in \mathbb{R}_+^* \mapsto \log(x)$ to the largest possible open, connected subset of the complex plane $\mathbb{C}$. The most natural way to proceed is to come back to the defining property of the real logarithmic function: it is the primitive of $x \in \mathbb{R}^* \mapsto 1/x$ that vanishes at $x=1$. We therefore define
\begin{equation}
\label{eq: complex log}
    \log(z) = \int_{\C(z)} \frac{\d z'}{z'}\,, \quad \text{for} \quad z \in \mathbb{C}^*\,,
\end{equation}
\begin{wrapfigure}{r}{0.35\textwidth}  
    ~\vspace{-12mm}\\ 
    \begin{center}
        \begin{tikzpicture}
            \draw[black, ->] (-1,0) -- (2,0) coordinate (xaxis);
		\draw[black, ->] (0,-1) -- (0,2) coordinate (yaxis);
		\node at (2.7, 0) {$\Re z'$};
		\node at (0, 2.3) {$\Im z'$};

            \draw[thick, pyblue, ->] (1, 0) .. controls (1.5, 0.5) and (0.5, 1.2) .. (-0.5, 1) node[pos=0.5, above right] {$\C_1(z)$};

            \draw[thick, pyred, ->] (1, 0) .. controls (0, -0.5) and (-1, 0) .. (-0.5, 1) node[pos=0.4, below right] {$\C_2(z)$};

            \draw[black, fill = black] (1, 0) circle (.03cm) node[below right] {$1$};
            \draw[black, fill = black] (-0.5, 1) circle (.03cm) node[above] {$z$};
        \end{tikzpicture}
    \end{center}
    \label{fig: complex log contours}
\end{wrapfigure}
where $\mathcal{C}(z)$ is a contour linking $1$ to $z$ in the complex plane. Eq.~(\ref{eq: complex log}) defines a function if every choice of the contour $\C(z)$ leads to the same value for $\log(z)$. Let us consider two contours $\C_1(z)$ and $\C_2(z)$ linking $1$ to $z$, but passing on either side of the origin, as depicted on the right. If we denote $\C$ the closed contour composed of $\C_1$ and $-\C_2$, then using the residue theorem, we obtain $\int_{\C(z)} \tfrac{\d z'}{z'} = 2i\pi$. This yields
\begin{equation}
    \int_{\C_1(z)} \frac{\d z'}{z'} = \int_{\C_2(z)} \frac{\d z'}{z'} + 2i\pi\,.
\end{equation}
It is therefore not possible to define the complex logarithm on the entire punctured complex plane $\mathbb{C} \backslash \{0\}$. To avoid such inconvenience, we restrict the argument of $z$, that we denote $\arg(z)$, to $\arg(z) \in (\alpha, \alpha+2\pi)$ with real $\alpha$. The line $\{z \, | \, \arg(z)=\alpha\}$ is a branch cut, with the origin $z=0$ being the branch point, i.e.~the common point to all branch cuts. Choosing $\alpha=-\pi$ so that the branch cut is the negative real axis $\mathbb{R}_-$ defines the principal branch of the complex logarithm. However, note that any other value for $\alpha$, defining a line $\mathfrak{C}$ linking the origin to infinity, would give a valid branch cut and a new complex logarithm can be defined on $\mathcal{C} \backslash \mathfrak{C}$. A natural consequence is if we consider two complex logarithms $\ell(z)$ and $\ell'(z)$ defined on $\mathbb{C} \backslash \mathfrak{C}$ and $\mathbb{C} \backslash \mathfrak{C}'$, respectively, the difference $\ell(z) - \ell'(z)$ is a multiple of $2i\pi$ for all $z \in \mathcal{C} \backslash \{\mathfrak{C} \cup \mathfrak{C}'\}$. 

\vskip 4pt
Importantly, a different point of view removes the need to introduce a branch cut to define the complex logarithm. In fact, when physicists' point of view is that of a multi-valued function, the mathematicians' one is that of a function defined on a Riemann surface. 

\vskip 4pt
\begin{wrapfigure}{r}{0.35\textwidth}  
    ~\vspace{-12mm}\\ 
    \begin{center}
        \includegraphics[scale=1]{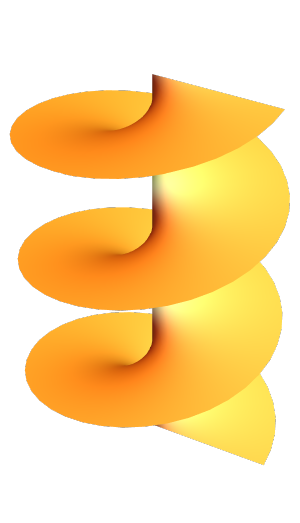}
    \end{center}
    \label{fig: log Riemann surface}
\end{wrapfigure}
Introducing a Riemann surface corresponds to replacing the punctured complex plane $\mathbb{C} \backslash \{0\}$, on which we tried without succeeding to define a complex holomorphic logarithm, by a larger connected domain. Intuitively, starting from the principal branch, rotating counter-clockwise once around the origin would pick a factor $2i\pi$ to the logarithm to end up on the next branch, and so on. We therefore can construct the following surface: take an infinite number of complex planes that we stack on top of each other. Each of these planes are cut on a branch cut, say the negative real axis $\mathbb{R}_-$, and the upper branch cut is glued to the lower one. The final result is a connected surface---a helicoid---resembling a spiralling parking garage with infinitely many levels, as shown on the right: it is the \textit{Riemann surface} of the complex logarithm where the $z$-axis is the imaginary part of $\log(z)$. A point on this Riemann surface $\mathfrak{R}$ can be thought of as a pair $(z, n)$, with $n\in\mathbb{Z}$ corresponding to a Riemann sheet for which the complex logarithm would give $\log(r) + i(\theta + 2\pi n)$, where $z=r e^{i\theta}$ with $r\in \mathbb{R}_+^*$ and $\theta = \arg(z)$. As such, the Riemann surface can be embedded in $\mathbb{C} \times \mathbb{Z}$. On $\mathfrak{R}$, the complex logarithm can be defined as follows: (i) we choose a contour $\C$ linking $(1, 0) \in \mathfrak{R}$ to the point $(z, n) \in \mathfrak{R}$, (ii) this contour can be projected vertically on the principal branch $n=0$, resulting in a contour joining $1$ to $z$, and (iii) we define $\log(z) = \int_\C \d z'/z'$. This way, the function does not depend on the contour any more as whenever we circle around the origin, we move to another branch and the logarithm picks up a factor $\pm 2i\pi$.

\paragraph{Complex square root.} 
\begin{wrapfigure}{r}{0.35\textwidth}  
    ~\vspace{-12mm}\\ 
    \begin{center}
        \begin{tikzpicture}
            \draw[black, ->] (-2,0) -- (2,0) coordinate (xaxis);
		\draw[black, ->] (0,-2) -- (0,2) coordinate (yaxis);
		\node at (2.7, 0) {$\Re z$};
		\node at (0, 2.3) {$\Im z$};

            \draw[-, decorate, decoration={zigzag, segment length=6, amplitude=2}, black, line width=0.2mm] (-2, 0) -- (0, 0);
            \draw[black, fill = black] (0, 0) circle (.05cm);

            \path[draw, line width = 0.8pt, gray, postaction = decorate, decoration={markings, mark=at position 1 with {\arrow[line width=1pt]{>}}}] (0.8, 0)  arc (0:165:0.8);

            \path[draw, line width = 0.8pt, gray, postaction = decorate, decoration={markings, mark=at position 1 with {\arrow[line width=1pt]{>}}}] (0.8, 0)  arc (0:-165:0.8);

            \node at (-1.5, 0.5) {$\pmm i\sqrt{r}$};
            \node at (-1.5, -0.5) {$\ppm i\sqrt{r}$};

            \node at (1.6, 1.5) {$\textcolor{pyblue}{f_>(z)}, \textcolor{pyred}{f_<(z)}$};
        \end{tikzpicture}
    \end{center}
\end{wrapfigure}
Another classic example of a function that presents challenges when extending its domain from the positive real axis to the complex plane is the square root. Indeed, defining $z = r e^{i\theta}$ with $r\in \mathbb{R}_+^*$ and $\theta = \arg(z)$, a naive definition of the square root would simply be $\sqrt{z} = \sqrt{r} e^{i\theta/2}$. However, the argument $\theta/2$ is only defined up to a factor $\pi$, resulting in a sign ambiguity on the sign of $\sqrt{z}$. This function is multi-valued: for a single value of $z$, there are two different choices for $\sqrt{z}$. We define these two choices by
\begin{equation}
    f_>(z) = \sqrt{r}e^{i\theta/2}\,, \quad f_<(z) = \sqrt{r}e^{i(\theta/2 + \pi)}\,,
\end{equation}
where $f_>$ and $f_<$ are the positive and negative branches of the square root, respectively. As we observed earlier with the complex logarithm, it is necessary to define a branch cut. From now on, we will adopt the principal branch cut, positioned along the negative real axis $\mathbb{R}_-^*$, i.e.~$\theta = \arg(z) \in (-\pi, \pi)$. Let us now look at the behaviour of each of the two branches $f_>(z)$ and $f_<(z)$ as $z$ approaches the branch cut from above and from below on a circle centred at the origin with a radius $r>0$. Taking the limit $\theta \to +\pi$, we obtain $f_>(z) = +i\sqrt{r}$ and $f_<(z) = -i\sqrt{r}$. Similarly, as $\theta \to -\pi$, we have $f_>(z) = -i\sqrt{r}$ and $f_<(z) = +i\sqrt{r}$. We represent these limits on the right for the \textcolor{pyblue}{positive} and \textcolor{pyred}{negative} branches. 

\vskip 4pt
\begin{wrapfigure}{r}{0.35\textwidth}  
    ~\vspace{-10mm}\\ 
    \begin{center}
        \includegraphics[scale=0.6]{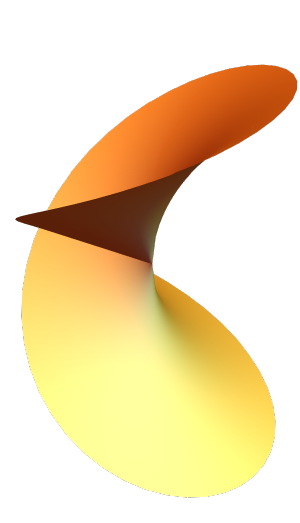}
    \end{center}
    \label{fig: sqrt Riemann surface}
\end{wrapfigure}
To define a function on the punctured complex plane $\mathbb{C}\backslash\{0\}$ that remains continuous along the branch cut, we observe that the sign is reversed after a single rotation around the origin but returns to its original value after two full rotations. 
We can e.g.~approach the branch cut from above using the positive branch $f_>(z)$ and exit from below using the negative branch $f_<(z)$. The corresponding Riemann surface for the complex square root is depicted on the right: when the square root takes a specific value on one Riemann sheet, it assumes the opposite value on the other sheet. We end up with a self-intersecting surface.

\paragraph{Constructing Riemann surfaces.} Whenever a function is multi-valued, it can be promoted to a single-valued function albeit extending the domain of validity to the corresponding Riemann surface. For pedagogical purposes, let us now explicitly construct the Riemann surface corresponding to the function describing the time-dependent frequency of a massive field scalar field in de Sitter: (dimensionless) frequencies $\omega$ such as
\begin{equation}
    \omega^2(z) = 1 + \frac{1}{z^2}\,.
\end{equation}
The function $\omega(z)$ has a pole at the origin $z=0$ and two branch points at $z=\pm i$. Taking the square root principal branch, we place the branch cut on the imaginary axis between $+i$ and $-i$. When trying to extend the domain of validity of the frequency $\omega(z)$ to the largest possible domain of the complex plane of $z$, we note that this function has two branches corresponding to the positive and negative branches of the square root, that we denote
\begin{equation}
    \omega_>(z) = +\sqrt{1 + \frac{1}{z^2}}\,, \quad \omega_<(z) = -\sqrt{1 + \frac{1}{z^2}}\,,
\end{equation}
that are defined on two distinct Riemann sheets. We now probe the behaviour of both branches on each sides of both branch cuts. On a small circle with infinitesimal radius $\epsilon>0$ centred at the origin, we have $\omega_>(\epsilon e^{i\theta}) \approx e^{-i\theta}/\epsilon \to \mp i/\epsilon$ as $\theta\to\pm\pi/2$. Similarly for the negative branch, we obtain $\omega_<(\epsilon e^{i\theta}) \approx e^{-i(\theta-\pi)}/\epsilon \to \pm i/\epsilon$ as $\theta \to \pm \pi/2$, which can be visualised as follows:
\begin{equation*}
    \begin{tikzpicture}
        \draw[black, ->] (-3,0) -- (3,0) coordinate (xaxis);
	\draw[black, ->] (0,-3) -- (0,3) coordinate (yaxis);
	\node at (3.6, 0) {$\Re z$};
	\node at (0, 3.3) {$\Im z$};

        \draw[-, decorate, decoration={zigzag, segment length=6, amplitude=2}, black, line width=0.2mm] (0, 2) -- (0, -2);
        \draw[black, fill = black] (0, 2) circle (.05cm);
        \draw[black, fill = black] (0, -2) circle (.05cm);

        \draw[black, fill = black] (0, 0) circle (.05cm);

        \path[draw, line width = 0.8pt, gray, postaction = decorate, decoration={markings, mark=at position 1 with {\arrow[line width=1pt]{>}}}] (0.8, 0)  arc (0:80:0.8);

        \path[draw, line width = 0.8pt, gray, postaction = decorate, decoration={markings, mark=at position 1 with {\arrow[line width=1pt]{>}}}] (0.8, 0)  arc (0:-80:0.8);

        \path[draw, line width = 0.8pt, gray, postaction = decorate, decoration={markings, mark=at position 1 with {\arrow[line width=1pt]{>}}}] (-0.8, 0)  arc (0:80:-0.8);

        \path[draw, line width = 0.8pt, gray, postaction = decorate, decoration={markings, mark=at position 1 with {\arrow[line width=1pt]{>}}}] (-0.8, 0)  arc (0:-80:-0.8);

        \node at (1, 1) {$\ppm i/\epsilon$};
        \node at (1, -1) {$\pmm i/\epsilon$};

        \node at (-1, 1) {$\pmm i/\epsilon$};
        \node at (-1, -1) {$\ppm i/\epsilon$};

        \node at (2.5, 2) {$\textcolor{pyblue}{\omega_>(z)}, \textcolor{pyred}{\omega_<(z)}$};
    \end{tikzpicture}
\end{equation*}
The Riemann surface can be constructed by noticing that we can continuously enter the upper portion of the branch cut from the left using the positive branch to exit the lower portion of the branch cut using the negative branch cut. However, when crossing the imaginary axis away from the branch cut, we remain on the same Riemann sheet. We obtain the Riemann surface used in the main text.

\subsection{Saddle Point Approximation of Complex Integrals}

Local behaviours of integrals in certain limits can completely determine their asymptotic forms without introducing free parameters. This is in sharp contrast with a local study of differential equations around singular points where solutions are always found up to arbitrary integration constants. Among these local analysis, the saddle point approximation---also called the method of steepest descent---is a method to find approximate solutions to complex integrals where one deforms the integration contour to pass a saddle (stationary) point in the direction of steepest descent or stationary phase. We invite the reader to consult e.g.~\cite{Bruijn,bender78:AMM} for more details.

\paragraph{Preliminary example.} The main idea of the saddle point approximation is to deform the integration contour in the complex plane so that the integrand has a {\it constant phase} on the new contour. Before formally exposing the method, we consider a preliminary example to motivate why it is helpful to shift complex contours to perform an asymptotic analysis. Let us consider the following integral:
\begin{equation}
\label{eq: saddle point example}
    I(\mu) = \int_0^1 \d t \, \log(t) e^{i \mu t}\,.
\end{equation}
\begin{wrapfigure}{r}{0.35\textwidth}  
    ~\vspace{-20mm}\\ 
    \begin{center}
        \begin{tikzpicture}
            \draw[black, ->] (-1,0) -- (2,0) coordinate (xaxis);
		\draw[black, ->] (0,-1) -- (0,2) coordinate (yaxis);
		\node at (2.7, 0) {$\Re t$};
		\node at (0, 2.3) {$\Im t$};

            \draw[thick, pyblue, ->] (0, 0) -- (1.5, 0);

            \draw[thick, pyred, ->] (0, 0) -- (0, 1.5) node[pos=0.4, left] {$\C_1$};

            \draw[thick, pyred, ->] (0, 1.5) -- (1.5, 1.5) node[pos=0.4, above] {$\C_2$};

            \draw[thick, pyred, ->] (1.5, 1.5) -- (1.5, 0) node[pos=0.4, right] {$\C_3$};

            \draw[black, fill = black] (1.5, 0) circle (.03cm) node[below] {$1$};
            \draw[black, fill = black] (0, 1.5) circle (.03cm) node[left] {$t_\star$};
        \end{tikzpicture}
    \end{center}
    ~\vspace{-15mm}
\end{wrapfigure}
As $\mu \gg 1$, the asymptotic expansion of $I(\mu)$ is non-analytic, which complicates finding an approximate solution. We therefore deform the integration interval in \textcolor{pyblue}{blue} to the complex contour $\C = \C_1 + \C_2 + \C_3$ as depicted in \textcolor{pyred}{red} on the right. By Cauchy's theorem, we have $I(\mu) = \int_{\C_1 + \C_2 + \C_3} \d z \, \log(z) \, e^{i\mu z}$. As $t_\star \to \infty$, the integral over $\C_2$ vanishes exponentially. The integral~(\ref{eq: saddle point example}) becomes
\begin{equation}
    I(\mu) = i \int_0^\infty \d z \, \log(iz)\, e^{-\mu z} - i\int_0^\infty \d z \, \log(1+iz) \, e^{i\mu(1+iz)}\,.
\end{equation}
The first integral can be performed exactly, yielding
\begin{equation}
    i \int_0^\infty \d z \, \log(iz)\, e^{-\mu z} = -i \, \frac{\log(\mu)}{\mu} - \frac{i\gamma + \pi/2}{\mu}\,,
\end{equation}
where we have used the principal branch of the complex logarithm $\log(iz) = \log(z)+i\pi/2$ and $\int_0^\infty\log(t)e^{-t}\d t = -\gamma$ with $\gamma = 0.5772\ldots$ the Euler's constant. The second integral can also be performed by Taylor expanding the logarithm and integrating each term in the series
\begin{equation}
    i\int_0^\infty \d z \, \log(1+iz) \, e^{i\mu(1+iz)} = ie^{i\mu} \, \sum_{n=1}^\infty \frac{(-i)^n (n-1)!}{\mu^{n+1}}\,.
\end{equation}
Importantly, notice that along the contours $\C_1$ and $\C_3$, the phase of $e^{i\mu t}$ is constant, with frequency $0$ and $1$, respectively. These contours are also the steepest-descent paths as $|e^{i\mu t}|$ decreases most rapidly (in fact exponentially) as $t$ ranges from $0$ and $1$ to $\infty$. This is no coincidence as we now review.

\paragraph{Steepest-descent complex contours passing through saddle points.} The method of steepest descent enables one to approximate {\it complex} integrals of the form
\begin{equation}
    I(\mu) = \int_\C \d z \, g(z) e^{\mu f(z)}\,,
\end{equation}
in the limit $\mu \gg 1$. The contour $\C$ can be deformed into a new contour $\C'$ without changing the integral provided no poles nor branch cuts are crossed. If one chooses a path that satisfies the following conditions: (i) the phase $\Im[f(z)] \equiv \uptheta$ is constant along $\C'$ and (ii) the contour $\C'$ passes through one or more saddle points $z_\bullet$ such that $f'(z_\bullet) = 0$, then the integral reads
\begin{equation}
\label{eq: Saddle point method formula}
    I(\mu) = e^{i\mu \Im[f(z)]} \int_{\C'} \d z \, g(z) \, e^{\mu \Re[f(z)]} \approx e^{i\mu \uptheta}  \sqrt{\frac{2\pi}{\mu \tfrac{\d^2}{\d z^2}\Re[f]|_{z_\bullet}}} \, g(z_\bullet) \, e^{\mu \Re[f(z_\bullet)]}\,,
\end{equation}
where the remaining integral is approximated using Laplace's method, which follows from Taylor expanding the function $\Re[f]$ around the stationary point $z=z_\bullet$ and finishing the Gaussian integral by extending the limits of integration by $\pm \infty$.\footnote{This creates only a small error as the exponential decays very fast away from the saddle point.} Extensions to arbitrary precision can be systematically found. In the presence of multiple saddles, one needs to sum their contributions. However, it is in general expected that the saddle points that are located the closest to the original integration contour contribute the most. Note that if the function $f$ is analytic, constant phase contours are equivalent to steepest-descent contours. Indeed, setting $z=x+iy$, using the Cauchy-Riemann equations, we obtain
\begin{equation}
    \frac{\partial \Re[f]}{\partial x} \frac{\partial \Im[f]}{\partial x} + \frac{\partial \Re[f]}{\partial y} \frac{\partial \Im[f]}{\partial y} = 0 \quad \Leftrightarrow \quad \nabla \Re[f] \cdot \nabla \Im[f] = 0\,,
\end{equation}
which shows that $\Im[f]$ is constant on contours whose tangents are parallel to $\nabla \Re[f]$. Two distinct steepest curves can only intersect at saddle points.

\paragraph{Illustrative example.} We now apply this method to a concrete example. Let us derive the asymptotic expansion of the integral
\begin{equation}
    I(\mu) = \int_\C \frac{\d s}{2i\pi} \, e^s s^{-\mu}\,,
\end{equation}
in the limit $\mu\gg 1$, where $\C$ is the Hankel contour that begins at $-\infty-i\epsilon$, encircles the branch cut lying on the real negative axis and ends up at $-\infty+i\epsilon$ with $\epsilon>0$. We first make the change of variables $s = \mu z$ to convert the movable saddle point that depends on $\mu$ to a fixed one, yielding
\begin{equation}
    I(\mu) = \frac{1}{\mu^{\mu-1}} \int_\C \frac{\d z}{2i\pi} \, e^{\mu(z - \log z)}\,.
\end{equation}
Defining $f(z) = z - \log z$, solving $f'(z) = 0$ gives a single saddle point at $z_\bullet = 1$. To study the structure of this saddle, we set $z = x+iy$ so that $f(z) = x - \log\sqrt{x^2+y^2} + i(y - \arctan y/x)$. Therefore, paths of constant phase emerging from $z_\bullet=1$ (as $f(z_\bullet)=1$) satisfy
\begin{equation}
    y = \arctan y/x\,.
\end{equation}
\begin{wrapfigure}{r}{0.35\textwidth}  
    \begin{center}
        \begin{tikzpicture}
            \draw[black, ->] (-2,0) -- (2,0) coordinate (xaxis);
		\draw[black, ->] (0,-2) -- (0,2) coordinate (yaxis);
		\node at (2.7, 0) {$\Re z$};
		\node at (0, 2.3) {$\Im z$};

            \draw[-, decorate, decoration={zigzag, segment length=6, amplitude=2}, black, line width=0.2mm] (-2, 0) -- (0, 0);
            \draw[black, fill = black] (0, 0) circle (.05cm);

            \draw[pyred, thick, postaction = decorate, decoration={markings,
			mark=at position 0.6 with {\arrow[line width=1pt]{>}}}, domain=-1.1:1.1, samples=200, variable=\t, -] plot ({1-1/20*atan(\t)*\t}, {\t}) node[above right] {$\C'$};

            \path[draw, line width = 0.8pt, pyblue] (0.2, 0)  arc (0:90:0.2);
		\path[draw, line width = 0.8pt, pyblue] (0.2, 0)  arc (0:-90:0.2);
            \path[draw, line width = 0.8pt, pyblue, postaction = decorate, decoration={markings,
			mark=at position 0.5 with {\arrow[line width=1pt]{>}}}] (0, 0.2)  -- (-2, 0.2) node[above right] {$\C$};
		\path[draw, line width = 0.8pt, pyblue, decoration] (-2, -0.2)  -- (0, -0.2);

            \draw[black, fill = black] (1, 0) circle (.03cm) node[above right] {$1$};
        \end{tikzpicture}
    \end{center}
\end{wrapfigure}
There are two solutions to this equation: $y=0$ and $x = y/\tan(y)$. Now we can deform the original contour $\C$ in \textcolor{pyblue}{blue} to $\C'$ represented in \textcolor{pyred}{red} that passes by the saddle.\footnote{It is not possible to continuously deform the original contour to the steepest-descent path spanning the positive real axis.} The constant-phase (or steepest-descent) curve $\C' = \{(t/\tan(t), t)|t \in \mathbb{R}\}$ is parallel to $\Im[z]=y$ around $z_\bullet=1$ and crosses the imaginary axis at $\pm i\pi/2$. Along this path, $|e^{\mu(z-\log z)}|$ reaches its maximum at the saddle and decays exponentially as $z\to -\infty \pm i \pi$. In the case of no saddle point, $|e^{\mu(z-\log z)}|$ would decrease in one direction along $\C'$ and increase in the other direction, and the integral would not be convergent. Note that if instead the steepest-descent path is finite and does not pass through a saddle point, then the maximum of the integrand must occur at an endpoint of the contour. To finish the integral, we examine the neighbourhood of $z_\bullet$ along $\C'$ by setting $z = 1 + it$ with $t\ll1$, obtaining
\begin{equation}
    I(\mu) \approx \frac{e^{\mu}}{\mu^{\mu-1}}\int_{-\epsilon}^{+\epsilon} \frac{\d t}{2\pi} \, e^{-t^2/2} \approx \frac{e^{\mu}}{\mu^{\mu-1}} \frac{1}{\sqrt{2\pi\mu}}\,,
\end{equation}
where we have let $\epsilon \to \infty$. Note that it is important to stay on the constant-phase contour as naively applying~(\ref{eq: Saddle point method formula}) would have given an overall wrong phase $e^{-i\pi/2}$.

\paragraph{Application to massive field in de Sitter.} We now treat in detail the simplest case of a massive scalar field in de Sitter with zero chemical potential $\kappa=0$ and unity sound speed $c_s=1$, studied in Sec.~\ref{subsubsec: Unit Sound Speed & No Chemical Potential}. After applying the leading-order WKB approximation to the massive mode function and discarding the negative-frequency integral as it is Boltzmann suppressed, we are left with the following integral
\begin{equation}
\label{eq: J12alpha integral}
    F_L^{(\alpha)} = \int_{-\infty^+}^0 \frac{\d\tau}{\sqrt{2\omega_s(\tau)}} \, e^{i\left(k_{12}\tau - \int_{\tau_i}^\tau \omega_s(\tau')\d\tau'\right)}\,,
\end{equation}
where $\tau_i$ is an arbitrary time in the infinite past, and the effective frequency $\omega_s(\tau)$ is defined by
\begin{equation}
    \omega^2_s(\tau) = s^2 + \frac{\mu^2}{\tau^2} \,.
\end{equation}
The saddle points are found to $\tau_\bullet = \pm \mu/\sqrt{k_{12}^2 - s^2}$ which lie on the real axis in the complex $\tau$ plane as $k_{12}> s$ by the triangle inequality. The closest saddle point to the original integration contour is the one located on the negative real axis, so we expect its contribution to dominate the integral. Let us now evaluate the Dingle's singulant function
\begin{equation}
    F_s(\tau) = -2i \int_{\tau_c}^\tau \omega_s(\tau')\d\tau'\,,
\end{equation}
with $\tau_c = +i\mu/s$ the critical point (with negative imaginary part) that makes the effective frequency vanish. Selecting the positive branch of the square root, we obtain
\begin{equation}
    \begin{aligned}
        F_s(\tau) &= -2i \left[\tau \sqrt{s^2 + \mu^2/\tau^2} + \mu \, \arctanh\left(\frac{-\mu}{\tau \sqrt{s^2 + \mu^2/\tau^2}}\right)\right]_{-i\mu/s}^\tau \\
        &= -2i \left(\tau \sqrt{s^2 + \mu^2/\tau^2} + \mu \, \arctanh\left(\frac{-\mu}{\tau \sqrt{s^2 + \mu^2/\tau^2}}\right)\right) + \pi \mu\,.
    \end{aligned}
\end{equation}
After splitting the integral as $\int_{\tau_c}^\tau = \int_{\tau_i}^\tau - \int_{\tau_i}^{\tau_c}$, the exponent in the integral~\eqref{eq: J12alpha integral} therefore can be written in terms of $F(\tau)$ as
\begin{equation}
\label{eq: saddle point exponent}
    f(\tau) = ik_{12}\tau + \tfrac{1}{2}F_s(\tau) - \pi\mu/2 \,,
\end{equation}
where we have discarded the contribution coming from $\tau_i$ as it is arbitrary. 
Let us now study the steepest-descent curve passing through the saddle $\tau_\bullet$. To do that, we Taylor expand the exponent in the neighbourhood of the saddle $\tau_\bullet$
\begin{equation}
    f(\tau_\bullet + \epsilon) \approx f(\tau_\bullet) - i \alpha \epsilon^2\,,
\end{equation}
with $|\epsilon| \ll 1$, where $f(\tau_\bullet) = -i\mu \, \arctanh\sqrt{k_{12}^2-s^2}/k_{12} \equiv i\beta$ and $\alpha \equiv (k_{12}^2-s^2)^{3/2}/2k_{12}\mu$. We set $\epsilon = x+iy$ and obtain
\begin{equation}
    f(\tau_\bullet + \epsilon) \approx  2\alpha x y + i [\beta - \alpha(x^2-y^2)]\,.
\end{equation}
Since $f(\tau_\bullet)$ is purely imaginary with a negative imaginary component, the constant-phase curve crossing the saddle must satisfy the following equation
\begin{equation}
    \beta - \alpha (x^2-y^2) = -\lambda\,,
\end{equation}
\begin{wrapfigure}{r}{0.35\textwidth}  
    \begin{center}
        \includegraphics[scale=0.6]{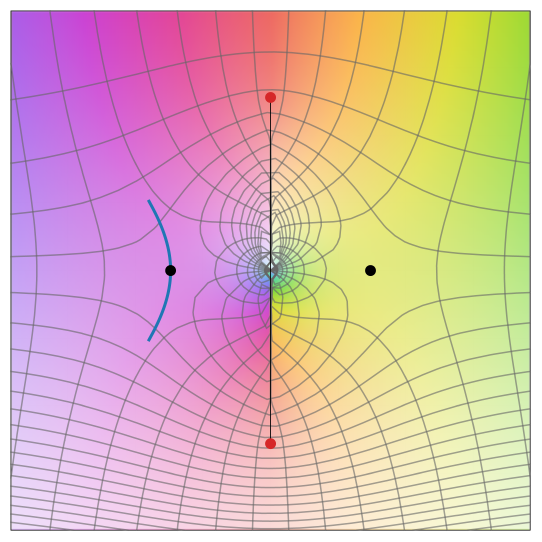}
    \end{center}
\end{wrapfigure}
where $\lambda >0$ is a free parameter that labels the different constant-phase curves. For $\tau_\bullet<0$, the solution to this equation is $x = -\sqrt{y^2 + (\beta+\lambda)/\alpha}$, and the constant-phase curves in the neighbourhood of the saddle are
\begin{equation}
    \gamma_\lambda(t) = \{(-\sqrt{t^2+(\beta+\lambda)/\alpha}, t)\,|\,t\in \mathbb{R}\}\,.
\end{equation}
The affine parameter where the curve crosses the saddle is $t_\bullet=0$ and the label selecting this particular curve is found to be $\lambda = \mu\sqrt{k_{12}^2-s^2}/2k_{12} + \mu \, \arctanh\sqrt{k_{12}^2-s^2}/k_{12}$. The tangent vector at the saddle point is found to be $\bm{T} = (0, 1)$. Therefore, the curve is parallel to the imaginary axis. We illustrate the complex $\tau$ plane on the right, where the two saddle points are depicted in black, the critical points in \textcolor{pyred}{red} along with the branch cut joining them, and the constant-phase curve passing by the saddle in \textcolor{pyblue}{blue} (whose expression is only found in the neighbourhood of $\tau_\bullet$). Additional constant-phase curves are shown in \textcolor{gray}{gray}. To expand the exponent~\eqref{eq: saddle point exponent} around the saddle point along the steepest-descent path, we set $\tau = \tau_\bullet - it$ with $t\ll 1$ as the original contour starts at $-\infty^+$. We obtain
\begin{equation}
    \begin{aligned}
        F_L^{(\alpha)} &= \frac{-i e^{-i\mu\, \arctanh\tfrac{\sqrt{k_{12}^2-s^2}}{k_{12}}}}{\sqrt{2\omega_s(\tau_\bullet)}} \int_{-\epsilon}^{+\epsilon} \d t \, e^{i\,\frac{(k_{12}^2-s^2)^{3/2}}{2k_{12}\mu}\, t^2} \\
        &\approx \frac{\sqrt{\pi\mu}}{(k_{12}^2-s^2)^{3/4}} e^{-i\pi/4} \exp\left[-i\mu \,  \arccosh\frac{k_{12}}{s}\right]\,,
    \end{aligned}
\end{equation}
where we have finished the Gaussian integral after setting $\epsilon\to \infty$. Note that integrating \textit{along} the constant-phase curve is essential to obtain the correct phase.

\newpage
\phantomsection
\addcontentsline{toc}{section}{References}
\small
\bibliographystyle{utphys}
\bibliography{references}

\end{document}